\documentclass[prc,nofootinbib]{revtex4}
\usepackage{graphicx,subfigure}  
\usepackage{dcolumn}   
\usepackage{bm} 
\usepackage{sidecap} 
\usepackage{amssymb}   
\usepackage{amsbsy}    
\usepackage{amsmath} 
\usepackage[utf8]{inputenc}

\begin{document}

\newcommand{\mnras}{MNRAS}
\newcommand{\cqg}{CQG}
\newcommand{\apjs}{ApJS}
\newcommand{\araa}{ARAA}
\newcommand{\aap}{A\&A}

\title{FRIB and the GW170817 Kilonova}

\author{A.~Aprahamian}
\affiliation{co-organizer, Department of Physics, University of Notre Dame, Notre Dame, IN 46556, USA}

\author{A.~Frebel}
\affiliation{co-organizer, Department of Physics, Massachusetts Institute of Technology, Cambridge, MA 02139, USA}

\author{G.~C.~McLaughlin}
\affiliation{co-organizer, Department of Physics, North Carolina State University, Raleigh, NC 27695, USA}
 
\author{R.~Surman}
\affiliation{co-organizer, Department of Physics, University of Notre Dame, Notre Dame, IN 46556, USA}

\author{A.~Arcones}
\affiliation{Institut f\"ur Kernphysik, Technische Universit\"at Darmstadt, Darmstadt 64289, Germany}
\affiliation{GSI Helmholtzzentrum f\"ur Schwerionenforschung GmbH, Planckstr.\ 1, Darmstadt 64291, Germany}

\author{A.~B.~Balantekin}
\affiliation{University of Wisconsin, Department of Physics, Madison, Wisconsin 53706 USA}

\author{J.~Barnes}
\altaffiliation{NASA Einstein Fellow}
\affiliation{Department of Physics and Columbia Astrophysics Laboratory, Columbia University, New York, NY 10027, USA}

\author{Timothy C. Beers}
\affiliation{Department of Physics, University of Notre Dame, Notre Dame, IN 46556, USA}
\affiliation{Joint Institute for Nuclear Astrophysics - Center for Evolution of the Elements,  USA}

\author{Maxime Brodeur}
\affiliation{Department of Physics, University of Notre Dame, Notre Dame, IN 46556, USA}

\author{Jolie A. Cizewski}
\affiliation{Department of Physics and Astronomy, Rutgers University, New Brunswick, NJ 08901, USA}

\author{Jason A. Clark}
\affiliation{Physics Division, Argonne National Laboratory, Argonne, IL  60439, USA}
\affiliation{Department of Physics and Astronomy, University of Manitoba, Winnipeg, Manitoba R3T 2N2, Canada}

\author{Benoit C\^ot\'e}
\affiliation{National Superconducting Cyclotron Laboratory, Michigan State University, East Lansing, MI 48824, USA}
\affiliation{Konkoly Observatory, Research Centre for Astronomy and Earth Sciences, Hungarian Academy of Scsoiences, Konkoly Thege Miklos ut 15-17, H-1121 Budapest, Hungary}
\affiliation{Joint Institute for Nuclear Astrophysics - Center for Evolution of the Elements,  USA}

\author{Sean M. Couch}
\affiliation{Department of Physics and Astronomy, Department of Computational Mathematics, Science, and Engineering, and National Superconducting Cyclotron Laboratory, Michigan State University, East Lansing, MI 48824, USA}
  
\author{M.~Eichler}
\affiliation{Insitut f\"ur Kernphysik, TU Darmstadt, D-64289 Darmstadt, Germany}

\author{Jonathan Engel}
\affiliation{Department of Physics and Astronomy, University of North Carolina, Chapel Hill, NC. 27599-3255, USA}

\author{Rana Ezzeddine}
\affiliation{Joint Institute for Nuclear Astrophysics, Center for the Evolution of the Elements, East Lansing, MI 48824, USA}
\affiliation{Department of Physics, Massachusetts Institute of Technology, Cambridge, MA 02139, USA}

\author{George M. Fuller}
\affiliation{Department of Physics, University of California, San Diego, La Jolla, CA 92093-0424, USA}

\author{Samuel A. Giuliani}
\affiliation{National Superconducting Cyclotron Laboratory, Michigan State University, East Lansing, MI 48824, USA}

\author{Robert Grzywacz}
\affiliation{Department of Physics and Astronomy, University of Tennessee, Knoxville, TN 37996, USA}

\author{Sophia Han} \affiliation{Department of Physics and Astronomy, University of Tennessee, Knoxville, TN 37996, USA}

\author{Erika M. Holmbeck}
\affiliation{Department of Physics, University of Notre Dame, Notre Dame, IN 46556, USA}
\affiliation{Joint Institute for Nuclear Astrophysics - Center for Evolution of the Elements,  USA}

\author{C.~J.~Horowitz}
\affiliation{Department of Physics and Center for the Exploration of Energy and Matter, Indiana University, Bloomington, IN 47405, USA}

\author{Anu Kankainen}
\affiliation{Department of Physics, University of Jyvaskyla, P.O. Box 35, FI-40014 University of Jyväskylä, Finland}

\author{Oleg Korobkin}
\affiliation{{CCS-7}, Los Alamos National Laboratory, P.O. Box 1663, Los Alamos, NM 87545, USA}
\affiliation{Center for Theoretical Astrophysics, Los Alamos National Laboratory, P.O. Box 1663, Los Alamos, NM 87545, USA}
\affiliation{Joint Institute for Nuclear Astrophysics, Center for the Evolution of the Elements}   

\author{A.~A.~Kwiatkowski}
\affiliation{TRIUMF, 4004 Wesbrook Mall, Vancouver, BC V6T 2A3, Canada}

\author{J.~E.~Lawler}
\affiliation{Dept. of Physics, Univ. of Wisconsin – Madison, 1150 University Ave., Madison, WI 53706 USA}

\author{Jonas Lippuner}
\affiliation{{CCS-2}, Los Alamos National Laboratory, P.O. Box 1663, Los Alamos, NM 87545, USA}
\affiliation{Center for Theoretical Astrophysics, Los Alamos National Laboratory, P.O. Box 1663, Los Alamos, NM 87545, USA}
\affiliation{Joint Institute for Nuclear Astrophysics, Center for the Evolution of the Elements}   

\author{Elena Litvinova}
\affiliation{Department of Physics, Western Michigan University, Kalamazoo, MI 49008, USA}
\affiliation{National Superconducting Cyclotron Laboratory, Michigan State University, East Lansing, MI 48824, USA}

\author{G.~J.~Mathews}
\affiliation{Department of Physics, Center for Astrophysics, University of Notre Dame, Notre Dame, IN 46556, USA}

\author{M.~R.~Mumpower}
\affiliation{Theoretical Division, Los Alamos National Laboratory, Los Alamos, NM, 87545, USA}
\affiliation{Center for Theoretical Astrophysics, Los Alamos National Laboratory, Los Alamos, NM, 87545, USA}
\affiliation{Joint Institute for Nuclear Astrophysics - Center for the Evolution of the Elements, USA}

\author{S.~Naimi}
\affiliation{RIKEN Nishina Center for Accelerator-Base Science, Wako, Saitama 351-0198, Japan}

\author{W.~Nazarewicz}
\affiliation{Department of Physics and Astronomy and FRIB Laboratory, Michigan State University, East Lansing, MI 48824, USA}

\author{Evan O'Connor}
\affiliation{Department of Astronomy and the Oskar Klein Centre, Stockholm
  University, AlbaNova, SE-106 91 Stockholm, Sweden}
  
\author{Brian W. O'Shea}
\affiliation{Department of Computational Mathematics, Science, and Engineering, Department of Physics and Astronomy, and National Superconducting Cyclotron Laboratory, Michigan State University, East Lansing, MI 48824, USA}
  
\author{Albino Perego}
\affiliation{Istituto Nazionale di Fisica Nucleare, Sezione Milano Bicocca, gruppo collegato di Parma, I-43124 Parma, Italia}
\affiliation{Universita degli Studi di Milano-Bicocca, Dipartimento di Fisica, Piazza della Scienza 3, 20126 Milano, Italia}

\author{G.~Perdikakis}
\affiliation{Department of Physics, Central Michigan University, Mt. Pleasant, MI 48859, USA}
\affiliation{National Superconducting Cyclotron Laboratory, Joint Institute for Nuclear Astrophysics, Michigan State University, East Lansing, MI 48823, USA}

\author{David Radice}
\affiliation{Institute for Advanced Study, 1 Einstein Drive, Princeton, NJ 08540, USA}
\affiliation{Princeton University, 4 Ivy Lane, Princeton, NJ 08544, USA}

\author{Sherwood Richers}
\affiliation{North Carolina State University, Raleigh, NC 27695, USA}

\author{Luke F. Roberts}
\affiliation{NSCL/FRIB, Michigan State University, East Lansing, MI 48824, USA.}

\author{Caroline Robin}
\affiliation{Institute for Nuclear Theory, University of Washington, Seattle, WA 98195, USA.}
\affiliation{JINA-CEE, Michigan State University, East Lansing, MI 48824, USA.}

\author{Ian U.\ Roederer}
\affiliation{Department of Astronomy, University of Michigan,
1085 S.\ University Ave., Ann Arbor, MI 48109, USA}
\affiliation{Joint Institute for Nuclear Astrophysics - Center for Evolution of the Elements,  USA}

\author{Daniel M. Siegel}
\altaffiliation{NASA Einstein Fellow}
\affiliation{Department of Physics and Columbia Astrophysics
  Laboratory, Columbia University, New York, NY 10027, USA
}

\author{Nicolas Schunck}
\affiliation{Nuclear and Chemical Sciences Division, Lawrence Livermore
  National Laboratory, Livermore, CA 94551, USA}

\author{T.~M.~Sprouse}
\affiliation{Department of Physics, University of Notre Dame, Notre Dame, IN 46556, USA}

\author{A.~Spyrou}
\affiliation{National Superconducting Cyclotron Laboratory, Department of Physics and Astronomy, Joint Institute for Nuclear Astrophysics, Michigan State University, East
Lansing, MI 48824, United States of America}

\author{Nicole Vassh}
\affiliation{Department of Physics, University of Notre Dame, Notre Dame, IN 46556, USA}

\author{Jinmi Yoon}
\affiliation{Department of Physics, University of Notre Dame, Notre Dame, IN 46556, USA}
\affiliation{Joint Institute for Nuclear Astrophysics - Center for Evolution of the Elements,  USA}

\author{Yong-Lin Zhu}
\affiliation{Department of Physics, North Carolina State University, Raleigh, NC 27695, USA}

\begin{abstract}
In July 2018 an FRIB Theory Alliance program was held on the implications of GW170817 and its associated kilonova for $r$-process nucleosynthesis. Topics of discussion included the astrophysical and nuclear physics uncertainties in the interpretation of the GW170817 kilonova, what we can learn about the astrophysical site or sites of the $r$ process from this event, and the advances in nuclear experiment and theory most crucial to pursue in light of the new data. Here we compile a selection of scientific contributions to the workshop, broadly representative of progress in $r$-process studies since the GW170817 event.
\end{abstract}
\smallskip
\maketitle

\section{Introduction and background}

Last year history was made with the first discovery of a binary neutron star merger event by LIGO/Virgo, GW170817. The merger was subsequently studied in over seventy ground- and space-based observatories, at wavelengths across the electromagnetic spectrum. Observations in the visible and infrared are most convincingly interpreted as an $r$-process kilonova: an electromagnetic counterpart to the merger event powered by the radioactive decay of some hundredths to tenths of a solar mass of newly-produced heavy nuclei. The GW170817 kilonova provides the first strong evidence for an astrophysical site of $r$-process element production and may even suggest that the majority of the $r$-process material in the galaxy originates in neutron star mergers. 

The latter claim, widely reported in the media, is extraordinary. The astrophysical site of production of the heaviest elements via $r$-process nucleosynthesis has been one of the major open questions in physics for decades. Thus each piece of the chain of evidence leading to this possible conclusion requires careful examination. 

The GW170817 kilonova signal is interpreted through comparisons with model signals. Realistic kilonova model signals require state-of-the-art merger simulations to determine the conditions for nucleosynthesis, nuclear network calculations to find the nuclei produced and their associated radioactive heating, and radiation transfer calculations to convert the heating rate into the observed electromagnetic counterpart. Large uncertainties are present in each step. Astrophysical modeling suggests multiple nucleosynthetic sites are present within the merger event, including very neutron-rich dynamical ejecta from the tidal tails of the merger, shock-heated ejecta from the merger interface, and wind ejecta from the post-merger accretion disk. However, the exact nature of the conditions in each of these environments and their relative contributions to the overall nucleosynthetic yield is unclear. Nuclear network calculations are subject to these astrophysical uncertainties as well as the considerable nuclear physics uncertainties of the thousands of neutron-rich nuclei that participate in a merger $r$ process.  The radiation transport calculations are additionally sensitive to the opacities of the elements produced, which rely on uncertain atomic physics properties. In order to reliably interpret GW170817 and future kilonovae, these aspects need to be discussed and addressed.

In addition to the direct interpretation of the kilonova signal, it is also crucial to examine how this new evidence fits with the available $r$-process data from complementary sources. One such source includes spectroscopic observations of $r$-process elements in low metallicity stars. From the elemental abundance patterns and $r$-process enhancements observed in these pristine stars, information can be gleaned on the frequency, distribution, and number of distinct types of $r$-process events. Another key set of data is the pattern of solar isotopic $r$-process residuals, which are known to high precision. The shape of the $r$-process pattern produced in a nucleosynthetic event is sensitive to both the nuclear physics and astrophysics of the event; detangling the nuclear physics uncertainties can thus result in a powerful probe of $r$-process conditions. 

In the summer of 2016, an ICNT/JINA-CEE joint program ``The $r$ process of nucleosynthesis: connecting FRIB with the cosmos" brought together over sixty observers, astrophysical modelers, nuclear experimentalists, and nuclear theorists to address the nuclear physics challenges associated with $r$-process nucleosynthesis. The outcome of this program was an extensive review article discussing promising $r$-process experiments, their likely impact, and their astrophysical, astronomical, and nuclear theory context \cite{Horowitz2018}. In July 2018, a follow-up program ``FRIB and the GW170817 kilonova" was sponsored by the FRIB Theory Alliance. In this document we collect the contributions of the participants  of this workshop, representing a cross-section of efforts and advances in $r$-process science since the GW170817 discovery. 

\section{Astrophysical simulations of nucleosynthetic environments}

\newcommand{\skynet}{\emph{SkyNet}}
\subsection{$\pmb{r}$-Process nucleosynthesis calculations with \skynet\ (Jonas Lippuner)}
\begin{figure}
\includegraphics[width=1.0\textwidth]{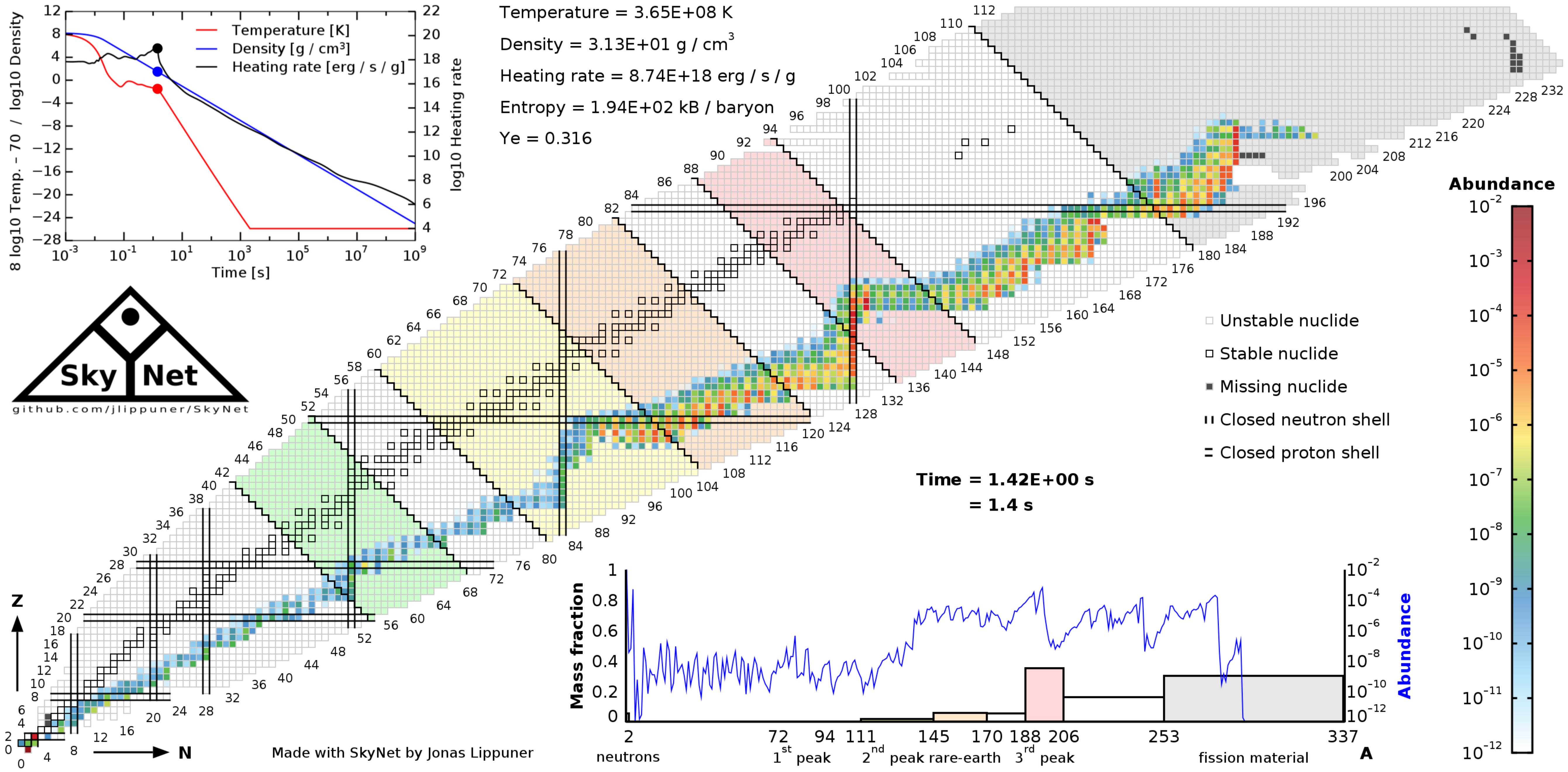}
\caption{A frame from the animation of the nucleosynthesis calculation of a very neutron-rich $r$-process. The frame shows the full extent of the $r$-process just when free neutrons get exhausted. The plot in the upper left corner shows the temperature,
density, and heating rate as function of time. The colored bands in the chart of nuclides correspond to the mass bins in the histogram at
the bottom. The histogram shows the mass fractions on a linear scale while the blue curve shows the abundances as a function of mass on a logarithmic scale. This figure appeared in \cite{lippuner:2015gwa}.}
\label{fig:skynet}
\end{figure}

$r$-Process nucleosynthesis calculations involve thousands of nuclides and over 100,000 nuclear reactions. Nuclear reaction networks are used to follow the abundances of the nuclear species under the influence of these reactions. \skynet\ is a state-of-the-art nuclear reaction network that is free and open-source available at \url{https://bitbucket.org/jlippuner/skynet} \cite{lippuner:17b}. 
\skynet\ was initially designed for evolving large reaction networks for
$r$-process nucleosynthesis calculations, but thanks to its modularity and
flexibility, \skynet\ can easily be used for nucleosynthesis computations
in many other astrophysical situations.  Besides correctness, the main design
goals behind \skynet\ are usability and flexibility, making
\skynet\ an easy to use and versatile nuclear reaction network that is
available for anyone to use. \skynet\ can evolve an arbitrary set of nuclear
species under various different types of nuclear reactions. 
\skynet\ can also compute Nuclear Statistical Equilibrium (NSE) compositions 
and switch between evolving NSE and the full
network in an automated and self-consistent way.
\skynet\ contains electron screening corrections and an
equation of state (EOS) that takes the entire composition into account. For
ease of use, \skynet\ can be used from within Python, and
\skynet\ can make movies of the nucleosynthesis evolution (see examples
at \url{http://stellarcollapse.org/lippunerroberts2015} and a frame from such a movie is shown in Fig.~\ref{fig:skynet}). \skynet\ has been used for
$r$-process nucleosynthesis calculations in different scenarios by various
authors: \cite{lippuner:2015gwa, radice:2016dwd, roberts:16b, lippuner:17a,
siegel:17, vlasov:17, fernandez:17a}.

\subsection{Incorporating experimental and theoretical nuclear data in astrophysical nucleosynthesis calculations (Trevor Sprouse)}

Calculations of nucleosynthesis in astrophysical environments rely on two broad categories of input data: astrophysical conditions, including the thermodynamic evolution and initial composition of the material undergoing nucleosynthesis, and nuclear data in the form of nuclear reaction rates, decay half-lives, masses, and fission properties.

The \textit{r} process poses a particular challenge to nucleosynthesis calculations. Because the neutron-rich nuclei involved in the \textit{r} process are short-lived, most are poorly understood experimentally. Therefore,  the theoretical predictions of nuclear models are needed to supply many of the nuclear data inputs required for nucleosynthesis calculations. Different nuclear models can vary widely in their predictions for the properties of \textit{r}-process nuclei, and these variations propagate to calculated \textit{r}-process abundances \cite{2015Mumpower_JPG,Mumpower2015,Mumpower2016}. The path to improving these calculations will follow the progress of both the nuclear theory and nuclear experiment communities. Current and planned future experimental campaigns continue to investigate properties of increasingly neutron-rich nuclei which can be used to benchmark and inform theoretical nuclear models, further enhancing their predictive capabilities when incorporated into nucleosynthesis calculations.

Portable Routines for Integrated nucleoSynthesis Modeling (PRISM) is a nuclear reaction network which employs an abstract form of the network equations, together with generalized data structures, in order to easily and efficiently incorporate a wide variety of nuclear datasets into its calculations. This approach enables an implementation of nuclear properties that is faithful to the nuclear models, experiments, or evaluations from which they originate.   
Furthermore, because theoretical nuclear datasets can so easily be interchanged, and new experimental data added, to its calculations, PRISM is ideally suited to investigate \textit{r}-process nucleosynthesis in the context of future advances in both nuclear experiment and theory that will emanate from FRIB. Calculations of the \textit{r} process with PRISM are included in, for example, \cite{Cote2018,holmbeck18b,zhu2018californium}.

\subsection{Numerical simulations of neutron star mergers (David Radice)}
\label{sec:David}

\begin{figure}
	\includegraphics[width=0.48\textwidth]{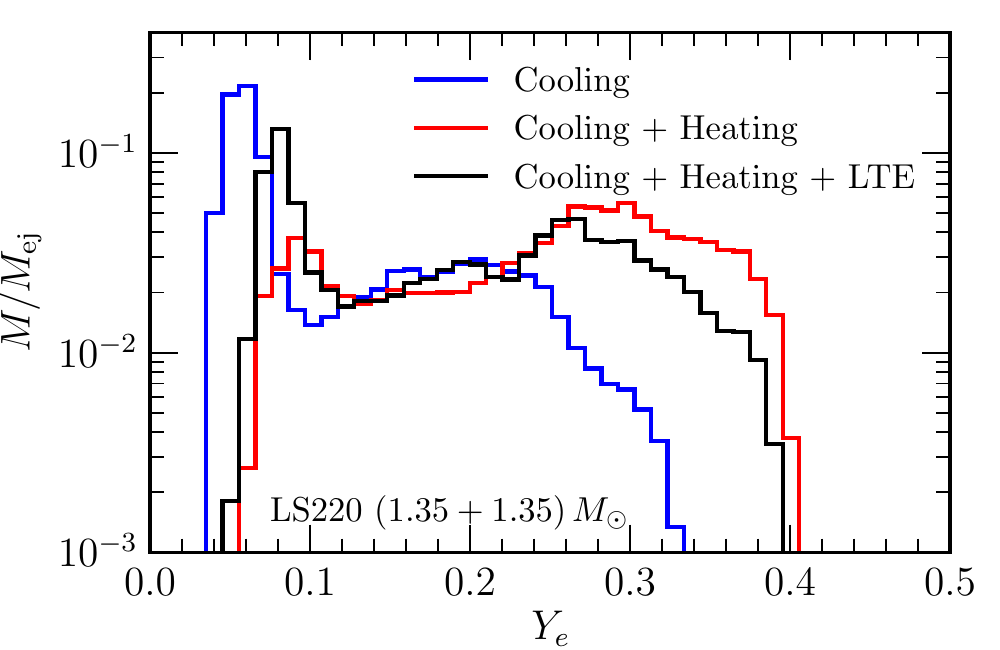}
    \hfill
	\includegraphics[width=0.48\textwidth]{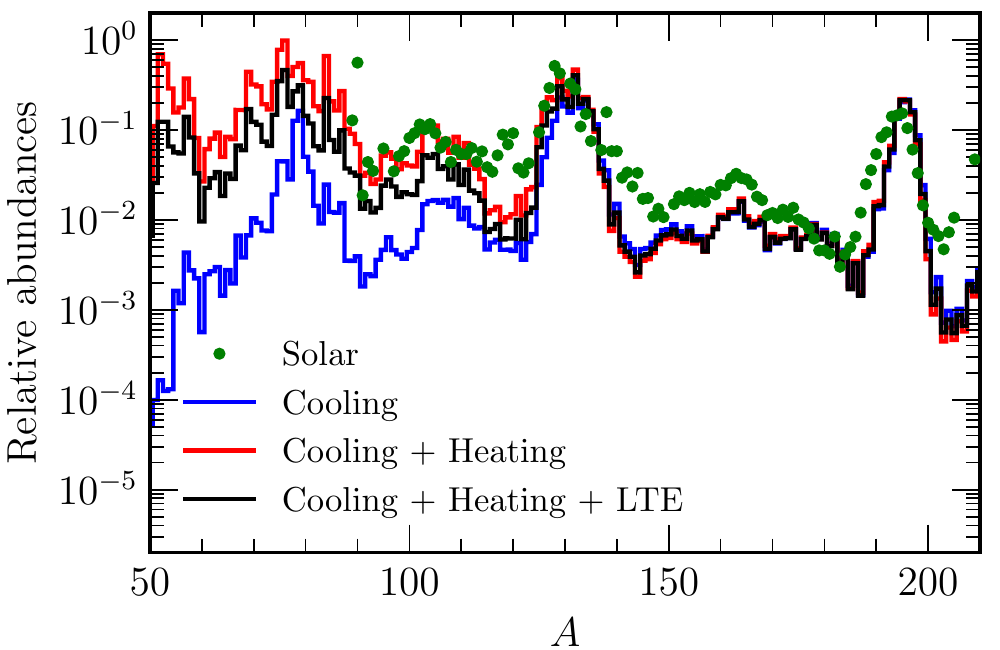}
    \caption{Dynamical ejecta composition (\emph{left panel}) and nucleosynthetic yields (\emph{right panel}) for a $(1.35 + 1.35)\, M_\odot$ binary neutron star system. The binary was simulated with the LS220 EOS \cite{lattimer:1991nc}. The ejecta composition was measured when the ejecta was still in nuclear statistical equilibrium and was crossing a coordinate sphere with radius $300 M_\odot G/c^2 \simeq 443\, {\rm km}$. The nucleosynthetic yields are computed to 32 years after the merger. The simulation was performed with the open source \texttt{WhiskyTHC} code \cite{radice:2012cu, radice:2013hxh, radice:2013xpa, radice:2015nva, radice:2016dwd}, while the nucleosynthesis was computed using the \texttt{SkyNet} code \cite{lippuner:2015gwa}. The ``Cooling'' simulation only included electron and positron captures, while the ``Cooling + Heating'' simulation also included compositional effects and heating due to the absorption of neutrinos. ``LTE'' indicates a simulation where the neutrino interaction cross-sections were computed assuming thermodynamical equilibrium between neutrino and matter. Non-LTE effects are approximatively included in the ``Cooling + Heating'' simulation. Neutrino re-absorption has a strong impact on the production of first peak r-process elements $A \lesssim 120$. Adapted from Radice, Perego, et al.~in prep.}
	\label{fig:radice}
\end{figure}

Numerical simulation are the cornerstone for the modeling of multimessenger signatures and nucleosynthetic yields from NS mergers. Early binary NS merger simulations either employed an approximate treatment for the gravitational field of the NSs \cite{ruffert:1995fs, rosswog:1998hy, rosswog:2001fh, rosswog:2003rv, rosswog:2003tn, oechslin:2006uk, rosswog:2012wb, korobkin:2012uy, bauswein:2013yna}, or included general relativistic (GR) effects, but compromised on the treatment of NS matter \cite{shibata:1999wm, shibata:2003ga, shibata:2005ss, baiotti:2008ra, kiuchi:2010ze, rezzolla:2010fd, rezzolla:2011da, hotokezaka:2011dh, hotokezaka:2012ze, palenzuela:2013hu, hotokezaka:2013iia, ruiz:2016rai, dietrich:2016hky, dietrich:2016lyp}. In the last few years, however, full-GR simulations with microphysical treatment of NS matter and with different levels of approximation for neutrino-radiation effects have also become available \cite{sekiguchi:2011zd, sekiguchi:2015dma, palenzuela:2015dqa, radice:2016dwd, lehner:2016lxy, sekiguchi:2016bjd, foucart:2016rxm, bovard:2017mvn}. These have clarified the mechanisms driving the mass ejection and highlighted the importance of including GR, a realistic description of dense matter, weak reactions, and neutrino radiation, all of which have been found to impact the mass ejection qualitatively and quantitatively. The sensitivity of the dynamical ejecta on the inclusion of neutrino heating, and the consequent nucleosynthetic yields are shown in Fig.~\ref{fig:radice}.

Neutron star merger simulations have also provided a new avenue to constrain the EOS of matter at extreme densities by providing a unified framework for the joint interpretation of EM and GW data from GW170817/AT2017gfo \cite{Bauswein:2017vtn, Radice:2017lry}. Bauswein et al.~\cite{Bauswein:2017vtn} showed that, under the assumption that GW170817 did not result in the prompt formation of a black hole (BH), it is possible to place a lower bound on the radius of a reference $1.6\, M_\odot$ NS, $R_{16}$. They used an empirical relation between $R_{16}$ and the threshold mass for prompt BH formation that they previously discovered using approximate-GR simulations \cite{Bauswein:2013jpa}. In a parallel work, we used a large set of full-GR simulations with detailed microphysics to constrain the tidal deformability of the binary as quantified by the parameter $\tilde\Lambda$ \cite{Abbott17a}. The latter was constrained by LIGO/Virgo observations to be less than $800$ with 90\% confidence. In our work, we discovered the existence of a critical value of $\tilde\Lambda$ below which a black hole is shortly formed after the merger and only a small amount of material, insufficient to explain the kilonova observations, is left outside of the horizon. On the one hand, this was expected because $\tilde\Lambda$ roughly scales as $(R/M)^6$, where $R$ and $M$ are the average masses and radiuses of the neutron stars in the binary \cite{De:2018uhw}. Consequently, binaries with small tidal deformabilities result in early BH formation. However, simulations were necessary to quantify this limit. On the basis of this simple observation, we were able to set a lower bound on the tidal parameter $\tilde\Lambda$ complementary to that of LIGO/Virgo. Our new constrain, in combination with that inferred from the analysis of the sole GW data, significantly reduces the range of allowed EOSs for NS and hybrid stars, as shown by Most et al.~\cite{Most:2018hfd}.

\subsection{Numerical simulations of the neutron star post-merger phase (Daniel Siegel)}\label{siegel}

Numerical simulations of the merger and post-merger phase are vitally important for identifying the astrophysical site or sites of r-process nucleosynthesis in neutron star (NS) mergers and to understand the conditions under which the r-process proceeds. Simulations have revealed several types of neutron-rich ejecta from NS mergers: dynamical ejecta including tidal and shock-heated components \citep{rosswog:1998hy,bauswein:2013yna,Hotokezaka2013a}, ejected on a timescale of $\sim\!\mathrm{ms}$ during the merger process; neutrino-driven and magnetically driven winds from a (meta-)stable remnant NS \cite{Dessart2009,Siegel2014,Ciolfi2017} (if the merger does not promptly collapse to a black hole), ejected on timescales of up to $\sim\!10\,\mathrm{ms}-1\,\mathrm{s}$ depending on the lifetime of the object; and outflows from a post-merger neutrino-cooled accretion disk on secular timescales of $\sim\!100\,\mathrm{ms}-1\,\mathrm{s}$ \cite{Fernandez2013,just:2015,Siegel2017a}. Several of these processes are at play in a typical NS merger event, generating ejecta material with different properties (amount of ejected material, composition, velocities). Therefore, in principle, kilonovae with multiple components are expected (see also Sec.~\ref{sec:Albino}).

The near-infrared emission of the kilonova associated with GW170817 provided strong evidence for the presence of lanthanides in the outflow material. The combination of a large amount of inferred neutron-rich ejecta ($\approx\!4-5\times 10^{-2}\,M_\odot$) and slow ejecta velocities $\approx\!0.1c$ (e.g., \cite{Villar2017} and references therein) is rather difficult to explain by dynamical (tidal) ejecta. However, assuming that the merger did not promptly lead to BH formation (see Sec.~\ref{sec:David}), a post-merger accretion disk naturally provides outflows with composition, velocities, and total mass that are consistent with the observationally inferred values \cite{Siegel2017a,siegel:17}.

Post-merger accretion disks can evaporate $30-40\%$ of their original disk mass into unbound outflows \cite{siegel:17,Fernandez2018}. These outflows are launched as thermal winds from a hot corona, which is the result of an imbalance between viscous heating from magnetohydrodynamic turbulence and cooling via neutrino emission at high latitudes off the disk midplane. Further acceleration due to $\alpha$-particle formation leads to an average outflow speed of $\approx 0.1c$. The composition of the outflows is controlled by a self-regulation mechanism based on electron degeneracy in the inner part of the disk, which keeps the mean electron fraction of the outflows well below a critical threshold of $Y_\mathrm{e}\approx 0.25$ for lanthanide and actinide production \cite{Siegel2017a,siegel:17}. Detailed nucleosynthesis calculations show that the production of the full range of r-process nuclei from the first to the third peak can be explained \cite{Wu2016,siegel:17}.

\subsection{Neutrino Microphysics in Compact Object Mergers (Evan O'Connor)}

As David Radice alluded to in Section~\ref{siegel}, neutrinos play a crucial role in mergers of compact objects.  Most importantly, neutrino are emitted from and absorbed onto the matter and can change the composition, i.e. the electron fraction. As Jonas describes above, the composition of the ejected material from mergers is critical as it sets the nucleosynthetic yields.  Research over the past few years has shown the impact that various approximate neutrino treatments can have. Pure neutrino leakage schemes \cite{rosswog:1998hy, rosswog:2003rv, neilsen:2014, deaton:2013} can capture the releptonization of matter that is shock heating and decompressed. However, they cannot capture neutrino irradiation, where neutrinos emitted elsewhere (say the central compact object) stream away and are reabsorbed elsewhere. One can attempt to include this reabsorption in a parameterized way, however the few geometric symmetries present in compact object mergers generally make this difficult \cite{perego:2014}.  Since this neutrino reabsorption is critical, it necessitates the use of neutrino transport methods that naturally capture this physics. The downside of transport methods is their expense. A commonly used technique, M1 transport \cite{shibata:2011, foucart:2015, shibata:2012, just:2015}, can increase the number of grid variables by 100s if energy-dependent, multispecies transport is used.  The large matter velocities and velocity gradients can also make evolving the transport equations difficult.  Typically what has been done in the literature so far is so-called gray transport.  In gray transport, the energy dependence of the neutrinos is ignored and suitably averaged neutrino emission rates and opacities are used.  A further improvement to this method is gray+ transport, where in addition to solving the gray transport equation for the neutrino energy, an equation for the neutrino number is also solved.  This gives a local estimate of the neutrino mean energy and allows for a better conservation of lepton number and a more accurate evolution of the composition \cite{foucart:2016rxm}. While many of these neutrino scheme and methods may give an accurate estimate of the total mass ejected, they are likely not enough to capture the composition of the ejecta.  For this, full energy-dependent transport will be needed going forward. 

\subsection{Neutrino microphysics in supernovae and SN neutrino detection (Chuck Horowitz)}

Unfortunately, because of the large distances, it is difficult to detect neutrinos from neutron star mergers.  However, neutrinos can be detected from galactic core collapse supernovae and the next galactic SN should provide a data set that could be of fundamental importance for nucleosynthesis both in SN and NS mergers.  Nucleosynthesis and the electron fraction $Y_e$ in neutrino driven winds is sensitive to small differences between electron neutrino and antineutrino spectra and fluxes.  Nucleon-nucleon correlations \cite{correlations_CJH,correlations2_CJH}, the presence of muons \cite{muons_CJH}, weak magnetism \cite{weakmag_CJH}, and binding energy shifts \cite{bindingE_CJH} can all impact neutrino spectra.  Neutrino oscillations in mergers and SN are uncertain because of complex matter and nonlinear effects.  This increases the importance of SN observations.  It is important to have neutrino detectors that can separately measure the flux and spectra of $\bar\nu_e$, $\nu_e$, and heavy flavor $\nu_x$ neutrinos and antineutrinos.  Existing water Cherenkov and liquid scintillator detectors such as Super Kamiokande should measure both the flux and spectra of $\bar\nu_e$ well.  It is important to also have a good $\nu_e$ detector.  This should be provided by the liquid Ar detector DUNE.  However, it would be very useful to calibrate DUNE for SN neutrinos by measuring the Ar charged current cross section with pion decay at rest neutrinos from, for example, the Spallation Neutron Source at Oak Ridge.  Finally a neutral current $\nu_x$ detector that could accurately measure the $\nu_x$ spectra would also be very useful.  One option would be a neutrino-nucleus coherent elastic scattering detector \cite{coherent2_CJH,coherent_CJH}.

\subsection{The multicomponent, anisotropic character of kilonovae (Albino Perego)} \label{sec:Albino}

Binary neutron star mergers are intrinsically multi-dimensional, multi-scale, and multi-physics processes. Sophisticated numerical models are necessary to model the dynamics of the merger and of its aftermath, and to predict the associated multimessenger observables. Weak interactions and neutrino transport are relevant in predicting the detailed composition of the ejecta and, up to a certain extend, its dynamics. In fact, the neutron abundance set by weak interactions is, together with the entropy and the expansion time scale, a crucial ingredient to compute reliable nucleosynthetic yields. The composition ultimately determines the matter opacity to photons \citep{Metzger2010,Roberts2011,Kasen2013,Tanaka2013,Wollaeger2018}. The recent detection of gravitational waves from the inspiral of two neutron stars (GW170817), followed by the discovery of several electromagnetic counterparts of the merger, including a gamma-ray burst and a kilonova (also called macronova), provided the first direct test for the present understanding of compact binary mergers \citep{Abbott17a,Abbott17b}. In particular, the kilonova signal was characterized by a quasi-thermal emission, peaking in the UV/optical frequencies around one day after the merger, followed by the reddening of the emission spectrum on a time scale of a few days (see, e.g.,\cite{Pian2017,tanvir17}). Interpretations of the photometric light curves revealed immediately that observations could not be explained in terms of a single component, where a component is defined as a spherically symmetric ejection characterized by a certain amount of mass, expansion velocity and photon opacity (see e.g. \citep{cowperthwaite17,Chornock2017,drout17,Villar2017,Tanaka2017,Shibata2017}). Models with at least two components succeeded in fitting the observed light curves. These models are often defined as the suitable combination of single, spherical models. Thus, they often subtend correlations between the properties of the ejecta within each channel. 

\begin{figure}
	\includegraphics[width=0.40\textwidth]{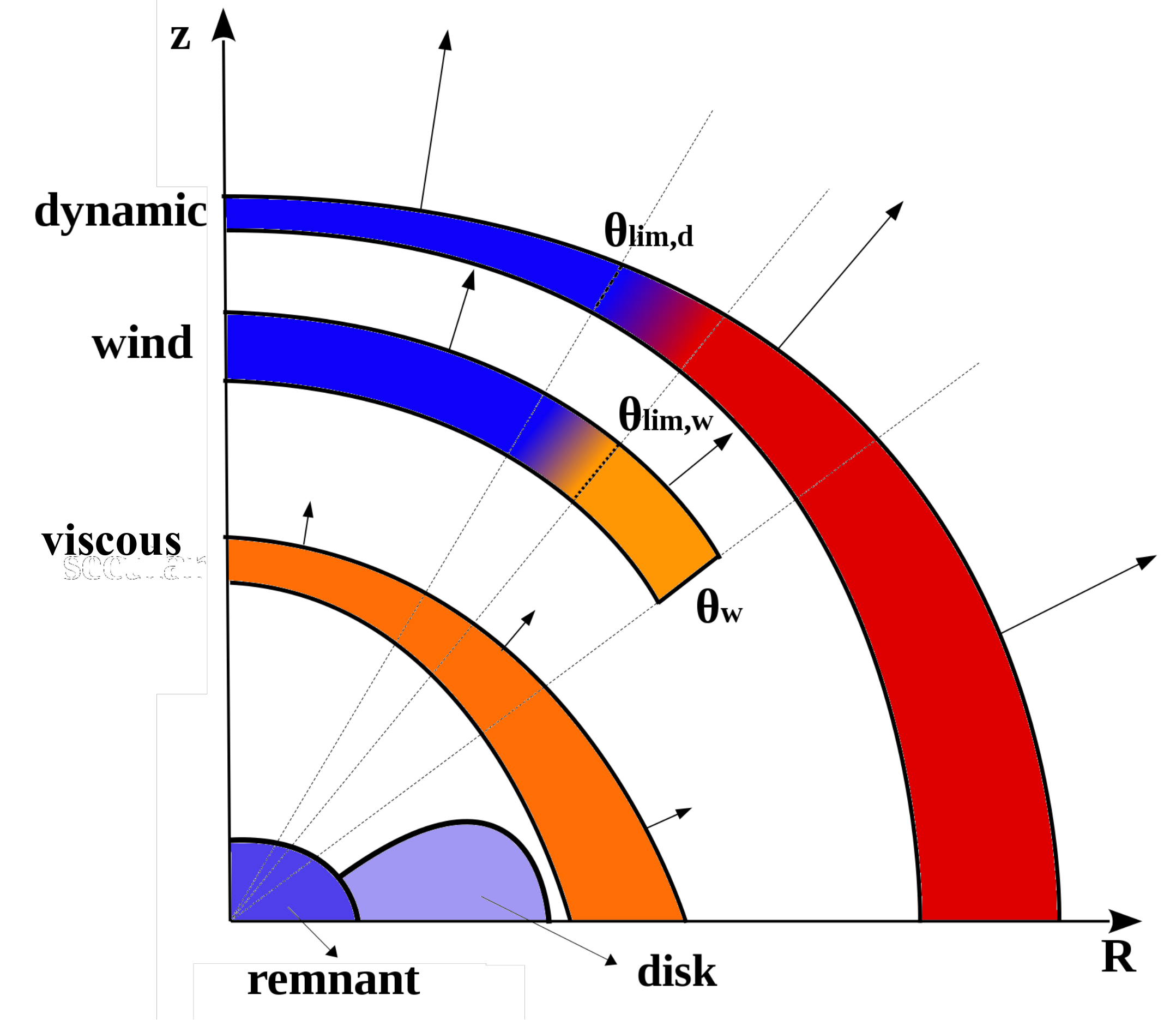}
    \hfill
	\includegraphics[width=0.52\textwidth]{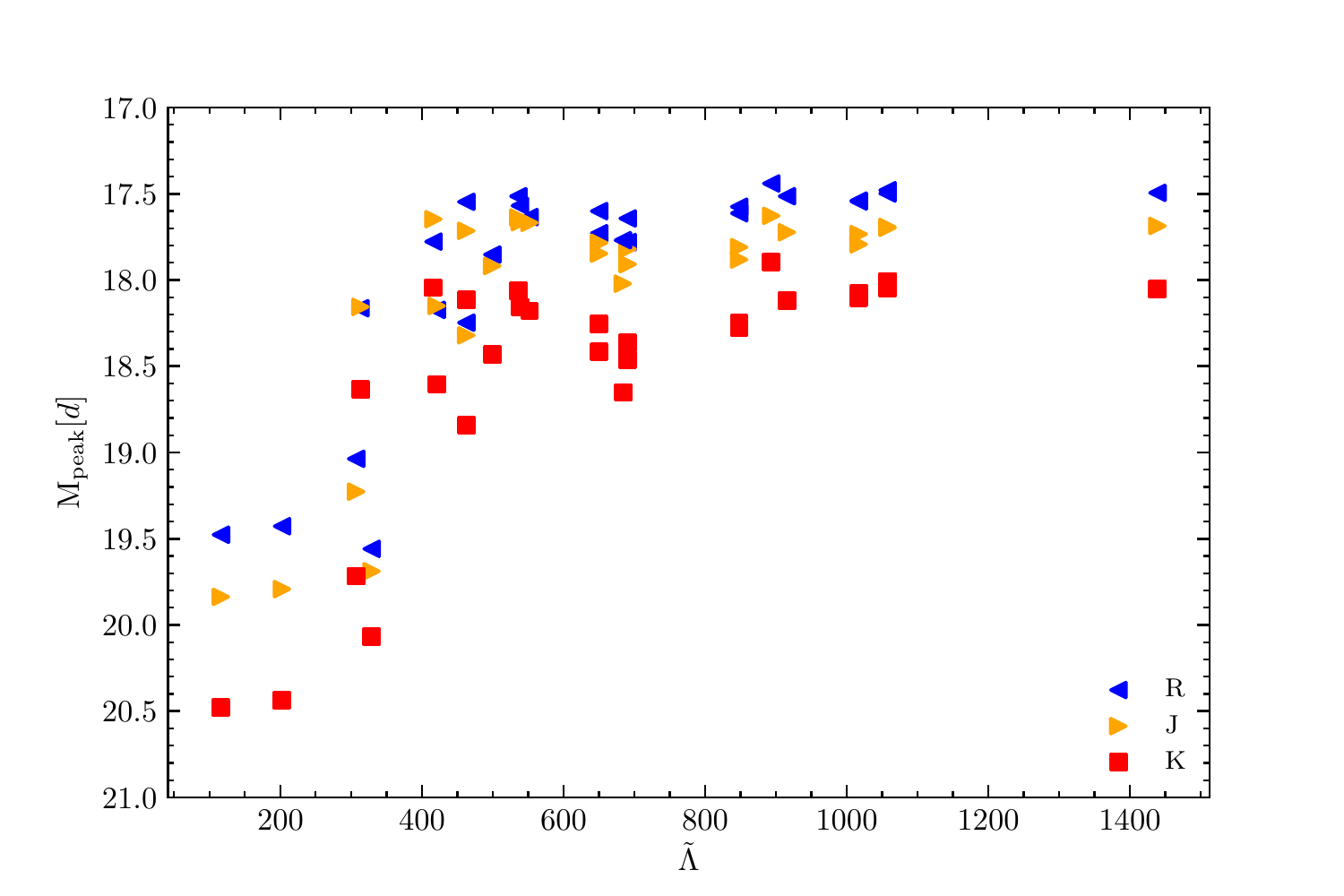}
    \caption{Left: Sketch of the multi-component, anisotropic kilonova model reported in \citep{Perego2017}. Different matter ejections lead to different components (dynamic, wind and viscous ejecta) with an explicit dependence on the polar angle. Right: Peaks of the kilonova light curves in three different bands ($R$, $J$ and $K$) as a function of the binary deformability parameter $\tilde{Lambda}$. The light curves were obtained by the model referring to the left plot, using dynamical ejecta and disk properties obtained for a wide range of binary neutron star mergers performed in full GR (see David's contribution, Radice et al, in preparation).]}
	\label{fig:perego}
\end{figure}

In a recent work \citep{Perego2017}, we have explored if the observed light curves could be explained in terms of the properties of the ejecta, based on their present understanding and modeling. State-of-the-art simulations of merger and of post-merger have revealed that matter ejection happens through different channels. Each channel is primarily characterized by different ejection mechanisms, which operate on different time scales and provide the expanding matter with qualitatively different properties. Broadly speaking, we can identify at least three different kind of ejecta: dynamical ejecta (e.g., \cite{korobkin:2012uy,bauswein:2013yna,radice:2016dwd,hotokezaka:2013iia,bovard:2017mvn,sekiguchi:2015dma}), wind ejecta expelled by neutrino or magnetic processes \citep{perego:2014,Siegel2014}, and viscous ejecta \citep{Fernandez2013,siegel:17,just:2015,Fujibayashi2018}. Each of these processes reveals a certain degree of anisotropy. Thus, we have set up a multi-component, anisotropic kilonova model, where we have discretized the polar angle in different slices. Within each slide, we have adopted the semi-analytical model presented in \citep{Grossman2014,Martin2015} and explicitly prescribed a dependence of the model properties on the polar angle. In addition, we have considered variations in the nuclear heating rate due to different nuclear compositions, time-dependent thermalization efficiency \citep{Barnes2016}, and the effect of irradiation of inner on outer photospheres. The computational efficiency of our scheme has allowed a broad parameter exploration. We found that a total ejecta of 0.03-0.06 $M_\odot$ is required by our model to explain the data. Moreover, the explicit dependence on the polar angle gave a constraint on the relative inclination between the merger axis and the light of sight of 30$^o$, compatible with measurements derived from gravitational wave and gamma-ray burst afterglow observations. Last but not least, we found that the presence of a suitable amount of fast expanding, low opacity (i.e. lanthanide-poor) ejecta at high latitude is required to fit the data. This result is a possible direct evidence of the role of neutrinos in setting the properties of the ejecta and in shaping the kilonova emission properties. The importance of multi-dimensional photon transport and the relevance of anisotropic ejecta properties have also been recently recognized. More detailed multi-dimensional radiative transfer models are necessary to address quantitative answer regarding the kilonova emission \cite{tanvir17,Wollaeger2018,Kawaguchi2018}, and to inform and to gauge computationally cheaper semi-analytical models. The latter are still a suitable tool to produce large numbers of models and to discover trends in large parameter space explorations.

\subsection{Many Facets of Kilonova Modeling (Oleg Korobkin)} \label{sec:kilonova-many-facets}
\begin{figure}
  \begin{tabular}{cc}
    \includegraphics[width=0.52\textwidth]{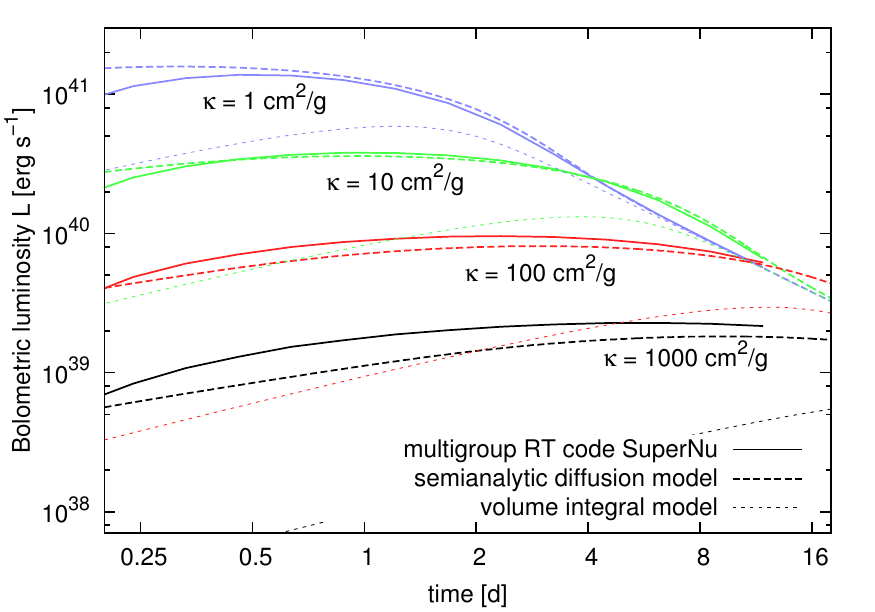} &
    \includegraphics[width=0.42\textwidth]{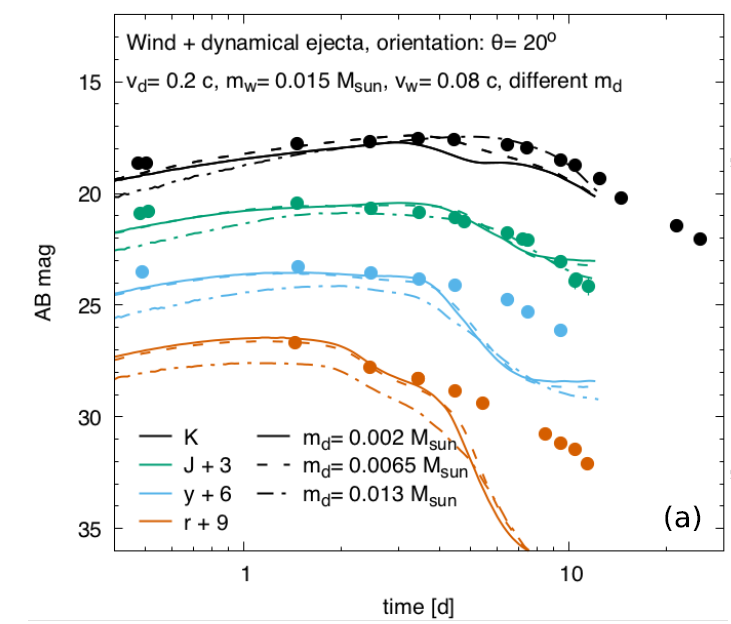} \\
    (a) & (b)
  \end{tabular}
   \caption{Left: Bolometric light curves for spherical gray-opacity models with three levels of approximation: full multigroup Monte Carlo radiative transfer models (solid lines), semianalytic diffusion model (dashed lines), and the much simpler model of {Grossman et al. (2014)} (’volume integral model’, dotted lines~\cite{rosswog18}).  Right: Optical and infrared light curves for the state-of-the art two-component, two-dimensional models with simulated morphologies, detailed opacities and effective nuclear heating, compared with observations~\cite{tanvir17}.}
\label{fig:kilonova-models}
\end{figure}

Interpreting kilonovae and linking observational data with fundamental properties of nuclear matter requires a comprehensive but pragmatic model which incorporates all essential ingredients.  The latter may include ejecta morphology, composition, detailed temperature and density-dependent opacities of heavy elements, and multifrequency, multidimensional radiative transport.

Among all these aspects of kilonova modeling, detailed opacities are perhaps the most difficult to tackle. Therefore, several studies of GW170817 used an effective, frequency-independent (gray) opacity as an intermediate parameter, which can be inferred from observations and later connected to a comprehensive theoretical radiative transfer model.  For the case of GW170817-associated kilonova, two components are required (see Section~\ref{sec:Albino}), with effective opacities 
$\kappa = 1\ {\rm cm}^2 {\rm g}^{-1}$ and 
$10\ {\rm cm}^2 {\rm g}^{-1}$. 
These correspond to what is expected from light (heavy) $r$-process, respectively.

As the LANL group demonstrated in \cite{Wollaeger2018}, gray opacity models produce Doppler-broadened blackbody spectra. A spectrum of such kind was observed during the first few days of GW170817, which motivates the use of gray opacity models for blue kilonovae. However, at late epochs, when the peak of emission shifts to near-IR, the spectrum deviates from blackbody quite significantly.  Nevertheless, bolometric light curves \citep{kasliwal17,rosswog18} can still be reconstructed and compared with theoretical effective nuclear heating rates, independently from radiative transfer. Note that in the case of spherical symmetry and constant gray opacity the radiative diffusion equation admits separation of variables and thus can be solved (semi-)analytically~\cite{pinto00a}.  Figure~\ref{fig:kilonova-models}a (taken from \cite{rosswog18}) demonstrates good agreement between a full radiative transfer code and the semianalytic model, but also shows that some more simplistic models --such as e.g. volume integral ones-- underestimate luminosity at early times and thus tend to overpredict the mass of ejecta.

Prior to GW170817, the LANL group has developed a unique multidimensional kilonova model, incorporating simulated morphologies, atomic opacities, effective nuclear heating and multigroup Monte Carlo radiative transfer~\citep{Wollaeger2018}.  When kilonova was discovered, we were able to apply our model with different parameters for two components to fit the light curves (see Figure~\ref{fig:kilonova-models}b).  The model produced estimates of masses and velocities which are more in line with predictions of numerical relativity~\cite{bovard:2017mvn}. It was recently confirmed independently in \cite{Kawaguchi2018}, where it was shown that multidimensional nature of kilonova changes the picture.

Another facet is nuclear heating: kilonova light curves are shaped by the effective heating rates much more sensitively than by ejecta mass and velocity. However, these rates still remain highly uncertain, varying by almost an order of magnitude~\cite{rosswog17}.  Recently, Zhu et al. (2018)~\cite{zhu2018californium} discovered a potential prominent imprint of an individual isotope ${}^{254}$Cf on the nuclear heating rate.

However, since this fissioning actinide decays on a timescale of about 60 days, it will only affect kilonova when it shifts to mid-IR regime.  In~\cite{zhu2018californium}, the LANL group has constructed an effective one-zone model for mid-IR light curves. Our theoretical predictions for JWST and Spitzer missions indicate that at certain conditions the presence of fissioning ${}^{254}$Cf makes a difference between confident detection and non-detection in mid-IR.

\subsection{Quantifying Uncertainties in Neutrino Transport Methods (Sherwood Richers)}

Neutrino transport in simulations of neutron star mergers is treated in the literature with a wide variety of methods. These can range from approximate leakage schemes to two-moment transport in dynamical calculations. However, more advanced transport schemes like Monte Carlo can be used to post-process simulations to get a handle on the errors introduced by using approximate transport schemes and the relevance of different neutrino interactions \cite{richers:2015,foucart:2018}. Monte Carlo transport calculations using on 3D snapshots of disks formed after a merger event \cite{Radice:2017lry} using the SEDONU code show that errors in the neutrino radiation pressure tensor are on the order of locations several percent of the total neutrino energy density, and errors in the third moment are on the order of a few percent of the total neutrino energy density. As Evan mentions, such errors in the transport method can significantly change the disk outflow mass and composition. It should also be noted that neutrinos can change flavor in merger environments in a way that could significantly affect the rate of energy deposition by neutrinos, and therefor modify winds \cite{zhu:2016,frensel:2017}. The effects of such neutrino oscillations on merger dynamics is an open question that needs to be addressed, but is computationally extremely challenging because of the resolution required to resolve neutrino oscillations.

\subsection{Neutron Star Mergers as the Source of the Actinide Boost (Erika M.\ Holmbeck)}
\begin{figure}[!th]
\begin{center}
  \includegraphics[width=0.5\textwidth]{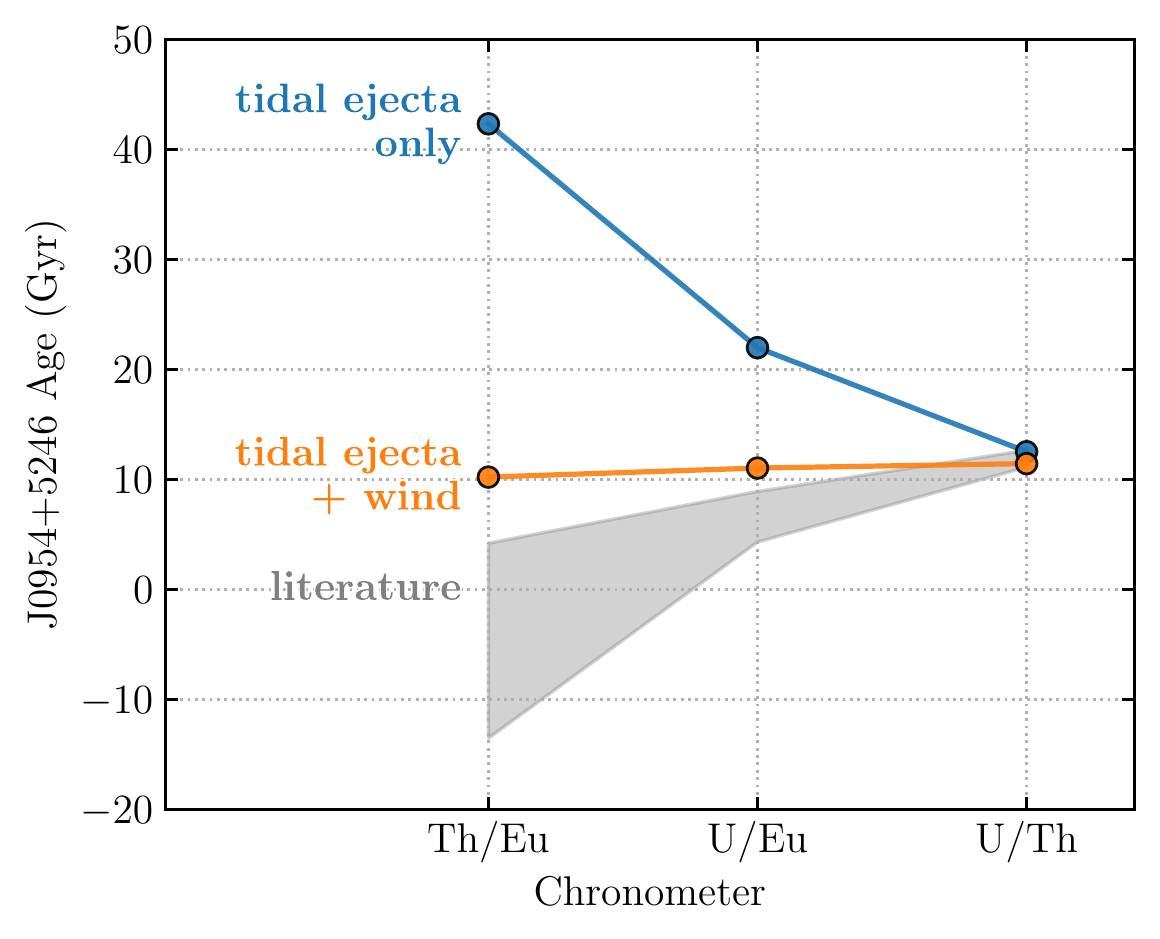}
  \caption{Age of the most actinide-boost star, J0954$+$5246 \citep{holmbeck18}, calculated using Th/Eu, U/Eu, and U/Th cosmochronometers. Literature production ratios that use supernovae conditions or waiting-point approximations \citep{Schatz2002,farouqi2010,wanajo2002} cannot reproduce the actinide-enhancement observed in some metal-poor, \emph{r}-process enhanced stars, often predicting negative ages. Using Th and U abundances created only in a tidal ejecta environment, the actinides are overproduced, and the estimated age of the star exceeds the age of the universe. Adding a high-$Y_e$ component to simulate a disk wind sufficiently dilutes the actinides so that the three chronometers agree.\label{fig:j095442_age}}
\end{center}
\end{figure}

The rapid-neutron-capture (``\emph{r}-") process is responsible for synthesizing many of the heavy elements observed in both the solar system and Galactic metal-poor halo stars.
Simulations of \emph{r}-process nucleosynthesis can reproduce abundances derived from observations with varying success, but so far fail to account for the observed over-enhancement of actinides, present in about 30\% of \emph{r}-process-enhanced stars.
We investigate actinide production in the dynamical ejecta of a neutron star merger and explore if varying levels of neutron richness can reproduce the actinide boost \citep{holmbeck18b}. 
The actinides are over-produced across a variety of nuclear physics choices (fission distribution, $\beta$-decay, and mass model) if the initial conditions are sufficiently neutron-rich for fission cycling.
Actinide production can be so robust in the dynamical ejecta that an additional lanthanide-rich, actinide-poor component is necessary in order to reproduce observations of actinide-boost stars. 
A simple actinide-dilution model that folds in estimated contributions from two nucleosynthetic sites within a merger event is sufficient to dilute the actinides to observed levels in metal-poor stars (see Fig.~\ref{fig:j095442_age}).
While the dynamical ejecta of a neutron star merger is a likely production site for the formation of the actinides, a significant contribution from another site or sites (e.g., the neutron star merger accretion disk wind) is required to explain abundances of \emph{r}-process-enhanced, metal-poor stars.

\subsection{Galactic Chemical Evolution as a Diagnostic Tool for r-Process Sites (Benoit C\^ot\'e)}

There are various types of observation that provide insights into the properties of neutron star mergers, and there are various type of numerical approaches that can be used to interpret them. To define whether or not neutron star mergers can be the dominant site of the main r-process (between the second and third r-process peaks), different pieces of puzzle must be assembled to create a coherent and consistent picture. This represents a multi-scale challenge that involves  diverse fields of research ranging from nuclear physics experiments to galaxy formation theories \cite{Horowitz2018}. In a work in preparation (C\^ot\'e, Eichler, Arcones, et al.), we address the origin of r-process elements by combining the chemical abundances of metal-poor stars \cite{2015ARA&A..53..631F,Frebel2018}, the chemical evolution trend of europium in the disk of the Milky Way \cite{2016A&A...586A..49B,2018MNRAS.478.4513B}, the detection of short gamma-ray bursts in different galaxies \cite{2014ARA&A..52...43B,2017ApJ...848L..23F}, and the gravitational wave detection GW170817 \cite{Abbott17a}. We then analyze each observational evidence from the point of view of nucleosynthesis calculations, compact binary population synthesis models \cite{2017arXiv170607053B}, and galactic chemical evolution simulations \cite{2003PASA...20..401G,2008EAS....32..311P,2013ARA&A..51..457N,2014SAAS...37..145M}. In this context, we do not use galactic chemical evolution to define whether neutron star mergers are the dominant r-process site, we rather use them to provide a clue that must be combined with other clues coming from other research areas. The outcome of this work is that, although neutron star mergers are prime candidates to be the dominant r-process site, we find inter-disciplinary inconsistencies when we assume that they are the only site. In particular, short gamma-ray bursts detections and population synthesis models currently agree on the functional form of the delay-time distribution of neutron star mergers (i.e., the probability of a neutron star binary to merge after a given time). But using this delay-time distribution in galactic chemical evolution simulations does not allow to reproduce the $\sim$~10\,Gyr-long evolution trend of europium in the disk (not the halo) of the Milky Way \cite{2017ApJ...836..230C,2018arXiv180101141H}.  However, most of the inter-disciplinary tensions disappear if we assume that there is a second astrophysical site of r-process elements in the early universe, at low metallicity.  This site, if it exists, should slowly fade away at later times, at higher metallicity.  This is assuming that the analyzed observational evidences are all representative.  This work aims to set the stage for more in-depth investigations.

\subsection{Predictive Galactic Chemical Evolution in the era of large surveys (Brian O'Shea)}

Modern astronomical surveys present significant challenges to the predominant theoretical approach used to study galactic chemical evolution -- namely, one- or few-zone models with simple analytic prescriptions for galaxy growth, gas inflow, and gas outflow.  Modern photometric and spectroscopic surveys (such as SDSS and its extensions, Gaia and Gaia-ESO, GALAH, RAVE, DES, and many others) provide many dimensions of both accurate and precise information about tremendous numbers of stars in the Milky Way and its satellites, including mass, age, metallicity, detailed chemical abundances, and 3D spatial and velocity information.  Furthermore, these surveys have selection functions that are both inclusive and relatively easy to model, giving a statistically robust sense of the underlying population out of with the stars are sampled.  The quantity and quality of this data will facilitate asking detailed questions about the evolution of the stellar IMF with redshift and environment, the nature of Population III and extremely low metallicity star formation, the properties of the site(s) of the r-process, and the relationship between stellar evolution and cosmological structure formation.  Chemical evolution models that can confront these types of datasets need to be comparable levels of sophistication to modern semi-analytical models of galaxy formation, including cosmological structure formation through merger trees generated from N-body simulations (which also provides spatial and kinematic information), prescriptions for star formation, stellar feedback, and exchange of materials with the circumgalactic and intergalactic media that are motivated by astrophysical theory and observations of galaxies at a wide range of masses and redshift.  \citep{2018ApJ...859...67C} They also require an understanding of the uncertainties inherent in the models and their inputs, and the consequences that these uncertainties have on predictions (and thus on the ability of the models to inform our understanding).  \citep{2016ApJ...824...82C,2017ApJ...835..128C} Furthermore, these models require sophisticated statistical tools, including Bayesian Markov Chain Monte Carlo sampling and Gaussian Process models, to effectively derive inference from a diverse range of multimessenger astronomical surveys.  \citep{2012ApJ...760..112G,2014ApJ...787...20G}  The JINA-NuGrid Galactic Chemical Evolution Pipeline has been developed over the past few years to pursue this goal.  \citep{2017nuco.confb0203C}  After rigorous testing, the first set of scientific questions that this pipeline will be used for will revolve around a deeper understanding of the site of the r-process, simultaneously using observations from the Galactic disk, stellar halo, and ultra-faint dwarf satellites as constraints on model parameters.

\subsection{Do actinide-boost stars have additional abundance anomalies? (Marius Eichler)}

About one third of the currently known r-II stars ([Eu/Fe~$ > +1.0$]) have an enhanced Th/Eu abundance ratio compared to the solar ratio \cite{Schatz2002,Roederer2009,Mashonkina2014,holmbeck18}, while the yields of the lighter r-process elements agree with the solar values. All of these so-called actinide-boost stars observed so far have a very low metallicity ([Fe/H~$ < -2.0$]). Recently, Ji \& Frebel (2018) \cite{JiFrebel2018} measured an unusually low Th/Eu ratio for a star in the dwarf galaxy Reticulum II, again accompanied by solar-like abundances in the rare-earth region. From a nucleosynthesis point of view, it is reasonable to assume that environments that lead to higher thorium abundances also produce more fissioning nuclei at the end of the r-process compared to environments with solar (or subsolar) Th/Eu ratios. Since most fission fragment distribution models predict the majority of fission fragments to be produced at or close to the second peak \cite{Eichler2015,Goriely2015}, this means that the shape of the second peak in actinide-boost stars should more closely resemble the fission fragment distribution than in the regular Th/Eu stars. This is illustrated in Figure, where we have performed r-process calculations for two different environments: one very neutron-rich trajectory \cite{Rosswog2013} that results in high thorium abundances (blue) and a trajectory from a magneto-hydrodynamically driven (MHD) supernova trajectory from Ref.~\cite{Winteler2012} with a moderate $Y_e$ (red). For both cases, we compare the abundances of the second peak of the full calculation (solid lines) with a case where we block fission after the fission freeze-out (dashed lines). The shaded regions therefore indicate the contribution of fission fragments to the second peak abundances. The fission fragment distribution model used here is ABLA07 \cite{Kelic2009}.

\begin{figure}[!ht]
\begin{center}
  \includegraphics[clip=false,width=0.40\textwidth]{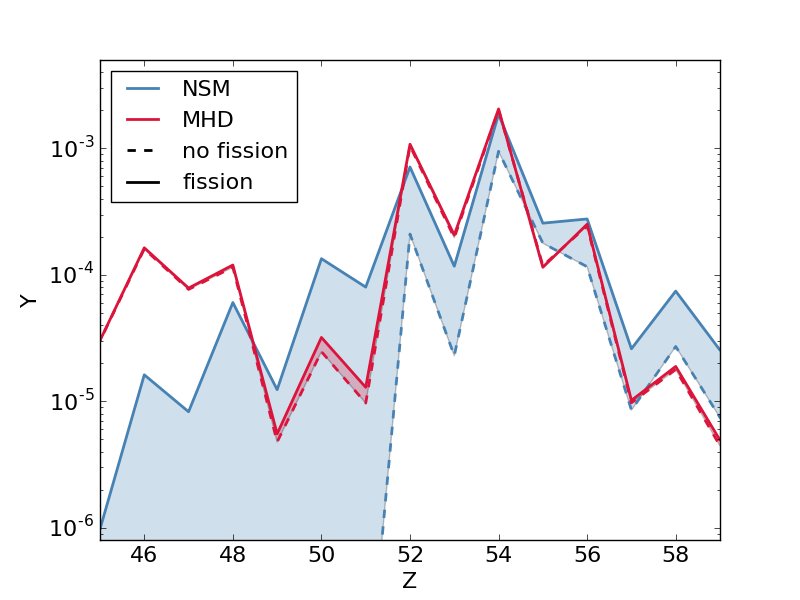}
  \caption{Abundances of the second r-process peak for two trajectories with low (blue) and moderate (red) initial $Y_e$. The dashed lines correspond to calculations where fission is blocked after the r-process freeze-out. See text for details. Figure adapted from C\^ot\'e et al. (in preparation). \label{fig:2ndpeak}}
\end{center}
\end{figure}

\subsection{The r-process in neutron star mergers and supernovae (Almudena Arcones)}

The r-process occurs in neutron star mergers. This has been confirmed by the observation of the decay of neutron rich radioactive nuclei as a kilonova light curve after the neutron star merger detected with gravitational waves GW170817. We have studied the nucleosynthesis and kilonova light curve of neutrino-driven winds after the merger of two neutron stars and combined it with the dynamical ejecta \cite{Martin.etal:2015}. The nucleosynthesis of the wind strongly depends on the time that the central massive neutron star survives and on the angle of ejection. Matter ejected close to the disk is more neutron rich allowing for the production of elements up to the second r-process peak, while the matter ejected perpendicular to the disk produces only the first r-process peak. This can have consequences for the observed kilonova. Moreover, depending on the mixing of this matter with the dynamical ejecta, one can produce different abundance patterns. If both ejecta are equally mixed, the abundance pattern has high enrichment of heavy r-process elements (from second to third peak) and it is similar to  the Sneden star. In contrast, if in some direction the mixing is not perfect and the wind ejecta is contaminated by a small amount of the dynamical ejecta, then the pattern is closer to the Honda star. Combining these two components of the merger ejecta one could explain abundances of many ultra metal poor stars \cite{Hansen.etal:2014}.

As pointed out above, galactic chemical evolution suggests that an additional astrophysical site has contributed to the r-process at low metallicities. A good candidate are magneto-rotational driven supernovae. In these explosions, part of the matter of the outer layer of the neutron star is rapidly ejected by the magnetic field before the neutrinos can change the neutron richness. This matter collimates along jets due to the magnetic fields and fast rotation leading to a jet-like explosion. Several groups have found r-process favourable conditions along these jets (e.g.,~\cite{Winteler2012, Nishimura2017, Moesta2017}). However, those simulation were done with simple neutrino treatment. Based on recent simulations with M1 neutrino transport \cite[]{ObergaulingerAloy2017}, we have explored the different components of the nucleosynthesis and found proton-rich ejecta, slightly neutron-rich, and very neutron rich. Therefore, we can conclude that these magneto-rotation explosions produce a variety  of nucleosynthesis going from standard neutrino-driven supernovae to the r-process. However, we find that the abundances of the third peak are relative low compared to the second peak. More simulations are necessary to understand this important contribution to the r-process at low metallicities.

\subsection{Core-collapse Supernovae as the Initial Conditions of Neutron Star Mergers (Sean Couch)}

All neutron stars are formed as the result of massive stellar core collapse and the supernova explosions that follow.
Direct multimessenger observations of NSM's \cite{Abbott17a} will reveal a wealth of information about the nature of their constituent neutron stars, but a complete picture of the {\it population} of NSM's will require a firm theoretical understanding of core-collapse supernovae (CCSNe), in which neutron stars are born. 
Simulating CCSNe requires many of the same physical inputs as for NSM's: high-resolution 3D magnetohydrodynamics, general relativity, microphysical nuclear equation of state, and accurate neutrino transport.
This is a difficult computational challenge, necessitating approximations of various kinds. 
Historically speaking, emphasis has been placed on neutrino transport over general relativity in the simulation of CCSNe relative to NSM's, but that gap is slowly closing.
Ultimately, identical simulation frameworks are likely to be used for simulating both NSM's and CCSNe.
Thus, technical progress in either area is naturally beneficial to the other.

The last several years have seen substantial progress in our theoretical understanding of CCSNe, driven in large part but the ability to execute high-fidelity 3D simulations of the CCSN mechanism (e.g., \cite{hanke:2013,lentz:2015,roberts:2016,melson:2015a,summa:2018,oconnor:2018b}).
3D simulations of the CCSN mechanism have, so far, revealed several interesting and important aspects of the problem.
First, in general, the conditions for explosion are somewhat less favorable than for 2D simulations \cite{tamborra:2013,couch:2014,lentz:2015}. 
This is a result of 2D being prone to exaggerated growth of key instabilities such as the standing accretion shock instability (SASI; \cite{blondin:2003}) and the unphysical behavior of turbulence in 2D.
3D simulations have also shown that turbulence plays a key role in achieving successful explosions \cite{murphy:2013,couch:2015a}. 
Recent work has revealed that the 3D structure of CCSN {\it progenitors} is also an important aspect to the problem and including such 3D structure in the initial conditions for CCSN simulations generally aids in achieving energetic explosions \cite{couch:2013b,couch:2015,muller:2015,muller:2016a,muller:2017}.
3D CCSN simulations have also uncovered the presence of new instabilities such as the lepton-number emission self-sustained asymmetry (LESA; \cite{tamborra:2014,oconnor:2018b}).

While it is now clear that the CCSN mechanism is fundamentally 3D, recent progress in 2D simulations has been, nevertheless, encouraging. 
There is emerging {\it quantitative} as well as {\it qualitative} agreement in the results for 2D simulations from different groups when simulating the same initial conditions \cite{bruenn:2016,summa:2016,burrows:2016,oconnor:2018}.
These works find similar explosion times and energies for a handful of progenitors in the mass range 12--25 M$_\odot$, with the only exception being \cite{bruenn:2016}, who find generally earlier and more robust explosions for all progenitors simulated.
Still, the broad qualitative agreement amongst results in 2D is encouraging.
New efforts are underway to carry out controlled code-to-code comparisons with the goal of understanding the origin of quantitative differences in results from different groups and simulation approaches \cite{oconnor:2018a,pan:2018a,just:2018,cabezon:2018}.

CCSNe arise from an incredible variety of initial conditions, a parameter spaces that includes dimensions of progenitor mass, metallicity, rotation rate, binarity, etc. 
In order to make more direct connection with observations, including those of NSM's, we must strive to make predictions of the {\it population} statistics of CCSNe, as well as their result neutron star and black hole remnants. 
In 3D, covering the entire parameter space with adequate sampling is still computationally unfeasible, but we are beginning explore these aspects more fully.
The near future is likely to see continued significant progress in our understanding of massive stellar death, as well as in the our technical ability to execute these quintessentially multiphysics calculations. 

\subsection{Searching for the origin of the $r$-process rare-earth abundance peak with neutron-rich measurements and Markov Chain Monte Carlo (Nicole Vassh)}

\begin{figure}
\includegraphics[scale=0.45]{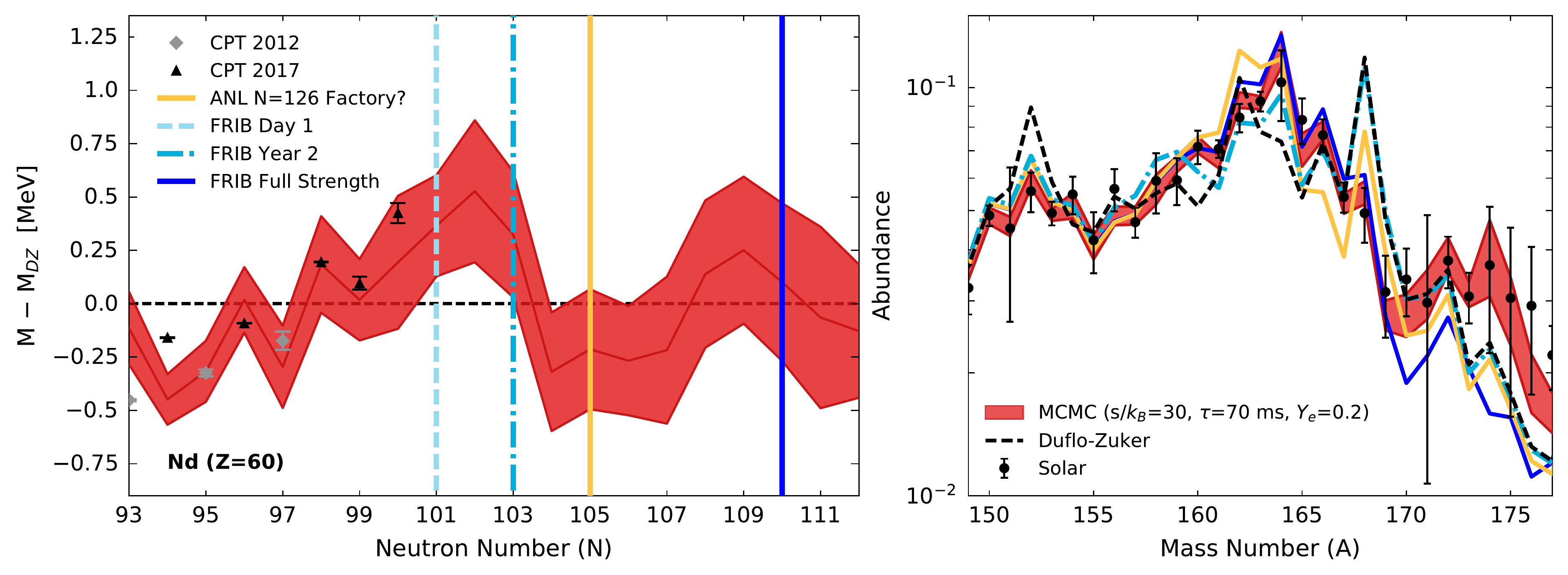}
\caption{(Left Panel) Comparison between CPT measurement values (black triangles) and theoretical predictions (red band) for the nuclear masses relative to the Duflo-Zuker mass model for neodymium isotopes in a merger accretion disk wind scenario ($s/k_B = 30$, $\tau=70$ ms, and $Y_e=0.2$). The vertical lines show the reach of future experiments with blue lines for the FRIB range at day 1, year 2, and full strength and a yellow line to represent an optimistic range for the ANL N=126 Factory. (Right Panel) The rare-earth peak abundances found when the masses values from the left panel are applied out to the experimental ranges described with Duflo-Zuker masses assumed for the neutron-rich isotopes beyond the indicated range.}
\label{fig:NVmassab}
\end{figure}

We now have observational indications that lanthanide material is synthesized during neutron star mergers from the electromagnetic counterpart of the multi-messenger event GW170817 \cite{Abbott17a,Abbott17b,cowperthwaite17}. For $r$-process nucleosynthesis within merger conditions, lanthanide production can vary significantly due to the astrophysical uncertainties (such as the exact neutron-richness) suggested by simulation. Uncertainties are further compounded by a limited knowledge of nuclear physics far from stability. For instance the feature suggesting an enhancement in lanthanide abundances at $A\sim164$ which is the rare-earth element peak is not robustly produced in $r$-process calculations. The possibility that formation occurs via a dynamic mechanism, with the $r$-process path encountering a local nuclear physics feature such as a subshell closure during the decay to stability \cite{Reb97,Matt12}, has remained open given nuclear physics unknowns. Promising advancements in experimental measurements, such as the recently reported neutron-rich mass measurements by the CPT at CARIBU collaboration \cite{OrfordVassh2018}, and future radioactive beam facilities such as FRIB could soon be in a position to evaluate the viability of the dynamic mechanism of peak formation as they explore the rare-earth region. Since both observational and experimental approaches relevant for $r$-process physics have made clear advancements, we should seek to advance the theoretical tools used to study heavy element nucleosynthesis as well. We make use of a promising, modern theoretical approach which employs the well-established statistical techniques of the Metropolis-Hastings algorithm and Markov Chain Monte Carlo (MCMC) \cite{REMM1,REMM2}. With these tools, we invert the traditional approach of evaluating nuclear mass models through their $r$-process abundance predictions and instead use the observational data \cite{solardata,goriely99} for the rare-earth peak to find the nuclear masses required to fit this region. For this procedure, we make use the mass parameterization $M(Z,N) = M_{DZ}(Z,N) + a_N e^{-(Z-C)^2/2f}$ to make predictions for the mass corrections to the Duflo-Zuker mass model (here we set C=60 and f=10). We model 28 $a_N$ parameters for $N=93-120$ and update separation energies, Q-values, $\beta$-decay rates, and neutron capture rates at each timestep in order to calculate the abundance prediction in a self-consistent manner. In Fig.~\ref{fig:NVmassab}, we show results for a merger accretion disk wind scenario (a hot wind with an entropy of 30 $k_B$/baryon, a dynamical timescale ($\tau$) of 70 ms, and electron fraction ($Y_e$) of 0.20) which undergoes a traditional, hot $r$-process dominated by neutron capture, photodissociation, and $\beta$-decay. Our error bars are determined from taking the standard deviation of the results from 50 independent, parallel MCMC runs. The dip in the red band mass surface of Fig.~\ref{fig:NVmassab} at $N=104$ produces an upward kink in the separation energy surface where the material which eventually forms the peak accumulates. As discussed in \cite{OrfordVassh2018}, the local features in the mass surface predicted by our reverse-engineering analysis given these neutron star merger wind conditions are consistent with the Penning trap mass measurement data recently found by the CPT at CARIBU (black triangles in Fig.~\ref{fig:NVmassab}). The vertical lines in the left panel of Fig.~\ref{fig:NVmassab} show the FRIB reach at day one, year 2, and full design strengths. Measurements from the projected year two FRIB range will test this peak formation scenario by further resolving the mass information influencing the left edge of the rare-earth peak, as can be seen in the right panel of Fig.~\ref{fig:NVmassab}. An optimistic range for the ANL N=126 Factory will reach the key $N=104$ feature we find responsible for material pile-up in this scenario. Should consistency with the predicted MCMC masses be found by FRIB at full design strength, the structure of the rare-earth peak could be almost fully resolved. Since this MCMC method is intended to be used to gain new insights into the astrophysical site of rare-earth peak production, we must perform this procedure for a variety of astrophysical conditions in order to differentiate between the trends in the mass surface required to fit the rare-earth solar data. We have completed calculations for both hot and cold very neutron-rich merger dynamical ejecta conditions, as well as accretion disk winds which are colder than those considered in Fig.~\ref{fig:NVmassab}, and look forward to direct comparisons with current and future experiment.

\subsection{Nucleosynthesis and Electromagnetic Transients from Neutron Star Mergers (Luke Roberts)}
After the observation of GW170817, it has become increasingly important to understand the composition of material ejected during neutron star mergers and how the composition of this material is impacted by non-equilibrium processes at high density. In particular, weak reactions occurring at high densities can strongly impact the neutron richness of material that is eventually ejected. The neutron richness of the material in turn impacts the final nuclei that are produced and the properties of any kilonova that is observed in coincidence with the merger. To produce lanthanides and actinides during the r-process, the ejecta must have $Y_e \lesssim 0.25$ \citep[e.g.][]{lippuner:2015gwa}. In material that reaches high temperatures before it is ejected, electron/positron captures can substantially raise $Y_e$ from its cold beta-equilibrated value. Similarly, strong neutrino irradiation can increase the $Y_e$ of the material in situations were the spectra of the electron neutrinos and antineutrinos are similar. Both of these processes can push significant amounts of material above $Y_e = 0.25$ and qualitatively impact the eventual nucleosynthesis products. Therefore, modeling these weak reactions in neutron star mergers is of paramount importance. Electrons and positrons are thermalized throughout the entire merger process, so electron and positron captures in the ejected material are easily tracked. On the other hand, charged-current neutrino interactions are more difficult to model since neutrinos are not thermalized and can travel macroscopic distances between interactions with the background medium. The treatment of neutrino transport in neutron star merger simulations to date has relied on approximate methods, at least compared to the state-of-the-art neutrino transport used in the supernova community. This is in part due to technical challenges that are present in mergers but not in core collapse supernovae, such as fluid velocities close to the speed of light and strong deviations from a spherically symmetric geometry. To date, most work has used energy averaged, or gray, transport either using leakage type prescriptions \citep{radice:2016dwd} or moment based transport \citep{sekiguchi:2012, foucart:2016rxm}. From this work, it is clear that the predictions of the electron fraction distribution in the ejected material is sensitive to the details of the transport descriptions used \citep{foucart:2016rxm}. Therefore, a significant effort must be made to improve the treatment of neutrino transport and include neutrino energy dependent transport, as Sherwood and Evan discussed above. The changes induced by the transport prescriptions can be particularly important for understanding the production of the first r-process peak in mergers.

\subsection{Primordial Black Holes and $r$-Process Nucleosynthesis (George M. Fuller)}

Detection of the gravitational wave in-spiral signal from GW170817 and observations of its accompanying kilonova dramatically demonstrate that $r$-process material can be ejected from a violent event in which neutron-rich matter is \lq\lq mined\rq\rq\ from initially cold neutron stars. These observations represent a spectacular achievement, perhaps heralding the advent of a golden age for multi-messenger, time domain astronomy. As to be expected with such impactful observations, many questions are answered, and new questions arise. One question posed is whether compact object mergers produce all of the $r$-process, in particular the heaviest $r$-process species, the actinides.

Simulations of binary neutron star mergers and (non-extreme mass ratio) black hole-neutron star binary mergers show that ejection of neutron-rich material is to be expected in these events. However, these events can be accompanied by significant neutrino radiation. Other than the \lq\lq tidal tail\rq\rq\ material tidally torn out of the stars before they touch, the other ejecta $r$-process sources, outflow directed perpendicular to the disk, and material outflowing along the disk {\it might} (depending on outflow speeds and neutrino flavor physics) have its neutron excess reduced by neutrino and antineutrino charged current captures on nucleons. We do not know whether this is problematic for production of the heaviest $r$-process species in these events. Nevertheless, it is interesting to consider other potential $r$-process sources and what the chemical evolution and transient astronomy implications of these may be.

We do know from observations of abundances on the surfaces of stars in Ultra-Faint Dwarf UFDs) galaxies that $r$-process nucleosynthesis is likely to be a relatively rare event (see Anna Frebel’s contribution in this volume). Though the binary neutron star merger rate and ejection mass properties may be consistent with producing the bulk of the $r$-process, other sources are not yet precluded.

For example, it has been argued by Fuller, Kusenko, and Takhistov, Ref.~\cite{rprocPBH} that a significant fraction of the $r$-process inventory, perhaps all of it, could be produced by rare captures of tiny primordial black holes (PBHs) by neutron stars. PBHs with masses in the range ${10}^{-14}\,{\rm M}_\odot < {\rm M}_{\rm PBH} < {10}^{-8}\,{\rm M}_\odot$, if they comprise a few percent or more of the Dark Matter, are ideal in this regard. A PBH impacting a neutron star, and then losing energy through Landau damping (dynamical friction) in the star’s dense nuclear matter, can be captured and sink to the center of the star. Through accretion of the dense matter at the center, this black hole will grow and eventually consume the entire neutron star. If this PBH capture process involves a millisecond-period neutron star, the resulting spin-up, as the star shrinks in radius, can result in centrifugal ejection of a few tenths of a solar mass of cold, ultra-neutron rich matter. This could result in a classic high neutron-to-seed ratio $r$-process with fission cycling. The details of the $r$-process abundance pattern in this scenario have yet to be simulated and probably depend on the centrifugal mass ejection history. Unlike a binary neutron star merger event, this PBH capture process likely will not result in significant neutrino emission, removing any worry about neutrino reprocessing of the neutron-to-proton ratio.

The ejection of this amount of neutron-rich matter likely would result in a kilonova-like electromagnetic display, albeit one {\it not} accompanied by a gravitational wave in-spiral signal. This feature would provide a means for observational discrimination between compact object mergers occurring within the $~ 200\,{\rm Mpc}$ sensitivity limit of aLIGO, aVIRGO, and KAGRA and, for example, the PBH capture scenario.

Interestingly, as shown in Ref.~\cite{rprocPBH}, millisecond pulsar statistics in the Galaxy and $r$-process chemical evolution histories, including constraints on these from the UFD observations, are consistent with this PBH-capture scenario. Moreover, this scenario is also consistent with the Dark Matter spatial distribution in the Galaxy and UFDs. We would then expect the bulk of $r$-process production in the PBH-Neutron Star capture scenario to take place where the Dark Matter and millisecond period neutron star densities were coincidentally high. For example, higher densities of both occur in the Galactic Center; but in globular clusters, where millisecond pulsars are numerous, there may not be much Dark Matter. It is interesting that two great mysteries, the origin of the $r$-process and the nature of the Dark Matter, may have a connection. 

\section{Observations of r-process elements}

\subsection{Abundances determination uncertainties in r-process stars (Rana Ezzeddine)}
Accurate determination of the chemical abundances of (r)apid neutron-capture enhanced stars hold important clues to our understanding of the different r-process enhancement populations (r-II, r-I, limited-r, etc ..) \cite{Frebel2018,Hansen2018}, which 
could essentially allow us to characterize their different formation channels. 

Determining the atmospheric stellar parameters of the star, including effective temperatures ($T_{\mathrm{eff}}$), surface gravities ($\log\,g$), metallicities ($\mbox{[Fe/H]}$) and micro-turbulent velocities ($\xi_{t}$) is a fundamental cornerstone in any abundance study. These parameters are commonly determined via spectroscopic methods which entails iteratively changing those parameters by removing any abundance trends of Fe\,I lines as a function of excitation potential ($\chi$) and line strengths ($\log(EW/\lambda)$, where $EW$ is the equivalent width of the line), as well as minimizing the average abundance differences between Fe\,I and Fe\,II. 
Chemical abundances in stars are then derived by calculating the emergent flux of
for a given $T_{\mathrm{eff}}$, $\log\,g$, $\mbox{[Fe/H]}$ and and chemical composition and comparing it with observations (spectra, or $EW$). 
Most models of radiative transfer used to determine stellar abundances adopt the assumption of Local Thermodynamic Equilibrium (LTE). However, departures from LTE conditions in the atmospheres of stars may significantly affect determinations of both stellar parameters and chemical abundances, in particular the determinations based on spectral lines of minority species \cite{Mashonkina2011}. 

\begin{SCfigure}
        \caption{Departure coefficients from Local Thermodynamic Equilibrium (LTE) for Sun and the metal-poor stars J\,0313$-$6708 and HE\,1327$-$2326 as a function of atmospheric optical depths ($\log\,\tau_{5000\,\text{\AA}}$), shown for the ground level populations of neutral Fe\,I. The deviation from LTE in the spectral line forming regions ($-2<\log\lambda_{5000\,\text{\AA}}<-1$) is more severe toward lower metallicity stars.}
	\includegraphics[width=0.6\textwidth]{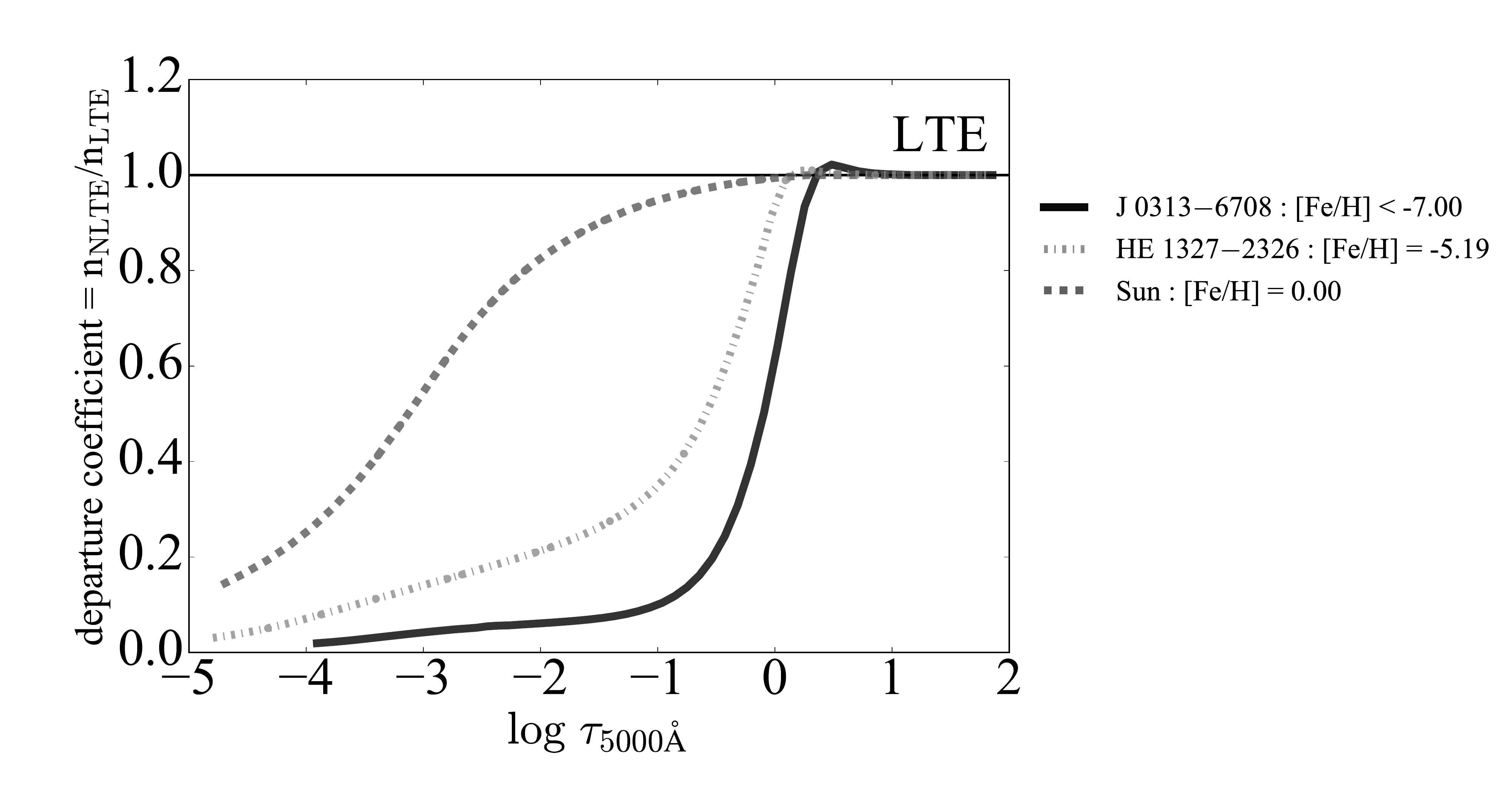}
	\label{fig:ezzeddine}
\end{SCfigure}

Low metallicity stars can suffer from large deviations from LTE. The departures coefficients from LTE of the ground Fe\,I level populations 
(defined by $b={n_{{\rm NLTE}}/{n_{\rm LTE}}}$), as a function of optical depths  ($\log\,\tau_{5000\,\text{\AA}}$) 
are shown in Figure\,\ref{fig:ezzeddine} for two iron-poor stars, as well as the Sun for comparison. 
The departures grow larger from LTE toward stars with lower metallicities and extended atmospheres \cite{Ezzeddine2017}. This is due to decreasing number of electrons donated by metals contributing to collisional rates which create "thermal equilibrium" conditions. Thus non-LTE modeling of the spectral lines of these stars becomes vital for r-process stars which are typically iron-poor ($\mbox{[Fe/H]}<-1.0$) \cite{Frebel2018}.

We thus present a comprehensive NLTE Fe\,I/Fe\,II grid computed with 1D MARCS model atmospheres \cite{Gustafsson1975,Gustafsson2008} in a $T_{\mathrm{eff}}$/$\log\,g$/$\mbox{[Fe/H]}$/$\xi_{t}$ parameter space, and using up-to-date accurate radiative and collisional atomic data. The grid is
defined by $T_{\mathrm{eff}}=[4000;6400, \Delta=50$\,K], $\log\,g=[0.5;5.0, \Delta=0.1$], $\mbox{[Fe/H]}=[-5.00;-0.50, \Delta=0.5$\,dex] and $\xi_{t}=[0.0;5.0, \Delta=0.5$\,km\,s$^{-1}$] (Ezzeddine et al, in prep). Given observed $EW$ measurements for a star, the grid can be used to determine 1D, NLTE atmospheric parameters as well as abundances, equivalent widths ($EW$) and departure coefficients for a large number of commonly used UV, optical and IR Fe\,I and Fe\,II lines. 

\subsection{The $R$-Process Alliance:  Progress and Preview (Timothy C. Beers)}

The $R$-Process Alliance (RPA) is a collective effort of observers, theoreticians, modelers, and nuclear experimentalists with the goal of identifying and understanding the astrophysical site(s) of the $r$-process.  The RPA is presently in the midst of a 5-year observational  effort to quadruple the number of known highly $r$-process-enhanced ($r$-II) stars in the halo of the Galaxy, increasing the numbers of such stars from $\sim$25 to $\sim$100.   

\begin{figure}
\centering
\includegraphics[width=0.7\textwidth]{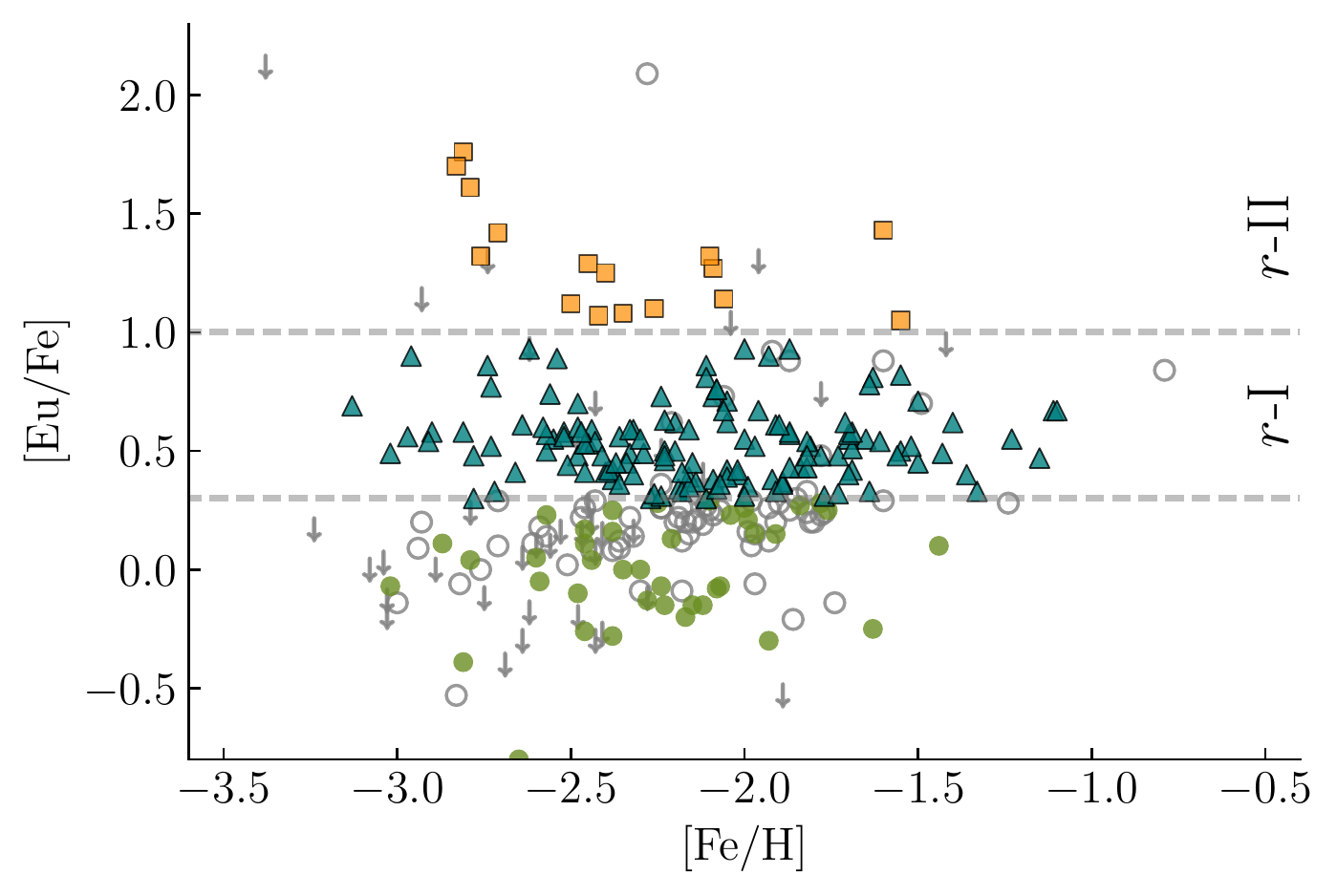}
\caption{Summary of the measured [Eu/Fe] ratios, as a function of [Fe/H], for the subset of snapshot spectra from Phase II of RPA that have been reduced and analyzed to date.  Regions that include 16 newly identified $r$-II and 135 $r$-I stars are marked.}\label{beers:results}
\end{figure}

The first phase of this project, now completed, was the identification of a large sample of bright ($V < 14.0$) stars with $\mbox{[Fe/H]} < -2.0$, and with temperatures suitable for the detection of the element europium (Eu) ($4500 < \mbox{T}_{\rm eff} < 5500$\,K).  Over 3000 such candidates have now been validated with atmospheric parameters calculated from medium-resolution spectroscopy.  Phase II of the RPA involves taking moderately high-resolution ($R \sim 30,000$), modest signal-to-noise ($S/N \sim  30$) ``snapshot” spectroscopy of these candidates, in order to identify the approximately 3-5\% of them that have $\mbox{[Eu/Fe]} > +1.0$, identifying them as $r$-II stars.  The Phase II pilot surveys have already identified 16 new $r$-II stars, with more expected once the data are fully reduced and analyzed  (\cite{hansen18}, \cite{sakari18}, Sakari et al., in prep, Ezzeddine et al., in prep.).  A total of over 950 snapshot spectra have been taken to date, roughly one-third of the eventual goal. Figure~\ref{beers:results} summarizes results from the reduced and analyzed snapshot spectra, which only represent a subset of the data in hand. 

Phase III of the RPA is to obtain higher resolution ($R \sim 60,000$), higher signal-to-noise ($S/N \sim 100$) ``portrait” spectroscopy of the most interesting $r$-process-enhanced stars identified during Phase II.  Papers describing a number of these stars have already been published or are in press (e.g., \cite{placco17} \cite{cain18}, \cite{gull18}, \cite{holmbeck18}), including several with measured abundances of both Th and U.  

It is estimated that the remaining Phase II observations will required another 2-3 years; Phase III is executed in parallel, so should be completed on roughly the same schedule.  The RPA has already begun the effort to account for the nature of the so-called ``actinide boost” stars (e.g., \cite{holmbeck18b}), and to explore the constraints that the observed frequencies and distribution of  [Eu/Fe] and [Eu/H] place on possible astrophysical origin(s) of the $r$-process, which is Phase IV of this effort.  It is expected that new measurements of isotopes involved with the production of $r$-process elements, Phase V of the RPA, will provide new constraints, and raise new questions, which can then be explored with the observations already collected during previous phases of the RPA.

\subsection{The Environment of the \textit{R}-Process (Ian U.\ Roederer)}

Major astronomical advances 
in understanding the nature of the \textit{r}-process
are a result of new environmental constraints that have 
become available in the last three years. Key examples include the detection of many highly \textit{r}-process-enhanced stars in the
ultra-faint dwarf galaxy \mbox{Reticulum~II}
\cite{Ji2016,roederer16ret}, analysis of the electromagnetic counterpart of 
a pair of merging neutron stars detected in
gravitational waves (GW170817; e.g., \cite{cowperthwaite17,drout17,kasen17,tanvir17}),
and the analysis of the kinematic properties 
of a large sample of highly \textit{r}-process-enhanced field stars
\cite{roederer18kinematics}. The occurrence frequency and initial baryonic mass of \mbox{Reticulum~II} favored a single, high-yield
($M_{\rm Eu} \sim 3 \times 10^{-5} M_{\odot}$;
\cite{Ji2016,beniamini16b}) event, like a neutron star merger.
The colors and delay timescale of the
electromagnetic counterpart of GW170817 
demonstrated that merging neutron stars can
host high-yield \textit{r}-process nucleosynthesis events.
Precise kinematic and orbital properties of 
\textit{r}-process-enhanced stars can now be calculated
thanks to the second data release from the \textit{Gaia} satellite.
Their kinematics indicate that these stars were likely 
formed in environments like low-mass dwarf galaxies
that were disrupted by the Milky Way, 
and that a single nucleosynthesis site
could dominate production of the \textit{r}-process elements
in all environments, including the disk of the Milky Way.
These three independent lines of evidence point to
low-frequency, high-yield events like neutron star mergers
as the dominant sites of \textit{r}-process nucleosynthesis.

\subsection{Stellar Spectroscopy and Possibilities for Kilonova Spectroscopy (J. E. Lawler)}
Analysis of $r$-process  elements in stellar photospheres has advanced with the use of accurate absolute atomic transition probabilities or $\log(gf)$s \cite{Sneden09}.  Extensive use of experimental techniques for radiative lifetime measurements from laser induced fluorescence in combination with emission branching fractions from Fourier transform spectroscopy has yielded many atomic transition probabilities with $\log(gf)$ uncertainties as small as a few $\times$ 0.01 dex \cite{Lawler09}.  Emergence of a consistent $r$-process abundance pattern in five different metal poor stars matching the Solar System $r$-process pattern indicates that $r$-process nucleo-synthesis has been stable across space and time \cite{Sneden09}.  The early UV portion of kilonova light curves will likely yield important new information on the $r$-process.  Although sharp line spectroscopy may not be practical due to large (first order) Doppler shifts from ejecta speeds ranging up to several tenths of the speed of light and from the emission of material in wide range of directions, there is still some possibility for spectroscopic structure from ion lines in the early UV portion of the light curve.   A reduction of Doppler shifts in a kilonova observation with favorable orientation is a possibility.   Kilonova spectra in the UV from more highly ionized material might also reveal structure if the relevant electron configurations are simpler.  Laboratory astrophysics on heavy elements that are few times ionized will be essential for interpretation of any structure and for opacity evaluations in the early UV portion of kilonova light curves.   The number of known energy levels of heavy element spectra  decreases rapidly as one moves from neutral, to singly ionized, to few times ionized species.  Analysis of the spectra to first identify the upper and lower levels of spectral lines (to classify the lines) and to label levels with appropriate configurations and terms (to assign the levels) is largely absent for few times ionized heavy elements.  Such a classical analysis of a spectrum is a first step before atomic transition probabilities are determined.  Neutron star merger statistics are likely to improve more rapidly than observations of kilonova light details.  GW detectors are steadily increasing in sensitivity and thus will yield both larger monitored volumes and better event statistics.  Sensitivity/distance scaling of a GW amplitude sensor is more favorable than that of telescopes and electromagnetic detectors which are energy sensitive.  The next generation of large space telescopes and 30 to 50 m ground based telescopes will be very important to studies of kilonova light details.

\subsection{Imprints of Single-Isotope Decay in Kilonova Light Curves (Jennifer Barnes)}

The radioactive decay of unstable nuclei synthesized by the r-process is the source of kilonova luminosity. However, radioactive energy initially takes the form of energetic photons and highly kinetic beta- and alpha-particles and fission fragments. These particles thermalize in the ejecta, upon which their energy is converted to heat and re-emitted as the thermal radiation that comprises the kilonova light curve. In recent years, these thermalizing processes have been studied both analytically \citep{Waxman.ea.2017.knEmission,KasenBarnes.2018.btherm} and numerically \citep{Hotokezaka.ea.2016.gtherm,Barnes2016}.
Though most radioactive \emph{energy} from the r-process is emitted in beta-decays, alpha-decay and fission can be important sources of kilonova \emph{luminosity} for two reasons. First, a typical alpha-decay (fission) releases a factor of 5-10 (${\sim}$100) more energy per decay than a typical beta-decay. Second, this energy thermalizes more efficiently in the merger ejecta \citep{Barnes2016}. This introduces the possibility that a single decaying isotope, if its decay energy and thermalization were sufficiently high, could supply a substantial fraction of the luminosity on timescales close to the isotope’s decay timescale.
The effect is likely to be particularly strong is the isotope is long-lived, in which case the energy the energy will be injected against a low-power background formed from beta-decays of decreasing energy and frequency.
Understanding these signatures is important for estimating ejecta masses, and may also be a useful probe of the composition burned in the merger that generated the kilonova.
One example of such in imprint was discussed in \cite{zhu2018californium} and is summarized in~\ref{subsec:YLZCf254}.

\subsection{Californium-254 and Kilonova Light Curves (Yonglin Zhu)} \label{subsec:YLZCf254}

In the 1950s, before it was known that supernova light curves were driven by the decay of nickel and cobalt, it was speculated that they might be driven by r-process nuclei, for example, the heavy, long-lived, neutron-rich Californium-254~\cite{baade1956supernovae, fields1956transplutonium}.  Californium-254 has a measured half-life of $60.5\pm0.2$ days~\cite{phillips1963spontaneous} with spontaneous fission as the dominant decay mode.

\begin{figure}[ht!]
    \centering
    \begin{subfigure}
            \centering
        \includegraphics[width=.48\textwidth]{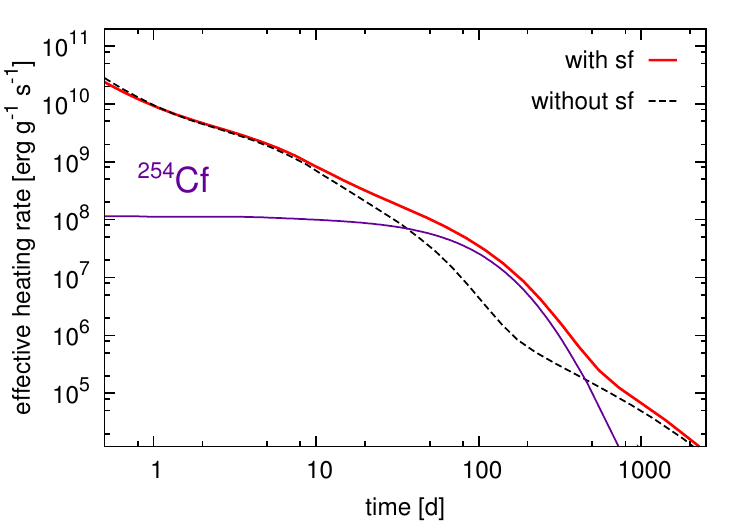}
            \end{subfigure}%
    ~ 
        \begin{subfigure}
            \centering
        \includegraphics[width=.48\textwidth]{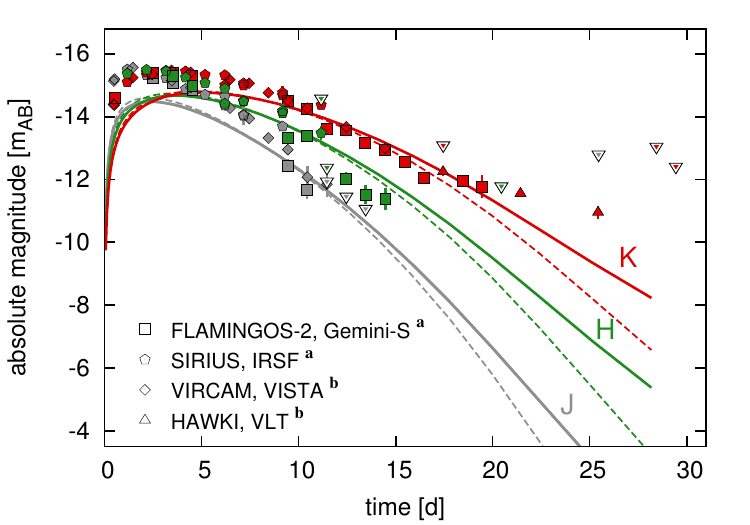}
            \end{subfigure}%
            \caption{Figures from \cite{zhu2018californium}. The panel on the left shows the effective heating rates in the case with and without the spontaneous fission of  Cf-254. The panel on the right shows observations in different bands of the near infrared (points).  Also shown in the panel on the right is a theoretical model with (solid lines) and without (dashed lines) the spontaneous fission of Californium.}
\label{fig:cf254}
\end{figure}

With the recent detection of the electromagnetic counterpart to the gravitation wave signal, GW170817, of two merging neutron stars (NSMs), the issue of the radioactive decay of Californium-254 in light curves has resurfaced \cite{zhu2018californium}. The observed signals in X-ray, UV, Optical, IR, and radio have been interpreted as coming mainly from freshly synthesized r-process nuclei by many authors (e.g.~\cite{kasen17}). Ref.~\cite{zhu2018californium} has explored several issues concerning the nuclear physics of Cf-254 production and its role in kilonova heating.  With its long half life and high $Q$-value of $200$ MeV, Cf-254 creates an unique imprint of the nucleosynthesis of actinide material. Comparing the resulting synthetic light curves to observational data in the right panel of Fig.~\ref{fig:cf254}, we see that the case without spontaneous fission results in dimmer light curves by almost two magnitudes in the near-infrared JHK-bands at twenty-five days after explosion. 
We conclude that the late time light curve is an important diagnostic for the production of the actinides.

\subsection{Is the $R$-Process the Only Source to Produce Barium at Extremely Low Metallicity? (Jinmi Yoon)}

\begin{figure*}
\centering
\subfigure[Cumulative frequencies]{\label{fig:a}\includegraphics[width=0.45\textwidth]{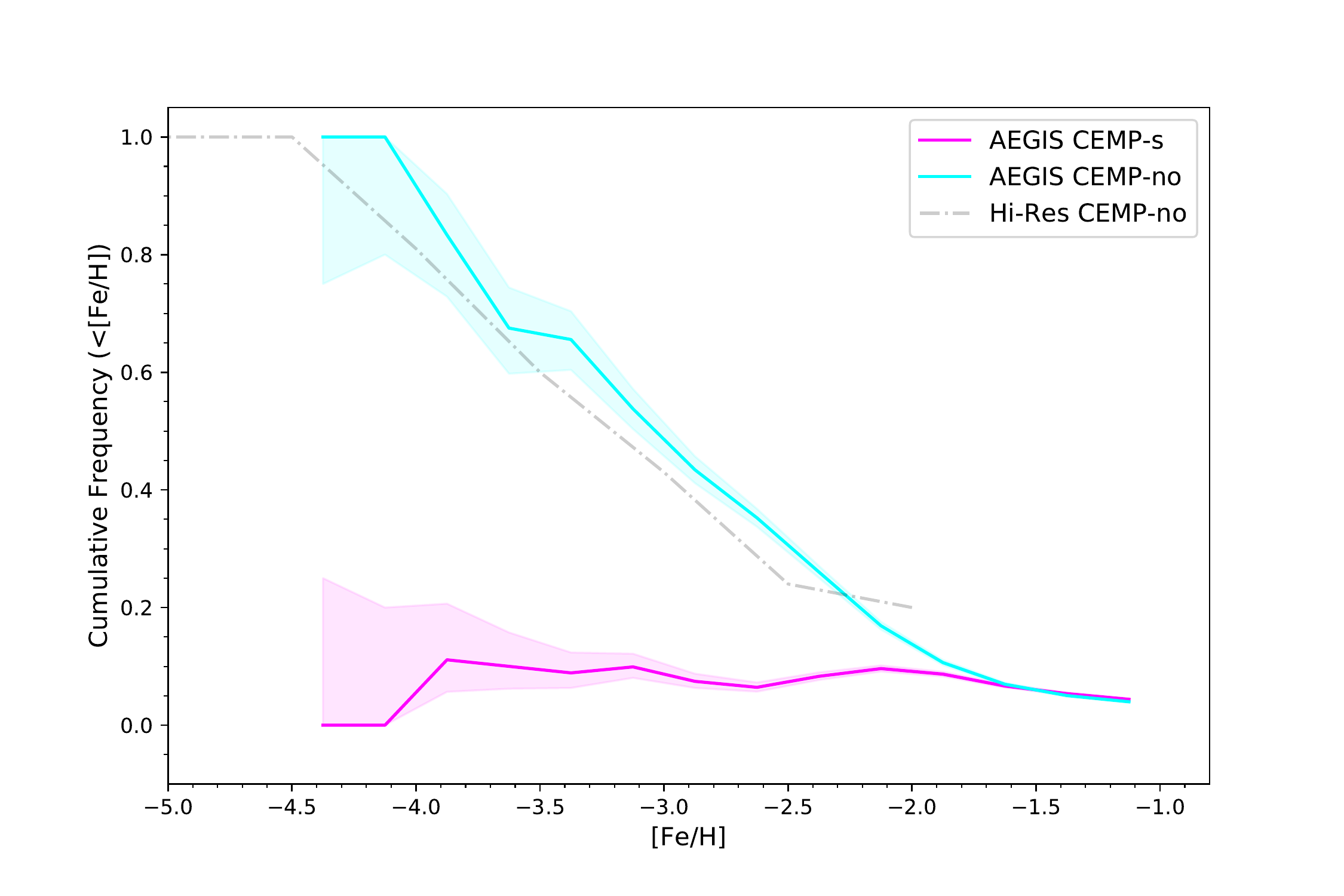}}
\subfigure[ Differential frequencies]{\label{fig:b}\includegraphics[width=0.45\textwidth]{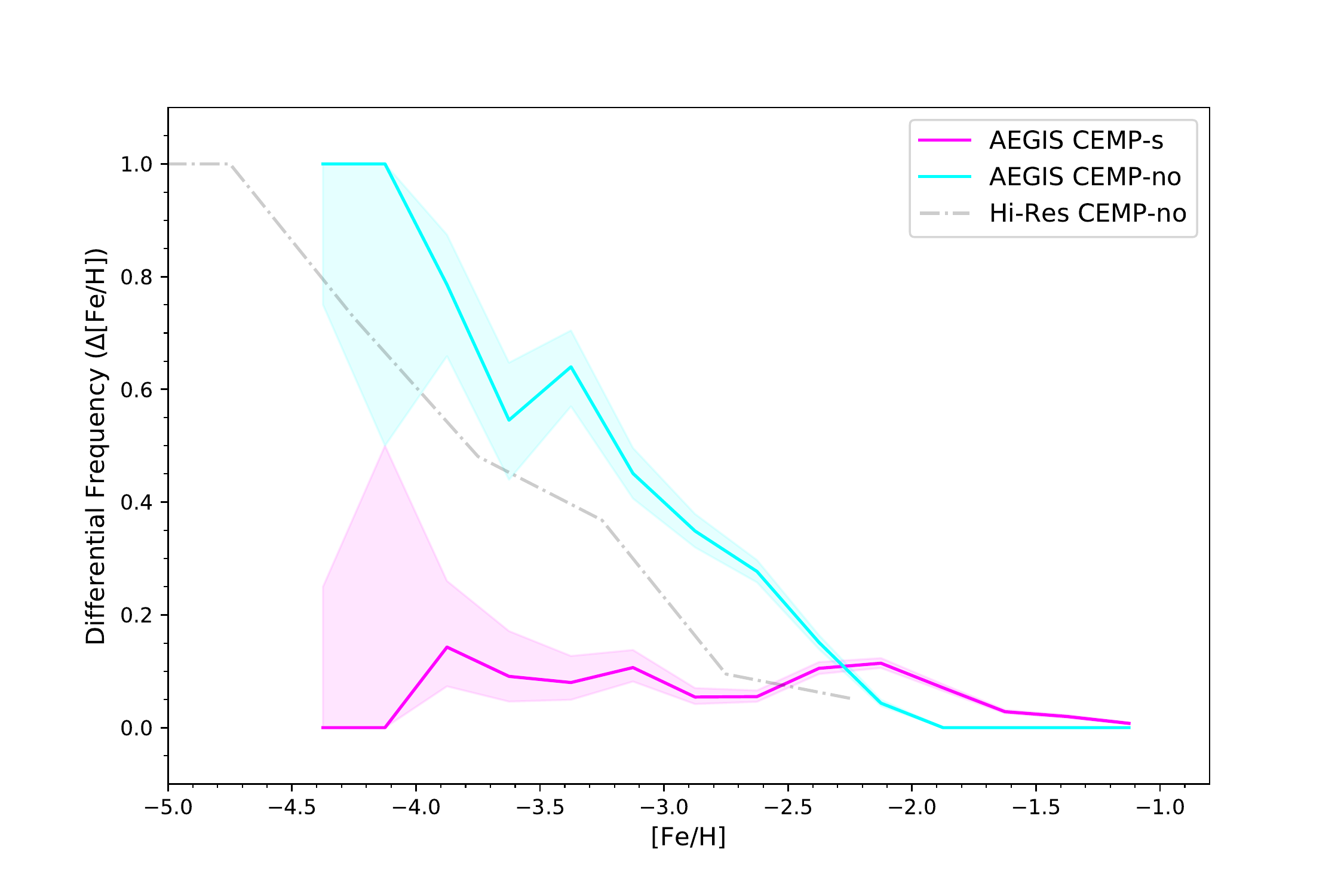}}
	\caption{Panels (a) and (b) represent the cumulative frequencies and differential frequencies of the AEGIS CEMP-no and CEMP-$s$ subgiant and giant (SG/G) stars, respectively. The frequencies are calculated relative to all SG/G stars.  The cyan and magenta lines represent the frequencies of CEMP-no only and CEMP-$s$ only stars, respectively. The lightly colored shaded areas around the solid lines indicate 1$\sigma$ Wilson approximation \cite{wilson1927} of confidence interval. The gray dash-dotted lines represent the high-resolution sample of CEMP-no stars from Placco et al. \cite{placco2014c}. \label{cemp-frequency}}
\end{figure*}

Carbon-enhanced metal-poor (CEMP \footnote{CEMP-$s$ : [C/Fe] $\geq$ +0.7, [Ba/Fe] $>$ +1.0, and
[Ba/Eu] $>$ +0.5 \newline CEMP-$r$ : [C/Fe] $\geq$ +0.7 and [Eu/Fe]$>$ +1.0
 \newline CEMP-$r/s$ or CEMP-$i$ : [C/Fe] $\geq$ +0.7 and 0.0 $<$ [Ba/Eu] $<$+0.5 
\newline CEMP-no : [C/Fe] $\geq$ +0.7 and [Ba/Fe] $<$ 0.0}) \cite{beers2005} stars are important astrophysical tracers of various nucleosynthetic processes, including the neutron-capture processes, that occurred in the early Universe. 
Here, we report on an indirect test of the role of that the $r$-process played on Ba production in the extremely metal-poor environment, based on the frequencies of 
CEMP stars as a function of metallicity, using a large data set from a medium-resolution ($R \sim$ 1300) spectroscopic survey (AEGIS survey, \cite{yoon2018}). We calculated the frequencies of one of the dominant CEMP sub-classes, the CEMP-$s$ stars  
(possibly including CEMP-$i$ stars), whose origin is associated with binary mass-transfer of asymptotic giant branch (AGB) stars \cite[i.e.,][]{herwig2005, bisterzo2011, abate2013, hampel2016} along with another dominant sub-class, the CEMP-no stars, whose progenitors are likely massive first stars \cite[i.e.,][]{meynet2010, nomoto2013, yoon2016}.  
We made use of the characteristically higher absolute carbon abundance, $A$(C)\footnote{ $A$(C) = $\log\,\epsilon$(C) = $\log\,$($N_{\rm C}/N_{\rm H}$)+12, where $N_{\rm C}$ and $N_{\rm H}$ represent number-density fractions of carbon and hydrogen, respectively.} $\geq$ 7.1 for CEMP-$s$ stars, to isolate them from CEMP-no stars with lower $A$(C) \cite{yoon2016}. This method is a much more efficient way to classify the dominant CEMP sub-classes than the conventional method of using high-resolution spectroscopic abundances of [Ba/Fe] $> +1.0$; it has been shown to be just as effective as the conventional method.
The frequencies CEMP-$s$ stars were calculated for the first time in this study, and are  shown along with those of the CEMP-no stars in Figure ~\ref{cemp-frequency}. 
The frequencies of the CEMP-$s$ stars are substantially lower than those of the CEMP-no stars and, interestingly, remain almost constant at a value of $\approx$10\% at [Fe/H] $\lesssim -$2.0, even at metallicities close to [Fe/H]$\sim -$4.0. This behavior may indicate that metallicity does not have a significant impact on the operation of the $s$-process. More importantly, the non-zero CEMP-$s$ frequency at the lowest metallicity challenges that notion that the $r$-process is the only source to contribute Ba production at [Fe/H]$<-$3.0. Perhaps the $s$-process and/or $i$-process may contribute to Ba production through AGB stars or rapidly rotating high-mass stars (spinstars, see \cite[i.e.,][]{meynet2010}), even before neutron star merger events had first happened. This may alter, at least to some degree, our current understanding of the role of the $r$-process in nearly chemically pristine environments.  However, at present, the number statistics at [Fe/H]$< -$3.5 is still too low to be clear on these points. We plan to use larger data sets such as SDSS/SEGUE to further explore whether or not this constant rate at low metallicity is a reliable estimate.

\section{Theory of neutrinos, neutron-rich nuclei, and dense matter}

\subsection{Constraining the Equation of State and the termination of the $r$-Process from GW170817/GRB170817A (Grant J. Mathews)}

Never in my lifetime has Nature revealed herself in such a magnificent way as the combined gravity waves from GW170817 \cite{Abbott17a} and the associated multimessenger \cite{Abbott17b} gamma-rays from GRB170817A, along with the observations in X-ray, UV, Optical, IR, and radio of the kilonova evolution.   In my mind there are two valuable revelations for nuclear physics that we have gleaned from this  event.  The deduced tidal polarizability \cite{Abbott17a} of the neutron stars has placed a valuable independent constraint on the compactness of neutron stars.  This in turn constrains the nuclear equations of state to those of modest radii  (e.g. \cite{Abbott17a, Fattoyev18}).  A follow-up analysis of the Chirp based on studies \cite{Suh17} we completed just months before the event suggests that the constraint on radii may even be stronger.  This may pose a new dilemma for nuclear physics.  The equations of state  that satisfy the compactness constraint can barely achieve a maximum mass of 2 M$_\odot$.  While at the same time there are a number of neutron stars in X-ray binaries that appear to have a mass of $\ge 2.5$ M$_\odot$ \cite{Lattimer12}.  Although the error bars are large, if the maximum mass turns out to be that large, then there is something very fundamental that is not understood about the nuclear equation of state.  One possibility is a transition between soft to stiff somewhere in the interior of neutron stars.  A second outcome of the observed kilonova that I find very intriguing is the possible evidence \cite{Smartt17} of Cs and Te absorption features in the kilonova spectrum in the days following the event.  These are exactly the elements one would expect if the $A=130$ peak were formed in the $r$-process.  In \cite{Shibagaki16} we pointed out that there would be a very different $r$-process abundance distribution emerging from  neutron star mergers depending upon whether the r-process path terminates near $A sim 285$ or 300.  Based upon the KTUY model we showed that there may be no $A=130$  $r$-process peak in the latter case for which symmetric fission is predicted, while the asymmetric fission yield distribution for a termination near $A \sim 285$ would lead to robust $r$-process peaks.  The presence of Cs and Te absorption features would confirm the existence of the $A=130$ peak confirming the termination of the $r$-process at lower mass numbers.  Clearly,  there is a need for better determination of fission barriers and yields as discussed at this workshop.  There is also a need for data from more neutron-star merger events and determinations of neutron-star masses to resolve the developing dilemma of the equation of state for neutron-star matter.

\subsection{Role of phase transition in the tidal parameters of binary neutron stars (Sophia Han)}

During the late stage of binary neutron star inspiral, while tidal effects are the largest to measure, the gravitational wave signal is too complex relying on the prediction from numerical simulations. Flanagan and Hinderer \cite{Flanagan:2007ix,Hinderer:2007mb} pointed out that in the early part of the phase evolution, a small but clean signature is also measurable, which can be characterized by the EoS-dependent tidal deformability $\lambda$ and Love number $k_2$. A recent binary neutron star (BNS) merger event GW170817 detected by the LIGO-Virgo (LV) collaboration \cite{LIGO:2017qsa} placed a first constraint on the dimensionless tidal deformability $\Lambda$, however in that analysis the EoS dependence of two individual neutron stars were treated uncorrelated. Employing the quasi-universal relation $\Lambda_{a} (\Lambda_{s}, q)$ \cite{Yagi:2016bkt,Chatziioannou:2018vzf}, Refs. \cite{LIGO:2018exr, LIGO:2018wiz} reanalyzed the data and claim limits e.g. on the $\Lambda$ value for $1.4\,{\rm M}_{\odot}$ neutron star to be $70\leq\Lambda_{1.4}\leq580$ (at $90\%$ confidence level) for low spin priors, but possible phase transitions were not taken into account.

It has been realized that assuming two individual neutron stars obey the same normal nuclear matter EoS, the weighted average tidal deformability $\tilde{\Lambda}$ in the BNS system as a function of the chirp mass ${\mathcal M}=(m_{1}m_{2})^{3/5}/(m_1+m_2)^{1/5}$ which can be accurately measured during the inspiral, is relatively insensitive to the unknown mass ratio $q=m_2/m_1$. We investigate how the degeneracy is altered when a sharp first-order phase transition from normal nuclear matter to quark matter takes place in the interior of neutron stars, and determine the most sensitive phase transition parameter to tidal deformation in the binary. For hadronic EoSs, there is a small spread of $k_2$ values, and the trend with varying neutron star masses are similar; for self-bound strange quark stars (SQS), $k_2$ behavior is qualitatively different with much higher values \cite{Hinderer:2009ca}. The generic feature for both $k_2(M)$ and $\Lambda(M)$ in the context of phase transition is the abrupt change above the transition pressure $p_{\rm trans}$, where a dense quark core emerges (if there is a stable branch). We find that the lowest value of the dimensionless tidal deformability for a neutron star with typical mass is given by the dense matter equation of state (EoS) that characterizes sharp phase transition in its interior around nuclear saturation density, transforming from soft hadronic matter to stiff quark matter. The two separate families, purely-hadronic stars and stable hybrid stars, could potentially be distinguishable in the future if more data from advanced LIGO of binary neutron star mergers were to even lower the upper limit of $\tilde{\Lambda}$, or provide a refined range estimate for given chirp mass statistically from multiple events. It is of great interest to study how possible phase transitions would modify theoretical predictions for e.g. post-merger GW signal, density and temperature evolution of the merger remnant \cite{Most:2018eaw}, ejecta mass, kilonova light curves, and r-process abundances.

\subsection{Neutrinos and nucleosynthesis (A.B. Balantekin)}

Astrophysical environments where various nucleosynthesis processes may take place also contain a very large number of neutrinos and antineutrinos. Energy spectra of electron neutrinos and electron antineutrinos determine neutron-to-proton ratio, a controlling parameter of nucleosynthesis. Hence neutrino properties, and especially processes which would alter their flavor content, directly impact element formation. The emergent phenomenon of collective neutrino oscillations arises from neutrino-neutrino interactions in environments with very large number of neutrinos. Since such environments are likely sites of heavy-element synthesis, understanding all aspects of collective neutrino oscillations seems to be necessary for a complete accounting of nucleosynthesis. The exact eigenvalues and the eigenstates of the Hamiltonian describing collective neutrino oscillations can be written in terms of the solutions of a set of Bethe ansatz equations and for the single-angle approximation, where the angles between the momenta of the scattering neutrinos is averaged over, these equations take a particularly simple form \cite{Pehlivan:2011hp}. Finding the roots of the Bethe ansatz equations, because of their nonlinear nature, to obtain an exact solution of the collective oscillations is very difficult. Typically calculations of the collective neutrino oscillations in the literature use the mean-field approximation: The two-body term in the Hamiltonian is replaced by a product of a one-body operator and an appropriately chosen average of a second one-body operator, i.e. a test neutrino interacting with a``mean-field" created by all the other neutrinos. A solution in the single-angle approximation was recently given without using the mean field approximation 
\cite{Birol:2018qhx}. Assuming that the conditions are perfectly adiabatic, it was shown that an initial state which consists of electron neutrinos and antineutrinos of an orthogonal flavor develops a spectral split at exactly the same energy predicted by the mean field formulation.

\subsection{Uncertainty quantification through Bayesian machine learning (W. Nazarewicz)}

In the context of many structural and astrophysical applications, the
challenge is to carry out reliable model-based extrapolations into the
regions where experimental data are not available.
To this end, one can improve a theory's predictive power by comparing model predictions
to existing data. Here, a powerful strategy is to estimate residuals by
developing an emulator for $\delta (Z,N)$ using a \textit{training} set of
known data. An emulator $\delta ^{\mathrm{em}}(Z,N)$ can, for instance, be
constructed by employing Bayesian approaches, such as Gaussian processes,
neural networks, and frequency-domain bootstrap \cite{Athanassopoulos2004,Bayram2017,Yuan2016,Utama16,Utama17,Utama18,Niu2018}. The unknown separation energies can then be estimated by combing theoretical predictions and estimated residuals.
It is worth noting that by developing reliable emulators $\delta ^{\mathrm{em}}(Z,N)$, which take into account correlations between data for different
nuclei, one can significantly refine mass predictions and estimate
uncertainties on predicted values.
Moreover, since the surface of residuals $\delta (Z,N)$ contains important
insights about model deficiencies, by studying the patterns of $\delta ^{\mathrm{em}}(Z,N)$ one can make progress in developing higher-fidelity
models.

In order to improve the quality of model-based predictions of nuclear properties  of rare isotopes far from stability, we considered the information contained in the  residuals  in the regions where the experimental information exist. As a case in point, in Ref.~\cite{Neufcourt2018}, we discussed two-neutron separation energies of even-even nuclei.
A similar strategy of correcting model predictions outside the training domain by estimated residuals  has recently  been applied in Ref.~\cite{Utama18}. The new aspect of our work is that we apply
the Bayesian method to provide a full quantification of the uncertainty surrounding the point estimate.

\subsection{Nuclear fission theory (Nicolas Schunck)}

\begin{figure}[!ht] 
\begin{center}
\includegraphics[clip=false,width=0.55\textwidth]{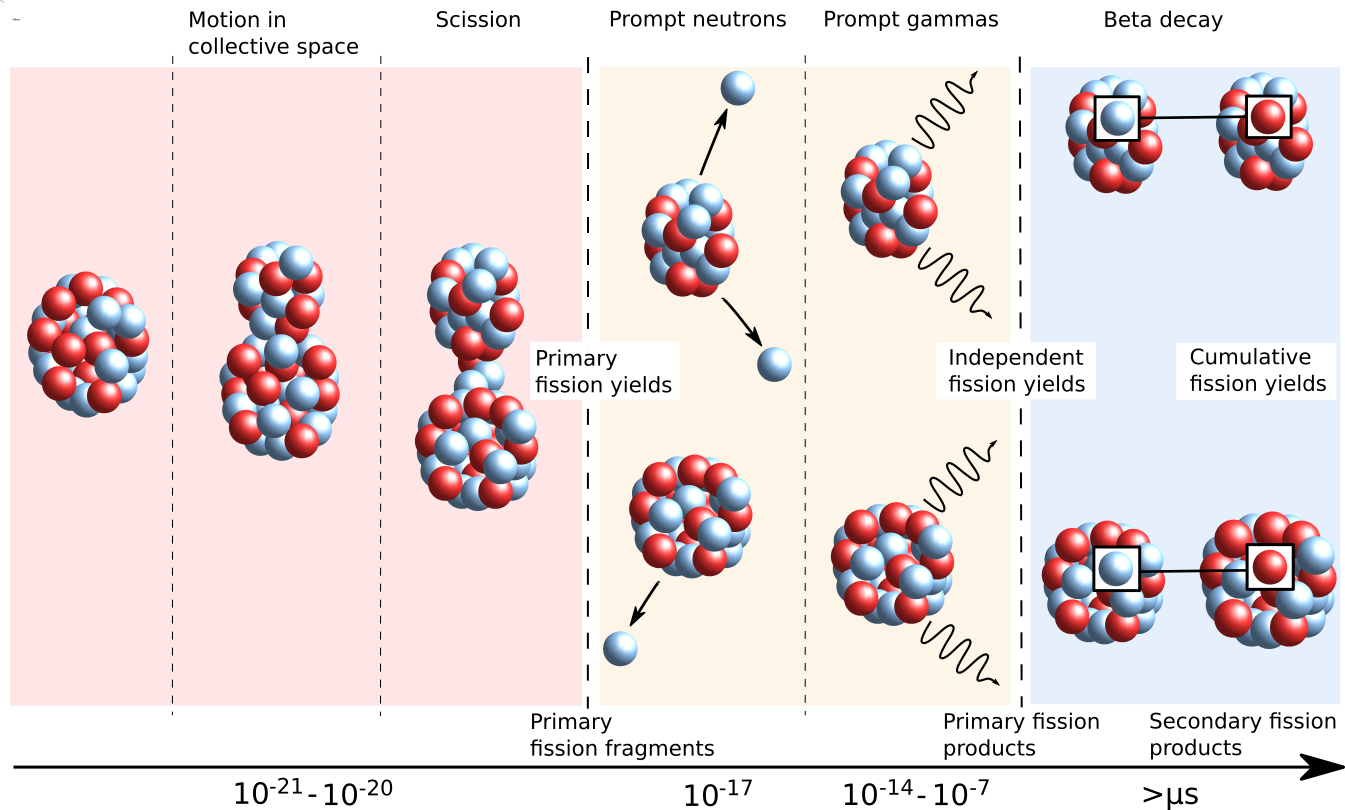}  
\includegraphics[clip=false,width=0.38\textwidth]{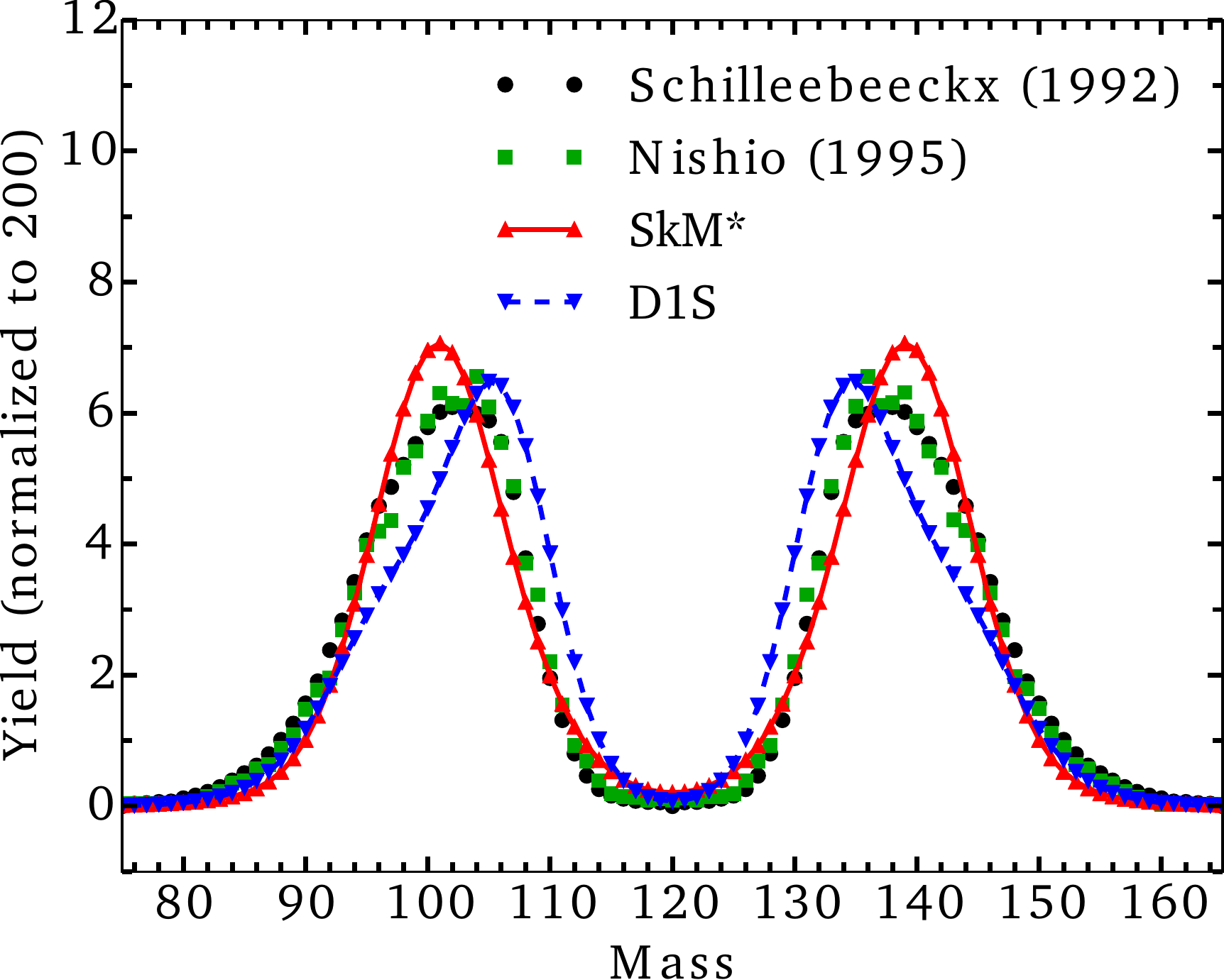}  
\caption{\label{schunck_fission} Left panel: schematic view of the fission process. The fissioning nucleus may be in its ground-state (spontaneous fission) or formed in a nuclear reaction (neutron capture, photofission, etc.). The de-excitation of the fission fragments through neutrons and $\gamma$ emission and their $\beta$-decay back to stability change the charge and mass distributions. Right: example of primary fission yields computed in the time-dependent generator coordinate method of nuclear density functional theory.}
\end{center}
\end{figure}

Nuclear fission remains one of the most complex scientific problems; see left panel of Fig.\ref{schunck_fission} for a schematic description. Early models of fission based on the macroscopic-microscopic approach to nuclear structure have been very successful in explaining a number of qualitative and semi-quantitative features of the fission process, such as the existence of fission isomers and asymmetric fission; see \cite{bolsterli1972,brack1972,bjornholm1980} for reviews. In recent years, these approaches have been complemented by direct simulations of fission dynamics based on various approximations to the stochastic Langevin equation \cite{randrup2011,randrup2011a,randrup2013,ishizuka2017,usang2017,ward2017}. In the short term, such models might have the greatest potential to improve on our understanding of the role of fission in the $r$-process because they lend themselves relatively well to large-scale calculations. However, this class of methods do not treat quantum many-body effects explicitly and mostly ignore nuclear interactions between nucleons. As a result, the theory is not consistent (there is a risk of over-fitting its various parameters) and its predictive power may be limited.

Nuclear density functional theory has emerged in recent years as a viable alternative for a comprehensive theory of nuclear fission \cite{schunck2016}. It is a fully quantum mechanical description of the nucleus. Information about in-medium nuclear forces is encoded in the energy density functional, which can be used to rigorously compute ground-state properties, response functions (as in $\beta$-decay or $\gamma$-emission), time-dependent phenomena, including fission. In the past 5 years alone, DFT simulations of fission have shown spectacular progress, including accurate predictions of fission fragment distributions in neutron-induced  \cite{regnier2016,zdeb2017} and spontaneous fission \cite{sadhukhan2016,sadhukhan2017}, spontaneous fission half-lives \cite{staszczak2013,baran2015}, real-time simulations of fission events and properties of the fission fragments \cite{goddard2015,scamps2015,goddard2016,bulgac2016,tanimura2017} or applications to $r$-process elements \cite{rodriguez-guzman2014,rodriguez-guzman2014a,giuliani2018}. The right panel of Fig.\ref{schunck_fission} shows but one single example of what is now possible. While DFT provides a consistent, global framework for nuclear calculations, putting everything together (masses, decay rates, fission fragment distributions, etc.) into a single calculation of $r$-process abundances will require very large computational capabilities and substantial effort on code development. Maintaining and developing US-based workforce with the right skills will also be essential.

\subsection{Termination of the $r$ process via fission (Matthew Mumpower)}\label{sec:Mumpower}

In the most extreme neutron-rich environments of merging neutron stars the extent of heavy element nucleosynthesis is limited by the process of fission where heavy nuclei break up into lighter fragments. 
It is of particular importance to determine the details of this termination and the extent to which the heaviest elements can be synthesized as the fragment yields may play a large role in determining the final abundances observed in the fission product region. 
Studies have shown that the primary fission channels in the $r$ process are neutron-induced and $\beta$-delayed fission \cite{Eichler2015, Mumpower2018a} which gives focus to theoretical efforts in the community in addition to the description of fission yields. 
The current state-of-the-art theoretical description of $\beta$-delayed fission relies on the initial population of excited states from Quasi-particle Random Phase Approximation (QRPA) as well as model inputs from statistical Hauser-Feshbach theory which include the neutron optical potential, $\gamma$-ray strength function and nuclear level density. 
Many of these theoretical input models can be constrained with experiments on lighter radioactive nuclides closer to stability offering a path forward to understanding the extrapolation of these models to the region of heavier nuclei. 
Observations also play a key role, for instance, the brightness of the kilonova at late-epochs in the near- and middle- infrared on the order of 100 days or longer may point to the production of actinide nucleosynthesis from the spontaneous fission of $^{254}$Cf \cite{zhu2018californium}. 
Thus the challenges faced by the various communities in nuclear astrophysics naturally have a symbiotic relationship.
This has been encapsulated by the stimulating discussions and exciting collaborations being formed at this workshop.
I look forward to the strengthening of this community in the wake of the light and gravitation waves from GW170817.

\subsection{A new set of neutron-induced fission rates from energy density functional calculations (Samuel~A.~Giuliani)}

A reliable calculation of the nuclear properties of $r$-process nuclei and their
stellar reaction rates is a crucial ingredient in the estimation of
nucleosynthetic yields from neutron star mergers (NSM) and the associated
kilonova event (see, for example, contributions from Arcones, Barnes, Eichler,
Liddick, Lippuner, Spyrou). In the case of tidal ejecta (Perego, Radice) and
accretion disk (Siegel), where material is predicted to be extremely neutron
rich, particular attention has to be paid to the fission properties of heavy and
superheavy nuclei (Mumpower, Vassh, Zhu). Due to the extreme difficulties in
modeling the fission process from pure microscopic arguments, several
approximations are introduced in the calculation of fission probabilities,
bringing a large uncertainty in the estimation of fission rates and
fission fragments distributions (Schunck). Moreover, the nuclear structure models employed
in the calculation of fission properties are usually different from those
employed in the estimation of nuclear masses. This inconsistency can potentially
limit our capability to extract reliable information from sensitivity studies of
r-process abundances on different nuclear input.

In order to address these problems we employ the energy density functional (EDF)
approach to compute the fission properties of r-process nuclei. A new set of
neutron-induced reaction rates is obtained using the fission barriers,
collective inertias and nuclear masses predicted by the BCPM functional combined
with the Hauser-Feshbach statistical model~\cite{giuliani2018}.  Compared to the
traditional stellar rates obtained from the FRDM mass model and Thomas Fermi
barriers~\cite{Panov2010}, we found that the BCPM rates favour the pile-up of
material around mass number $A\sim 280$~\cite{Giuliani2017}.  This accumulation is
due to the combination of larger fission barrier and shell gap energies
predicted by BCPM around the neutron magic number $N=184$. In a further effort
to provide a consistent nuclear input suited for $r$-process calculations, we
employed the stochastic Langevin framework to compute fission fragments
distributions using the potential energy surfaces and collective inertias
predicted by different EDF's~\cite{sadhukhan2017}. Using the $^{294}$Og as a
test case we find that a highly-deformed fission decay, also known as cluster
emission, is predicted for this nucleus independently of the choice of the
functional. This result suggests that the EDF+Langevin scheme used for the
calculation of fission yields is rather robust, giving a certain confidence for
its employment to the case of r-process nuclei.

\subsection{Neutron capture rates away from stability for astrophysics applications (Georgios Perdikakis)}

Efforts to solve the puzzle of the synthesis of elements heavier than iron depend critically on the micro-physics input to astrophysics models. Ideally, a reliable set of experimentally measured neutron capture rates for most of the nuclei involved in the r-process is required. Due to the technological limitations that prevent as from developing a reaction target made out of neutrons or some other equivalent accelerator apparatus, we can not currently use the available radioactive beams to directly measure neutron capture reactions on  short-lived nuclei. Hence, neutron capture rates for r-process currently come from theoretical calculations that contain a large number of not adequately constrained parameters. It is the consensus of the community that these calculations infer large uncertainties to astrophysics calculations. 

Great promise for experimental constraints is demonstrated by two experimental developments. Indirect techniques that aim at deducing the neutron capture cross section from transfer reactions such as the surrogate reaction technique (see contribution from J. Cizewski) and the $\beta$-Oslo technique that is able to directly deduce statistical nuclear properties experimentally (see contribution from A. Spyrou). Both techniques are to a certain extent model dependent and the development of reliable reaction theory would benefit future developments in this field. 

To evaluate the yield outcome of various astrophysics scenarios we need to be able to reproduce in nucleosynthesis calculations, complex features of abundance yield patterns. For such comparisons to be meaningful, uncertainties in the nuclear input that affect nucleosynthesis calculations have to be identified, and their influence evaluated. To address this need we take a look at the sources of uncertainty that are most influential to the extrapolation of Hauser-Feshbach calculations away from stability and trace them back to the theoretical description of model ingredients that mostly influence neutron capture reaction rates, namely the level density, and the gamma-ray strength distribution. We calculate reaction rates using a number of adequate level density and gamma strength models for the neutron-rich isotopes of elements from oxygen to uranium. Then we compare the results of different calculations for each reaction rate and we calculate the ratio of minimum to maximum result for temperatures up to 10GK. We find results that vary up to a few orders of magnitude for each reaction rate. We show how the combined effect of inconsistent model predictions for the level density and the $\gamma$-ray strength create increased uncertainty and reduce the reliability of neutron capture rates away from stability. Based in these results it is clear that improvements in the current reaction theory and in particular the development of better microscopic models for gamma strengths and level densities is imperative as long as we are using the Hauser-Feshbach theory to calculate neutron capture rates.

\subsection{Nuclear transitions for the r-process in Relativistic Nuclear Field Theory (Caroline Robin, Elena Litvinova)}

The modeling of astrophysical processes requires the knowledge of excitation spectra, decay and reaction rates of a large number of nuclei, especially those at the drip-lines and, therefore, relies on theoretical models of nuclei which should be as universal, consistent, precise and predictive as possible. This area of nuclear physics is becoming particularly exciting as FRIB will enable the production of a large portion of unstable nuclei, and will be a great test of the theoretical predictions.
The r-process nucleosynthesis, which consists of an alternation of beta-decay and neutron capture, requires a consistent description of electromagnetic and spin-isospin transitions, first of all, electric dipole and Gamow-Teller in the (p,n) channel, in mid-mass to heavy nuclei. In this context, methods based on the density functional theory (DFT), or mean field approximation, such as the (quasiparticle) random phase approximation ((Q)RPA) \cite{RingSchuck} are largely used as they are applicable to a wide range of masses. Such methods, however, typically suffer from a lack of inter-nucleon correlations in their description of nuclei, leading to a poorly detailed description of nuclear excitation spectra.
While the (relativistic) QRPA ((R)QRPA) describes nuclear vibrations as superpositions of one-particle-one-hole (1p-1h), or two-quasiparticle (2 qp), excitations on top of the ground state, the relativistic nuclear field theory, which has been developed during the last decade (see \textit{e.g.} References\cite{Litvinova2007,Litvinova2008,Marketin2012,Robin2016,Robin2018}), goes a step further by considering the coupling between single nucleons and collective vibrations of the nucleus. Such coupling allows the resummation of meson-nucleon dynamics up to infinite order in the nucleonic self-energy, and introduces complex 2p-2h, or 4 qp, configurations in the nuclear response.
This method has been applied to the description of electric dipole transitions and neutron-capture rates in Ref. \cite{Litvinova2009}. Compared to the RQRPA level, the fragmentation of the strength distribution caused by the quasiparticle-vibration coupling (QVC) mechanism typically leads to an important improvement when comparing to the available experimental data.
Neutron-capture rates were found to be very sensitive to the position of the dipole Pygmy resonance which is affected by complex correlations of the nucleons.
On the charge-exchange side, Gamow-Teller (GT) transitions in open-shell nuclei were investigated in Refs. \cite{Robin2016,RobinINPC16,RobinNMP17,Robin2018}. As in the neutral channel the spreading mechanism caused by QVC systematically improves the comparison with the experimental data. The fragmentation and shift of the low-energy Gamow-Teller strength also brings the beta-decay half-lives in much better agreement with the experimental ones, considering a bare value of the weak axial coupling constant. The next step would be the calculation of consistent beta-decay and neutron-capture rates of a large quantity of nuclei, to investigate their impact on r-process simulations.
The latest development of the formalism in the proton-neutron channel include the treatment of complex ground-state correlations induced by QVC, as well as the introduction of the proton-neutron interaction in the pairing channel derived from the same meson-exchange interaction already used in the particle-hole channel, without introducing any extra free parameter. Such effects appear important when going towards $N=Z$, or when calculating GT transitions in the (n,p) channel in nuclei with $N>Z$. These new correlations are therefore expected to have a great impact on {\it e.g.} the calculation of electron-capture rates in core-collapse supernovae.

Recently, the response theory including the particle-vibration coupling effects was extended to the case of finite temperature 
\cite{LitvinovaWibowo2018}. For this purpose, the time blocking approximation to the time-dependent part of the in-medium nucleon-nucleon interaction amplitude was adopted for the thermal (imaginary-time) Green’s function formalism. We found, in particular, that introducing a soft blocking, instead of a sharp blocking at zero temperature, brings the Bethe-Salpeter equation to a single frequency variable equation also at finite temperature. The method is implemented self-consistently in the framework of Quantum Hadrodynamics and designed to connect the high-energy scale of heavy mesons and the low-energy domain of nuclear medium polarization effects in a parameter-free way. In this framework, we investigated the temperature dependence of dipole spectra in even-even medium-heavy nuclei with a special focus on the giant dipole resonance’s (GDR’s) width problem and on the low-energy dipole strength distribution. The obtained results are consistent with the existing experimental data on the GDR's width and with the Landau theory, explain the critical phenomenon of the disappearance of the GDR at high temperatures and predict the evolution of the low-energy dipole strength with temperature taking into account spreading effects. For instance, a noticeable  enhancement of the low-lying dipole strength has been found at the conditions corresponding to the radiative neutron capture. The temperature dependence of the Gamow-Teller resonance was also investigated and found to be even more sensitive to moderate temperatures. For more predictive conclusions about the potential impact of the finite-temperature effects on the r-process nucleosynthesis, continuum and superfluid pairing should be included into the framework of the thermal relativistic nuclear field theory. These developments will be addressed by future efforts.

\subsection{Global calculations of $\beta$ decay (Jonathan Engel)}

Reference \cite{Mus14} reported the development of the Finite-Amplitude Method \cite{Nak07,Avo11} for solving the charge-changing QRPA equations with Skyrme functionals, and Ref.\ \cite{Mus16} applied the method to create a table containing $\beta$-decay rates for essentially all even-even isotopes. After the publication of the table, Ref.\ \cite{Sha16} reported calculations, in the equal-filling approximation, in the rare-earth region of the decay of odd and odd-odd nuclei. These calculations are now being extended to the creation of a table of rates for all isotopes, including those with odd $N$, odd $Z$, or both.  At the same time, the calculations are being improved to obtain decay rates at finite-temperature to incorporate the chiral two-body current operators that appear to be in large part responsible for the "quenching" of $g_A$, a phenomenon that most calculations must include in an ad hoc way.

The result of all this work should be a ``best" set of QRPA rates, but the method is an adiabatic approximation to the density-functional-theory (DFT) response and therefore has a limited accuracy.  Extending the theory requires a time-dependent functional.  These do not exist but one can derive reasonable time-dependent extensions of Skyrme functionals by assuming that density-dependent Hamiltonians generate the functionals.  That assumption in turn creates a response function that can be treated in a better approximation than QRPA, for example in second QRPA.  The generalization of the Finite-Amplitude Method to second QRPA or some other ``beyond-QRPA'' approach is an important priority and perhaps the best hope for a qualitative improvement in global calculations of transition rates.

\section{Recent and planned experiments with radioactive ions}

\subsection{Decay studies of r-process nuclei at FRIB (Robert Grzywacz)}

With the opportunities provided by the construction and operation of new generation experimental facilities, the experimental studies of many of the r-process nuclei became feasible. The decay studies can perhaps reach the most neutron-rich isotopes due to their abilities to perform experiments at the very low production rates, as low as $10^-4 pps$ for lifetime measurements. The delayed neutron branching ratios have been identified as another important set of observables, which will impact the nucleosynthesis models, this is because all of the r-process nuclei are beta-delayed neutron emitters. However, because it may not be possible to measure all of the required isotopes, a broader scope of measurements beyond just measuring the most basic decay properties is needed to deepen our comprehension of the underlying physics, which determines the properties of exotic nuclei. With the goal in mind, the Decay Station is planned to be constructed at FRIB to provide an experimental tool to exploit the opportunities to study exotic isotopes comprehensively and efficiently. The instrument will consist of multiple detection systems, which will be very efficient and sensitive to various types of radiation. For example, the neutron energy measurements and neutron counting will have to be combined with the very sensitive gamma-ray detection setup to disentangle the complex decay patterns. In decay chain of one r-process nucleus which can be a beta-delayed one/two or three neutron emitter as many as thirty isotopes may be populated with various probabilities and each decay will be associated with the emission of beta particles, neutrons, and gammas. The principal goal of these studies will be to map the beta decay strength distribution for ranges of isotopes through total gamma absorption and neutron energy measurements. The experimental measurements of the strength distribution can be compared directly to models of the beta decays. The FRIB Decay Station will be a complex instrument capable of measuring decays of tens of isotopes in one experiment. The construction, operation and data analysis will be a complex operation and will provide the large volume of data on many isotopes and will engage researchers from multiple universities and national laboratories. 

\begin{figure}[!ht] 
\begin{center}
  \includegraphics[clip=false,width=0.60\textwidth]{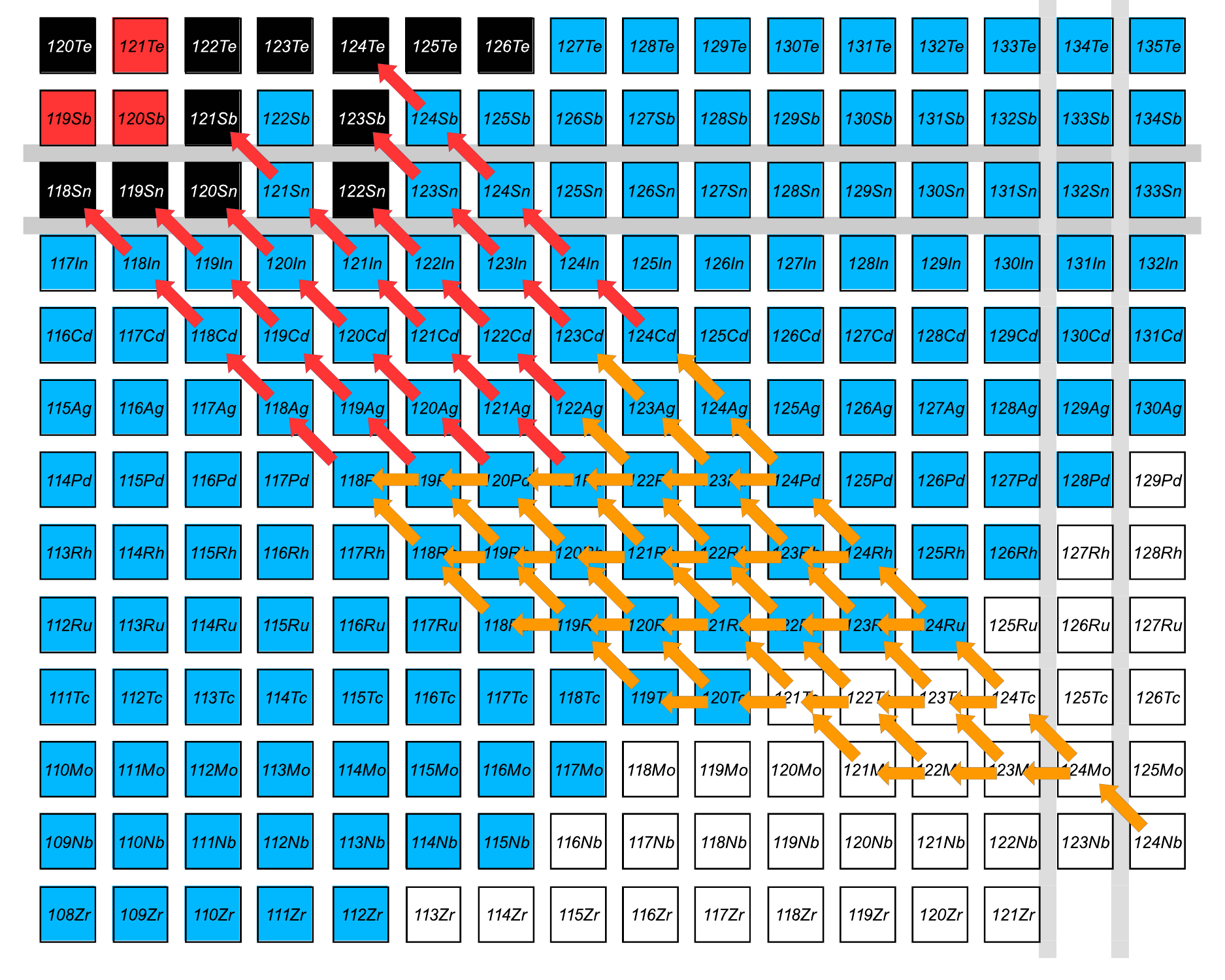}  
   \caption{\label{rg-decay} Decay of the $^{124}Nb$ and its complex decay chain. }
\end{center}
\end{figure}

\subsection{Experimental program for the $r$ process at the IGISOL facility (Anu Kankainen)}\label{sec:Kankainen}

The Ion Guide Isotope Separator On-Line (IGISOL) facility \cite{Moore2013} in the JYFL Accelerator Laboratory at the University of Jyv\"askyl\"a in Finland provides versatile possibilities to study properties of neutron-rich nuclei for the $r$-process. Neutron-rich nuclides are produced using proton-induced fission on uranium or thorium targets, and the fission fragments are thermalized in helium gas cell (typically at around 300 mbar) before they are extracted out, accelerated to 30 keV and mass-separated by a dipole magnet. The IGISOL method is universal and enables measurements of measurements of isotopes of any element. The production of neutron-rich nuclei at IGISOL is currently limited to fission fragments roughly located at around $70 \leq A \leq 170$. To pursue production of heavier nuclides relevant for the $r$ process, multinucleon-transfer reactions will be tested in future. 

One of the key instruments at the IGISOL facility is the JYFLTRAP double Penning trap mass spectrometer \cite{Eronen2012}. Nuclear masses are essential for $r$-process calculations \cite{Mumpower2015,Mumpower2016}, also after the GW170817 event \cite{Cote2018}. Recently, we have measured masses of several new neutron-rich rare-earth isotopes with JYFLTRAP \cite{Vilen2018}. These masses are relevant for the formation of the rare-earth abundance peak \cite{Surman1997,Mumpower2012}. The newly measured masses indicated that neutron pairing effect gets smaller when approaching the midshell and is weaker than predicted by commonly used theoretical mass models. The impact on the calculated $r$-process abundances was up to $\approx 25$~\%. The obtained pattern was found to be smoother than obtained in the baseline calculations performed with FRDM and AME16 \cite{Vilen2018}. In addition to the rare-earth region, mass measurements have been performed close to $^{78}$Ni and $^{132}$Sn at JYFLTRAP. For example the masses of $^{70}$Co and $^{74,75}$Ni have been measured for the first time. These are important for the core collapse supernovae and the composition of the neutron star crust, therefore affecting the seeds for neutron star mergers.  

The JYFLTRAP Penning trap has also been widely used as a high resolution beam purifier for post-trap decay spectroscopy (see e.g. the review in Ref.~\cite{Eronen2016}). One of the recent highlights is the first determination of beta-delayed multiple neutron emission beyond $A=100$ through direct neutron measurement using the BELEN detector at IGISOL. The $P_{2n}$ value of $^{136}$Sb was determined for the first time and found to be a factor 20 smaller than predicted by the FRDM+QRPA model used for $r$-process calculations \cite{Caballero2018}. Another example of post-trap decay spectroscopy at IGISOL is the observed enhanced gamma-ray emission from neutron-unbound states populated in beta decay in $^{87,88}$Br and $^{94}$Rb \cite{Tain2015}. Such an increase in the photon strength function can increase the $(n,\gamma)$ cross section that would have an impact on $r$-process abundance calculations. In addition, eight modules of the NuSTAR at FAIR MOdular Neutron time of flight SpectromeTER (MONSTER) \cite{Martinez2014} are currently at IGISOL for testing and experiments on beta-delayed neutron spectra. 

IGISOL is a good example of a smaller facility that can provide flexible possibilities and beamtime to explore properties of neutron-rich nuclei for the $r$-process as well as design and test new equipments. Therefore, the work carried out at IGISOL both in research and researcher training are essential for the future radioactive ion beam facilities like FRIB and FAIR.

\subsection{Mass measurements with the Canadian Penning trap mass spectrometer at CARIBU (J.A. Clark)}\label{sec:Clark}

The CARIBU (CAlifornium Rare Isotope Breeder Upgrade) facility \cite{sa08nimb} at Argonne National Laboratory provides neutron-rich nuclides as supplied from the spontaneous fission of a $^{252}$Cf source.  The source is situated within a large-volume gas catcher which contains high-purity helium gas to thermalize the emitted fission products.  A combination of gas flow and electric fields guide the fission fragments to an exit nozzle where the fragments are cooled through a radiofrequency quadrupole (RFQ) ion guide and subsequently mass separated by a normally conducting magnet with a mass resolving power approaching 20,000.  These ions are then bunched with an RFQ buncher, and are periodically ejected and sent into a Multi-Reflection Time-Of-Flight (MR-TOF) mass separator \cite{hi16nimb} with a resolving power reaching 100,000.  After being ejected from the MR-TOF, the ions are guided to the Canadian Penning Trap (CPT) mass spectrometer \cite{sc13prl} where the new phase-imaging ion-cyclotron-resonance (PI-ICR) technique \cite{el13prl} is being used to precisely determine the masses of neutron-rich isotopes provided by CARIBU.

Recently, a number of neutron-rich isotopes have been measured by the CPT to investigate the formation of the rare-earth peak \cite{OrfordVassh2018}.  Initially, measurements of neutron-rich neodymium and samarium isotopes were conducted and compared to the predictions of 'reverse-engineered' Markov Chain Monte Carlo (MCMC) calculations which determined the mass surface in the rare-earth region that would be necessary to reproduce the abundances of the rare-earth peak elements.  It was found that the masses measured by the CPT were consistent with that predicted by the MCMC calculations for a hot $r$-process in a neutron star merger wind.  Subsequent mass measurements of neutron-rich praseodymium, promethium, europium, and gadolinium isotopes are also consistent with these MCMC predictions.  The next stages in this work is to continue to measure the most neutron-rich nuclides possible with the CPT at CARIBU and examine other astrophysical trajectories with the MCMC technique, with the hope that comparisons between experiment and theory will serve to constrain certain astrophysical environments as possible sources of the rare-earth peak production.

Upon completion of the mass measurement program at CARIBU, the CPT will be moved to the $N$ = 126 beam factory as described in Sec.~\ref{sec:Brodeur}.  Here, the CPT will measure the masses of isotopes to investigate the production of the third $r$-process peak.  Although the CPT at that time will no longer measure masses provided by CARIBU, the CARIBU facility will be providing neutron-rich nuclides to a new low-energy experimental area where other experiments will take advantage of the isotopes available for study.  An ambitious program to examine the $\beta$-decay properties of these neutron-rich isotopes promises to provide much valuable information for $r$-process studies.

\subsection{Reaching the last abundance peak of the r-process with the $N$ = 126 beam factory (Maxime Brodeur)}\label{sec:Brodeur}

This workshop proceeding is a testament that the past few years has seen a large surge in the measurements of nuclear properties of relevance for the r-process. However most of these measurements were concentrated on the first two abundance peaks and the rare-earth peak (Sec.~\ref{sec:Clark} and ~\ref{sec:Kankainen}), leaving the third abundance peak mostly unexplored. This is because of the difficulty in producing these nuclides. First, fission reactions, which has been a successful mean of producing $N$ = 82 and rare-earth nuclei, cannot produce nuclei near $N$ = 126. Then, cross-sections for the production of nuclei via the commonly-used projectile fragmentation method falls sharply as more protons are removed from $^{208}$Pb \cite{Kurtukian2014}. As a result, only next-generation facilities such as FRIB will have the necessary primary beam intensities to allow sufficient production of exotic nuclei of relevance for the r-process south of $^{208}$Pb. Multi-nucleon transfer (MNT) reactions of $^{136}$Xe on $^{198}$Pt however present significantly larger cross-sections in this region \cite{Watanabe2015}. A unique facility that will make use of such reactions, called the $N$ = 126 beam factory, is currently under construction at Argonne National Laboratory. This facility will first comprise a large-volume gas cell to catch and thermalize the divergent fast recoils produced by the MNT reactions. After the gas cell the radioactive ions will be accelerated electrostatically to energies of up to 60 keV and non-isobaric contaminants will be removed by a mass analyzing magnet with a resolving power of 1000. Then, the continuous beam will be decelerated, accumulated, cooled and bunched using a radio-frequency quadrupole ion trap. A multi-reflection time-of-flight (MR-ToF) mass spectrometer \cite{Schultz2016}, currently being commissioned off-line at the University of Notre Dame, will remove isobaric contaminations with a resolving power of near 50,000, before sending the purified bunches to various experimental stations including the Canadian Penning Trap (CPT) for mass measurements and a beta-decay station for half-life measurements. Besides the $N$ = 126 region, different NMT reactions can be used to populate nuclei of interest for the r-process in other regions of the chart of the nuclides that are difficult of access. For example, $^{136}$Xe on $^{164}$Dy can be used to produce isotopes above and near Dy that cannot be reached via fission. The mass measurement of such isotopes would refine abundance calculations of the heavy side of the rare-earth peak as seen by recent measurements using JYFLTRAP \cite{Vilen2018}. That same reaction would also produce $N$ = 104 nuclei above Nd that would test reverse-engineered mass model mass-excess predictions \cite{OrfordVassh2018}. Finally, MNT reactions of $^{136}$Xe beams on actinides would produce very neutron rich isotopes that cannot otherwise be produced even via projectile-fragmentation. One possibility under study would be the use of $^{136}$Xe reaction on $^{251}$Cf to produce nuclei near $^{254}$Cf, which is a nuclei of great importance found to regulate the brightness of kilonova at late time \cite{zhu2018californium}.

\subsection{Surrogate neutron-capture reaction prospects for r-process nuclei (Jolie A. Cizewski)}

Understanding abundances from a neutron star merger r process is sensitive to unknown neutron capture $(n,\gamma)$ rates, especially near shell closures with low level density and weakly bound nuclei.  A specific example is the N=80 isotone $^{130}$Sn where ref.~\cite{Mumpower2016} has shown that the unknown $(n,\gamma)$ rate could have significant impact on predicting final abundances.  Near shell closures neutron capture is a competition between direct-semi-direct (DSD) capture and formation and decay of a compound nucleus (CN).  The neutron transfer $(d,p)$ reaction with radioactive ion beams is an excellent reaction to probe the single-neutron character of excitations in the final nuclei and is also a promising candidate for a surrogate for $(n,\gamma)$ when the gamma-decay radiation is measured in coincidence with reaction protons as a function of excitation energy.  The DSD capture cross sections have been deduced following $(d,p)$ reaction measurements with radioactive ion beams of $^{132,130,128,126}$Sn and stable $^{124}$Sn~\cite{Kozub2012,Manning2018}.  The deduced $^{132}$Sn DSD $(n,\gamma)$ cross section is consistent with previous calculations~\cite{Chiba2008} that also predict that DSD should be dominant for $^{132}$Sn.  However, the DSD $(n,\gamma)$ cross sections for lighter Sn isotopes are lower than previous studies~\cite{Chiba2008}, which assumed pure single-neutron character for the states with unknown excitation energies.  More importantly, the formation and subsequent statistical gamma decay of the compound nucleus is expected to dominate for $A<132$~Sn isotopes.  

The $(d,p\gamma)$ reaction has been validated as a surrogate for $(n,\gamma)$ following $^{95}$Mo$(d,p\gamma$) reaction studies in normal kinematics~\cite{Ratkiewicz2018}.  Theoretical formation cross sections and as a function of transferred angular momentum of the $^{96}$Mo CN following breakup of the deuteron~\cite{Potel2015} were combined with measured proton-gated gamma-decay probabilities for deducing the decay of the compound nucleus $G^{CN}(E_x,J,\pi)$.  These $G^{CN}$ probabilities were input to a Hauser-Feshbach calculation~\cite{Escher2018} of the $(n,\gamma)$ cross section as a function of neutron energy that very well reproduces the previously measured and evaluated $^{95}$Mo$(n,\gamma)$ cross sections.

To realize $(d,p\gamma)$ reaction studies with radioactive ion beams, the Gammasphere ORRUBA: Dual Detectors for Experimental Structure Studies (GODDESS)~\cite{Pain2017} was commissioned with the Oak Ridge Rutgers University Barrel Array (ORRUBA) of position-sensitive silicon strip detectors augmented with annular arrays of end cap detectors.  Preliminary results~\cite{Lepailleur2018} with the N=80 $^{134}$Xe beam showed population of a candidate for the $3p_{3/2}$ 3/2$^-$ state important for DSD capture and the gamma-decay probability data at high excitation energy that would be needed for a surrogate $(n,\gamma)$ analysis, should statistics be sufficient.  

The study of the $(d,p\gamma)$ reaction with $^{130}$Sn beams and as a surrogate for both DSD and CN $(n,\gamma)$ reactions will have to wait for FRIB.  In the near term $(d,p\gamma)$ measurements with $^{143}$Ba (with CARIBU at ATLAS) and $^{80}$Ge (at NSCL) beams have been approved.  The $^{143}$Ba$(d,p\gamma)$ measurements would be the first surrogate $(n,\gamma)$ study with a rare-earth fission fragment beam and would constrain predictions of $(n,\gamma)$ cross sections near the r-process path~\cite{Mumpower2016}.  $^{80}$Ge$(n,\gamma)$ is important for constraining $A\sim80$ synthesis in neutron-star mergers~\cite{Surman2014}.

\subsection{Neutron-capture reactions and $\beta$-decay strength distributions for the r process (Artemis Spyrou)}

Among the many nuclear physics properties that play an important role in r-process calculations, neutron-capture and $\beta$-decay rates are two of the most important ones.  Together with nuclear masses and fission properties, they define the path along the nuclear chart that the r process will take, and as a result they define the final abundance distribution. 

Neutron capture reactions on short-lived nuclei are extremely challenging to measure experimentally. For this reason, indirect techniques have been developed that can significantly constrain the reaction rate. These techniques often provide constraints on the nuclear structure ingredients needed in statistical model (n,$\gamma$) calculations, such as the nuclear level density (NLD) and the $\gamma$-ray strength function ($\gamma$SF). An indirect technique that is applicable far from stability is the $\beta$-Oslo method \cite{Spy14}. The technique is based on the traditional Oslo method \cite{Gut87}, which has been used for over 30 years to extract NLD and $\gamma$SF for nuclei along the valley of stability. For the $\beta$-Oslo method, the compound nucleus of interest is populated using $\beta$ decay, providing access to nuclei further away from stability that are available at current facilities only at very low beam intensities. In my talk I discussed our recent results on constraining neutron captures for three reactions: $^{68,69}$Ni(,n$\gamma)^{69,70}$Ni and $^{73}$Zn(n,$\gamma)^{74}$Zn \cite{Lid16, Spy17, Lew18}. In addition, I presented recent calculations that showed the impact of neutron-capture on isomeric states, and highlighted the importance of including isomers in r-process calculations. 

Global QRPA calculations are typically used to provide $\beta$-decay properties for r-process models. These are compared to experimental data for half-lives and $\beta$-delayed neutron emission probabilities. However, both of these are integral quantities, and a more sensitive comparison can be done using $\beta$-decay strength distributions. In my talk I showed recent results for nuclei in the mass 70 and 100 regions taken with the SuN detector at the NSCL. These results were compared to commonly used QRPA calculations, showing a large discrepancy between theory and experiment. In addition, the new experimental $\beta$-decay intensities were used to extract the fraction of energy released in the form of $\gamma$ rays. I showed that due to the phenomenon known as “Pandemonium” the existing literature underestimates significantly the emission of $\gamma$ rays in favor of $\beta$ and $\nu$ emission.  These can affect the energy release during the kilonova event.

\subsection{Planned mass measurements using the Rare-RI Ring at RIBF/Riken (Sarah Naimi)}

\begin{figure}[!ht] 
\begin{center}
  \includegraphics[clip=false,width=0.7\textwidth]{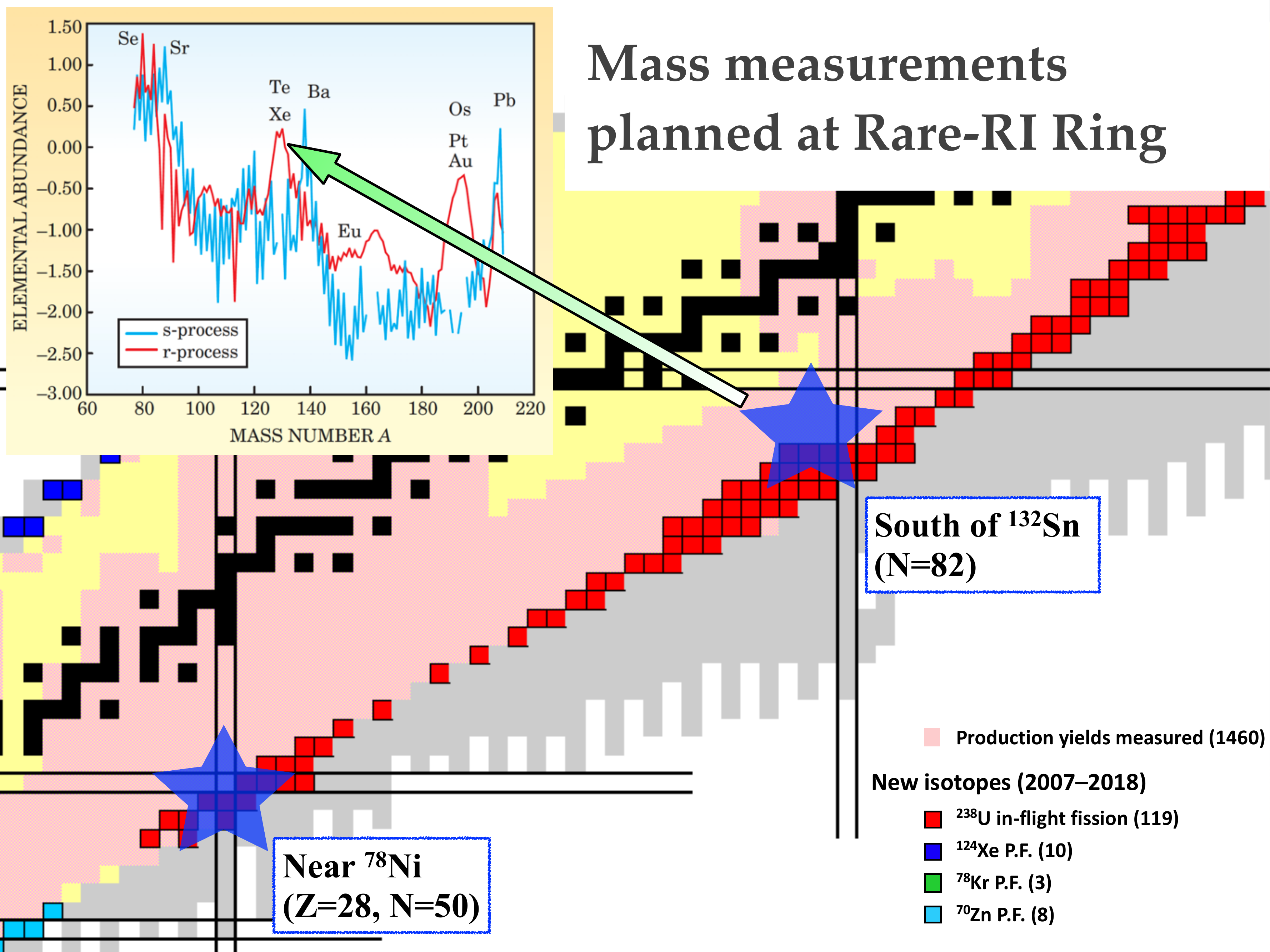}  
   \caption{\label{naimi_fig} Mass measurements relevant for the r-process planned at RIBF using the Rare-RI Ring}
\end{center}
\end{figure}

The Rare-RI Ring is a newly commissioned mass spectrometer at RIBF/Riken. 
The mass measurement determination is a based on Isochronous Mass Spectrometry \cite{2012Ozawa}. 
This technique allows reaching mass precision of $10^{-6}$ in an extremely short measurement time ($\sim$1 ms). 
The unique combination of the Rare-RI Ring with the Superconducting Ring Cyclotron (SRC) at RI Beam Factory (RIBF) at Riken has its own challenges but gives the opportunity to measure masses of r-process nuclei.
We have overcome most the technical challenges and the Rare-RI Ring is now ready to perform mass measurement of the most exotic nuclei. 
We have developed single-ion mass determination technique using the unique feature of event-by-event velocity correction in storage ring. This will also allow mass measurement of extremely low yield neutron-rich nuclei. 

Figure \ref{naimi_fig} shows the first planned mass measurements using the Rare-RI Ring. 
We will focus mainly on two key regions of the nuclear chart: near $^{78}$Ni region and south of $^{132}$Sn region. 
The latter has large impact on the second peak of the r-process abundance, where nuclear masses will determine the flow to higher mass region due to its vicinity to the $N=82$ shell gap \cite{2015Mumpower_JPG}. 
To determine the strength of the $N=82$ shell gap will require mass measurement beyond the $N=82$. 
The Isoltrap experiment at CERN/ISOLDE has measured Cd isotopes beyond the $N=82$ \cite{2015Atanasov}.  
We plan to measure even lower $Z$, namely Ag and Pd isotopes, which were already produced and studied at RIBF \cite{2015Lorusso,2018Shimizu}. 

Nuclear deformation plays also an important role in the r-process.  
According to the FRMD mass model \cite{MOLLER20161}, the Palladium isotopes present strong deformation before $N=82$.
By measuring these masses we will clarify the role of deformation before the $N=82$ shell gap on the r-process abundance.

\section*{Acknowledgements}
The `FRIB and the GW170817 kilonova' workshop was supported by the U.S. Department of Energy through the FRIB Theory Alliance under Contract No.\ \textrm{DE-SC}0013617.

A. Aprahamian and M.B. are supported by the U.S. National Science Foundation under grant number PHY-1713857 and the University of Notre Dame.

A. Arcones and M.E. are supported by the ERC starting grant “EUROPIUM” (Grant No.\ 677912). 

The research of A.B.B. is supported in part
by the U.S. National Science Foundation Grants No.\ PHY-1514695 and PHY-1806368.

T.C.B., E.M.H. and J.Y. acknowledge partial support
from grant PHY 14-30152; Physics Frontier Center/JINA Center for the
Evolution of the Elements (JINA-CEE), awarded by the US National Science
Foundation.

The work of J.A.~Cizewski is supported in part by U.S. Department of Energy National Nuclear Security Administration Stewardship Science Academic Alliances under cooperative agreement \textrm{DE-NA}0002132 and the National Science Foundation NSF-PHY-1404218.

J.A.~Clark is supported by the U.S. Department of Energy, Office of Nuclear Physics, under Contract No. DE-AC02-06CH11357. Mass measurements with the CPT used resources of ANL's ATLAS facility, which is a DOE Office of Science User Facility.

B.C. was funded by the ERC Consolidator Grant (Hungary) funding scheme (project RADIOSTAR, G.A. n. 724560) and by the National Science Foundation (USA) under grant No.\ PHY-1430152 (JINA Center for the Evolution of the Elements).

J.E. was supported in part by the U.S. Department of Energy under Contract DE-FG02-97ER41019 and by the US Department of Energy, Office of Science, Advanced Scientific Computing Research and Nuclear Physics under Contract \textrm{DE-SC}0018223.

M.E. acknowledges the support of the Swiss National Foundation, project number P2BSP2\_172068.

G.F. acknowledges the support of the U.S. National Science Foundation Grant PHY-1614864 at UCSD.

S.A.G. acknowledges support from the U.S. Department of Energy under Award Number DOE-DE-NA0002847 (NNSA, the Stewardship Science Academic Alliances program).

R.G. acknowledges the support of the Office of Nuclear Physics, U.S. Department of Energy under Award No. DE-FG02-96ER40983 and DE-AC05-00OR22725, and by the National Nuclear Security Administration under the Stewardship Science Academic Alliances program through DOE Award No.\ \textrm{DE-NA}0002132.

S.H. is supported by Chandra Award TM8-19002X and NSF grant PHY-1554876. 

C.J.H. supported in part by U.S. Department of Energy grants DE-FG02-87ER40365 and \textrm{DE-SC}0018083.

A.K. acknowledges the support from the Academy of Finland under grant No.\ 275389 and from the European Union’s Horizon 2020 research and innovation program under grant agreement No.\ 771036 (ERC CoG "MAIDEN").

J.E.L. acknowledges support from NASA grant NNX16AE96G.

The work of E.L. was funded by the US-NSF Grant No.\ PHY-1404343 and by the US-NSF Career Grant No.\ PHY-1654379.

Theoretical work at the University of Notre Dame (G.J.M., R.S., T.S., and N.V.) is supported by the U.S. Department of Energy under Nuclear Theory Contract No.\ DE-FG02-95-ER40934 (G.J.M. and R.S.), \textrm{DE-SC}0013039 (R.S. and T.S.), and DE-AC52-07NA27344 for the topical collaboration Fission In R-process Elements (FIRE) (N.V. and R.S.), and SciDAC Contract No.\ \textrm{DE-SC}0018232 (T.S. and R.S.), and by the U.S. National Science Foundation under grant number PHY-1630782 Focused Research Hub in Theoretical Physics: Network for Neutrinos, Nuclear Astrophysics, and Symmetries (N3AS) (R.S.). Part of this work utilized the computational resources of the University of Notre Dame Center for Research Computing (ND CRC) and the Laboratory Computing Resource Center at Argonne National Laboratory (ANL LCRC). 

G.C.M., S.R., and Y.Z. are supported in part by the U.S. Department of Energy under Contract No.\ DE-FG02-02ER41216 (G.C.M. and Y.Z.) and DE-AC52-07NA27344 for the topical collaboration Fission In R-process Elements (FIRE) (G.C.M. and Y.Z.), and by the U.S. National Science Foundation under grant number PHY-1630782 Focused Research Hub in Theoretical Physics: Network for Neutrinos, Nuclear Astrophysics, and Symmetries (N3AS) (G.C.M. and S.R).

M.M. was supported in part by the U.S. Department of Energy (DOE) under Contract No.\ DE-AC52-07NA27344 for the topical collaboration Fission In R-process Elements (FIRE) and under the auspices of the National Nuclear Security Administration of the U.S. DOE at Los Alamos National Laboratory (LANL) under Contract No.\ DE-AC52-06NA2539.

W.N. was supported by the U.S. Department of Energy, Office of Science, Office of Nuclear Physics, Grant No.\ \textrm{DE-SC}0013365.

A.P. acknowledges support from the INFN initiative ``High  Performance  data  Network" funded by CIPE.

D.R. acknowledges support from a Frank and Peggy Taplin Membership at the Institute for Advanced Study and the Max-Planck/Princeton Center (MPPC) for Plasma Physics (NSF PHY-1523261). D.R.'s computations were performed on the supercomputers Bridges, Comet, and Stampede (NSF XSEDE allocation TG-PHY160025), on NSF/NCSA Blue Waters (NSF PRAC ACI-1440083), and Marconi (PRACE proposal 2016153522).

C.R. acknowledges support from the Institute for Nuclear Theory under US-DOE Grant DE-FG02-00ER41132 and from JINA-CEE under US-NSF Grant PHY-1430152.

I.U.R. acknowledges support from NSF grants AST~1613536, AST~1815403, and PHY~1430152 (Physics Frontier Center/JINA-CEE).

D.M.S. acknowledges support from the National Aeronautics and Space Administration through Einstein Postdoctoral Fellowship Award Number PF6-170159 issued by the Chandra X-ray Observatory Center, which is operated by the Smithsonian Astrophysical Observatory for and on behalf of the National Aeronautics Space Administration under contract NAS8-03060.

The work of N.S. was performed under the auspices of the U.S. Department of 
Energy by the Lawrence Livermore National Laboratory under Contract 
DE-AC52-07NA27344 and was partially supported by the U.S. Department of Energy 
``Fission in R-process Elements" Topical Collaboration in Nuclear Theory. 
Computing support for this work came from the Lawrence Livermore National 
Laboratory (LLNL) Institutional Computing Grand Challenge program.

A.S. was supported by the National Science Foundation under Grants No.\ PHY 1102511 (NSCL),  No.\ PHY 1430152 (Joint Institute for Nuclear Astrophysics), and PHY 1350234 (CAREER). This material is based upon work supported by the Department of Energy/National Nuclear Security Administration under Award Numbers \textrm{DE-NA}0003221, \textrm{DE-NA}0000979 and \textrm{DE-NA}0002132.

This work was supported in part by the US Department of Energy through the Los
Alamos National Laboratory and has been assigned report number 
LA-UR-18-28022. Los Alamos National Laboratory is operated by Los
Alamos National Security, LLC, for the National Nuclear Security
Administration of US Department of Energy (Contract DEAC52-06NA25396).

\bibliographystyle{unsrt}
\bibliography{bibliography.bib}

\begin{thebibliography}{100}

\bibitem{Horowitz2018}
C.~J. {Horowitz}, A.~{Arcones}, B.~{C{\^o}t{\'e}}, I.~{Dillmann},
  W.~{Nazarewicz}, I.~U. {Roederer}, H.~{Schatz}, A.~{Aprahamian},
  D.~{Atanasov}, A.~{Bauswein}, J.~{Bliss}, M.~{Brodeur}, J.~A. {Clark},
  A.~{Frebel}, F.~{Foucart}, C.~J. {Hansen}, O.~{Just}, A.~{Kankainen}, G.~C.
  {McLaughlin}, J.~M. {Kelly}, S.~N. {Liddick}, D.~M. {Lee}, J.~{Lippuner},
  D.~{Martin}, J.~{Mendoza-Temis}, B.~D. {Metzger}, M.~R. {Mumpower},
  G.~{Perdikakis}, J.~{Pereira}, B.~W. {O'Shea}, R.~{Reifarth}, A.~M. {Rogers},
  D.~M. {Siegel}, A.~{Spyrou}, R.~{Surman}, X.~{Tang}, T.~{Uesaka}, and
  M.~{Wang}.
\newblock {r-Process Nucleosynthesis: Connecting Rare-Isotope Beam Facilities
  with the Cosmos}.
\newblock {\em ArXiv e-prints}, May 2018.

\bibitem{lippuner:17b}
J.~{Lippuner} and L.~F. {Roberts}.
\newblock {SkyNet: A Modular Nuclear Reaction Network Library}.
\newblock {\em The Astrophysical Journal Supplement}, 233:18, December 2017.

\bibitem{lippuner:2015gwa}
Jonas Lippuner and Luke~F. Roberts.
\newblock {r-Process Lanthanide Production and Heating Rates in Kilonovae}.
\newblock {\em Astrophys. J.}, 815(2):82, 2015.

\bibitem{radice:2016dwd}
David Radice, Filippo Galeazzi, Jonas Lippuner, Luke~F. Roberts, Christian~D.
  Ott, and Luciano Rezzolla.
\newblock {Dynamical Mass Ejection from Binary Neutron Star Mergers}.
\newblock {\em Mon. Not. Roy. Astron. Soc.}, 460(3):3255--3271, 2016.

\bibitem{roberts:16b}
L.~F. {Roberts}, J.~{Lippuner}, M.~D. {Duez}, J.~A. {Faber}, F.~{Foucart},
  J.~C. {Lombardi}, Jr., S.~{Ning}, C.~D. {Ott}, and M.~{Ponce}.
\newblock {The influence of neutrinos on r-process nucleosynthesis in the
  ejecta of black hole-neutron star mergers}.
\newblock {\em \mnras}, 464:3907, February 2017.

\bibitem{lippuner:17a}
J.~{Lippuner}, R.~{Fern{\'a}ndez}, L.~F. {Roberts}, F.~{Foucart}, D.~{Kasen},
  B.~D. {Metzger}, and C.~D. {Ott}.
\newblock {Signatures of hypermassive neutron star lifetimes on r-process
  nucleosynthesis in the disc ejecta from neutron star mergers}.
\newblock {\em \mnras}, 472:904--918, November 2017.

\bibitem{siegel:17}
D.~M. {Siegel} and B.~D. {Metzger}.
\newblock {Three-dimensional GRMHD Simulations of Neutrino-cooled Accretion
  Disks from Neutron Star Mergers}.
\newblock {\em The Astrophysical Journal}, 858:52, May 2018.

\bibitem{vlasov:17}
A.~D. {Vlasov}, B.~D. {Metzger}, J.~{Lippuner}, L.~F. {Roberts}, and T.~A.
  {Thompson}.
\newblock {Neutrino-heated winds from millisecond protomagnetars as sources of
  the weak r-process}.
\newblock {\em \mnras}, 468:1522--1533, June 2017.

\bibitem{fernandez:17a}
R.~{Fern{\'a}ndez}, F.~{Foucart}, D.~{Kasen}, J.~{Lippuner}, D.~{Desai}, and
  L.~F. {Roberts}.
\newblock {Dynamics, nucleosynthesis, and kilonova signature of black hole -
  neutron star merger ejecta}.
\newblock {\em \cqg}, 34(15):154001, June 2017.

\bibitem{2015Mumpower_JPG}
M~Mumpower, R~Surman, D~L Fang, M~Beard, and A~Aprahamian.
\newblock The impact of uncertain nuclear masses near closed shells on the r
  -process abundance pattern.
\newblock {\em Journal of Physics G: Nuclear and Particle Physics},
  42(3):034027, 2015.

\bibitem{Mumpower2015}
M.~R. Mumpower, R.~Surman, D.-L. Fang, M.~Beard, P.~M\"oller, T.~Kawano, and
  A.~Aprahamian.
\newblock Impact of individual nuclear masses on $r$-process abundances.
\newblock {\em Phys. Rev. C}, 92:035807, Sep 2015.

\bibitem{Mumpower2016}
M.R. Mumpower, R.~Surman, G.C. McLaughlin, and A.~Aprahamian.
\newblock The impact of individual nuclear properties on r-process
  nucleosynthesis.
\newblock {\em Progress in Particle and Nuclear Physics}, 86:86 -- 126, 2016.

\bibitem{Cote2018}
B.~{C{\^o}t{\'e}}, C.~L. {Fryer}, K.~{Belczynski}, O.~{Korobkin},
  M.~{Chru{\'s}li{\'n}ska}, N.~{Vassh}, M.~R. {Mumpower}, J.~{Lippuner}, T.~M.
  {Sprouse}, R.~{Surman}, and R.~{Wollaeger}.
\newblock {The Origin of r-process Elements in the Milky Way}.
\newblock {\em The Astrophysical Journal}, 855:99, March 2018.

\bibitem{holmbeck18b}
E.~M. {Holmbeck}, R.~{Surman}, T.~M. {Sprouse}, M.~R. {Mumpower}, N.~{Vassh},
  T.~C. {Beers}, and T.~{Kawano}.
\newblock {Actinide Production in Neutron-Rich Ejecta of a Neutron Star
  Merger}.
\newblock {\em ArXiv e-prints}, July 2018.

\bibitem{zhu2018californium}
Y.~{Zhu}, R.~T. {Wollaeger}, N.~{Vassh}, R.~{Surman}, T.~M. {Sprouse}, M.~R.
  {Mumpower}, P.~{M{\"o}ller}, G.~C. {McLaughlin}, O.~{Korobkin}, T.~{Kawano},
  P.~J. {Jaffke}, E.~M. {Holmbeck}, C.~L. {Fryer}, W.~P. {Even}, A.~J.
  {Couture}, and J.~{Barnes}.
\newblock {Californium-254 and Kilonova Light Curves}.
\newblock {\em The Astrophysical Journal Letters}, 863:L23, August 2018.

\bibitem{ruffert:1995fs}
M.~H. Ruffert, H.~T. Janka, and Gerhard Schaefer.
\newblock {Coalescing neutron stars: A Step towards physical models. 1:
  Hydrodynamic evolution and gravitational wave emission}.
\newblock {\em Astron. Astrophys.}, 311:532--566, 1996.

\bibitem{rosswog:1998hy}
S.~Rosswog, M.~Liebendoerfer, F.~K. Thielemann, M.~B. Davies, W.~Benz, and
  T.~Piran.
\newblock {Mass ejection in neutron star mergers}.
\newblock {\em Astron. Astrophys.}, 341:499--526, 1999.

\bibitem{rosswog:2001fh}
S.~Rosswog and M.~B. Davies.
\newblock {High resolution calculations of merging neutron stars I: model
  description and hydrodynamic evolution}.
\newblock {\em Mon. Not. Roy. Astron. Soc.}, 334:481--497, 2002.

\bibitem{rosswog:2003rv}
Stephan Rosswog and M.~Liebendoerfer.
\newblock {High resolution calculations of merging neutron stars. 2: Neutrino
  emission}.
\newblock {\em Mon. Not. Roy. Astron. Soc.}, 342:673, 2003.

\bibitem{rosswog:2003tn}
Stephan Rosswog, Enrico Ramirez-Ruiz, and Melvyn~B. Davies.
\newblock {High Resolution Calculations of Merging Neutron Stars. 3. Gamma-Ray
  Bursts}.
\newblock {\em Mon. Not. Roy. Astron. Soc.}, 345:1077, 2003.

\bibitem{oechslin:2006uk}
Roland Oechslin, H.~T. Janka, and A.~Marek.
\newblock {Relativistic neutron star merger simulations with non-zero
  temperature equations of state. 1. Variation of binary parameters and
  equation of state}.
\newblock {\em Astron. Astrophys.}, 2006.
\newblock [Astron. Astrophys.467,395(2007)].

\bibitem{rosswog:2012wb}
S.~Rosswog, T.~Piran, and E.~Nakar.
\newblock {The multi-messenger picture of compact object encounters: binary
  mergers versus dynamical collisions}.
\newblock {\em Mon. Not. Roy. Astron. Soc.}, 430:2585, 2013.

\bibitem{korobkin:2012uy}
O.~Korobkin, S.~Rosswog, A.~Arcones, and C.~Winteler.
\newblock {On the astrophysical robustness of neutron star merger r-process}.
\newblock {\em Mon. Not. Roy. Astron. Soc.}, 426:1940, 2012.

\bibitem{bauswein:2013yna}
A.~Bauswein, S.~Goriely, and H.~T. Janka.
\newblock {Systematics of dynamical mass ejection, nucleosynthesis, and
  radioactively powered electromagnetic signals from neutron-star mergers}.
\newblock {\em Astrophys. J.}, 773:78, 2013.

\bibitem{shibata:1999wm}
Masaru Shibata and Koji Uryu.
\newblock {Simulation of merging binary neutron stars in full general
  relativity: Gamma = two case}.
\newblock {\em Phys. Rev.}, D61:064001, 2000.

\bibitem{shibata:2003ga}
Masaru Shibata, Keisuke Taniguchi, and Koji Uryu.
\newblock {Merger of binary neutron stars of unequal mass in full general
  relativity}.
\newblock {\em Phys. Rev.}, D68:084020, 2003.

\bibitem{shibata:2005ss}
Masaru Shibata, Keisuke Taniguchi, and Koji Uryu.
\newblock {Merger of binary neutron stars with realistic equations of state in
  full general relativity}.
\newblock {\em Phys. Rev.}, D71:084021, 2005.

\bibitem{baiotti:2008ra}
Luca Baiotti, Bruno Giacomazzo, and Luciano Rezzolla.
\newblock {Accurate evolutions of inspiralling neutron-star binaries: prompt
  and delayed collapse to black hole}.
\newblock {\em Phys. Rev.}, D78:084033, 2008.

\bibitem{kiuchi:2010ze}
Kenta Kiuchi, Yuichiro Sekiguchi, Masaru Shibata, and Keisuke Taniguchi.
\newblock {Exploring binary-neutron-star-merger scenario of short-gamma-ray
  bursts by gravitational-wave observation}.
\newblock {\em Phys. Rev. Lett.}, 104:141101, 2010.

\bibitem{rezzolla:2010fd}
Luciano Rezzolla, Luca Baiotti, Bruno Giacomazzo, David Link, and Jose~A. Font.
\newblock {Accurate evolutions of unequal-mass neutron-star binaries:
  properties of the torus and short GRB engines}.
\newblock {\em Class. Quant. Grav.}, 27:114105, 2010.

\bibitem{rezzolla:2011da}
Luciano Rezzolla, Bruno Giacomazzo, Luca Baiotti, Jonathan Granot, Chryssa
  Kouveliotou, and Miguel~A. Aloy.
\newblock {The missing link: Merging neutron stars naturally produce jet-like
  structures and can power short Gamma-Ray Bursts}.
\newblock {\em Astrophys. J.}, 732:L6, 2011.

\bibitem{hotokezaka:2011dh}
Kenta Hotokezaka, Koutarou Kyutoku, Hirotada Okawa, Masaru Shibata, and Kenta
  Kiuchi.
\newblock {Binary Neutron Star Mergers: Dependence on the Nuclear Equation of
  State}.
\newblock {\em Phys. Rev.}, D83:124008, 2011.

\bibitem{hotokezaka:2012ze}
Kenta Hotokezaka, Kenta Kiuchi, Koutarou Kyutoku, Hirotada Okawa, Yu-ichiro
  Sekiguchi, Masaru Shibata, and Keisuke Taniguchi.
\newblock {Mass ejection from the merger of binary neutron stars}.
\newblock {\em Phys. Rev.}, D87:024001, 2013.

\bibitem{palenzuela:2013hu}
Carlos Palenzuela, Luis Lehner, Marcelo Ponce, Steven~L. Liebling, Matthew
  Anderson, David Neilsen, and Patrick Motl.
\newblock {Electromagnetic and Gravitational Outputs from Binary-Neutron-Star
  Coalescence}.
\newblock {\em Phys. Rev. Lett.}, 111(6):061105, 2013.

\bibitem{hotokezaka:2013iia}
Kenta Hotokezaka, Kenta Kiuchi, Koutarou Kyutoku, Takayuki Muranushi, Yu-ichiro
  Sekiguchi, Masaru Shibata, and Keisuke Taniguchi.
\newblock {Remnant massive neutron stars of binary neutron star mergers:
  Evolution process and gravitational waveform}.
\newblock {\em Phys. Rev.}, D88:044026, 2013.

\bibitem{ruiz:2016rai}
Milton Ruiz, Ryan~N. Lang, Vasileios Paschalidis, and Stuart~L. Shapiro.
\newblock {Binary Neutron Star Mergers: a jet Engine for Short Gamma-ray
  Bursts}.
\newblock {\em Astrophys. J.}, 824(1):L6, 2016.

\bibitem{dietrich:2016hky}
Tim Dietrich, Maximiliano Ujevic, Wolfgang Tichy, Sebastiano Bernuzzi, and
  Bernd Bruegmann.
\newblock {Gravitational waves and mass ejecta from binary neutron star
  mergers: Effect of the mass-ratio}.
\newblock {\em Phys. Rev.}, D95(2):024029, 2017.

\bibitem{dietrich:2016lyp}
Tim Dietrich, Sebastiano Bernuzzi, Maximiliano Ujevic, and Wolfgang Tichy.
\newblock {Gravitational waves and mass ejecta from binary neutron star
  mergers: Effect of the stars' rotation}.
\newblock {\em Phys. Rev.}, D95(4):044045, 2017.

\bibitem{sekiguchi:2011zd}
Yuichiro Sekiguchi, Kenta Kiuchi, Koutarou Kyutoku, and Masaru Shibata.
\newblock {Gravitational waves and neutrino emission from the merger of binary
  neutron stars}.
\newblock {\em Phys. Rev. Lett.}, 107:051102, 2011.

\bibitem{sekiguchi:2015dma}
Yuichiro Sekiguchi, Kenta Kiuchi, Koutarou Kyutoku, and Masaru Shibata.
\newblock {Dynamical mass ejection from binary neutron star mergers:
  Radiation-hydrodynamics study in general relativity}.
\newblock {\em Phys. Rev.}, D91(6):064059, 2015.

\bibitem{palenzuela:2015dqa}
Carlos Palenzuela, Steven~L. Liebling, David Neilsen, Luis Lehner, O.~L.
  Caballero, Evan O'Connor, and Matthew Anderson.
\newblock {Effects of the microphysical Equation of State in the mergers of
  magnetized Neutron Stars With Neutrino Cooling}.
\newblock {\em Phys. Rev.}, D92(4):044045, 2015.

\bibitem{lehner:2016lxy}
Luis Lehner, Steven~L. Liebling, Carlos Palenzuela, O.~L. Caballero, Evan
  O'Connor, Matthew Anderson, and David Neilsen.
\newblock {Unequal mass binary neutron star mergers and multimessenger
  signals}.
\newblock {\em Class. Quant. Grav.}, 33(18):184002, 2016.

\bibitem{sekiguchi:2016bjd}
Yuichiro Sekiguchi, Kenta Kiuchi, Koutarou Kyutoku, Masaru Shibata, and Keisuke
  Taniguchi.
\newblock {Dynamical mass ejection from the merger of asymmetric binary neutron
  stars: Radiation-hydrodynamics study in general relativity}.
\newblock {\em Phys. Rev.}, D93(12):124046, 2016.

\bibitem{foucart:2016rxm}
Francois Foucart, Evan O'Connor, Luke Roberts, Lawrence~E. Kidder, Harald~P.
  Pfeiffer, and Mark~A. Scheel.
\newblock {Impact of an improved neutrino energy estimate on outflows in
  neutron star merger simulations}.
\newblock {\em Phys. Rev.}, D94(12):123016, 2016.

\bibitem{bovard:2017mvn}
Luke Bovard, Dirk Martin, Federico Guercilena, Almudena Arcones, Luciano
  Rezzolla, and Oleg Korobkin.
\newblock {$r$-process nucleosynthesis from matter ejected in binary neutron
  star mergers}.
\newblock {\em Phys. Rev.}, D96(12):124005, 2017.

\bibitem{lattimer:1991nc}
James~M. Lattimer and F.~Douglas Swesty.
\newblock {A Generalized equation of state for hot, dense matter}.
\newblock {\em Nucl. Phys.}, A535:331--376, 1991.

\bibitem{radice:2012cu}
David Radice and Luciano Rezzolla.
\newblock {THC: a new high-order finite-difference high-resolution
  shock-capturing code for special-relativistic hydrodynamics}.
\newblock {\em Astron. Astrophys.}, 547:A26, 2012.

\bibitem{radice:2013hxh}
David Radice, Luciano Rezzolla, and Filippo Galeazzi.
\newblock {Beyond second-order convergence in simulations of binary neutron
  stars in full general-relativity}.
\newblock {\em Mon. Not. Roy. Astron. Soc.}, 437:L46--L50, 2014.

\bibitem{radice:2013xpa}
David Radice, Luciano Rezzolla, and Filippo Galeazzi.
\newblock {High-Order Fully General-Relativistic Hydrodynamics: new Approaches
  and Tests}.
\newblock {\em Class. Quant. Grav.}, 31:075012, 2014.

\bibitem{radice:2015nva}
David Radice, Luciano Rezzolla, and Filippo Galeazzi.
\newblock {High-Order Numerical-Relativity Simulations of Binary Neutron
  Stars}.
\newblock {\em ASP Conf. Ser.}, 498:121--126, 2015.

\bibitem{Bauswein:2017vtn}
Andreas Bauswein, Oliver Just, Hans-Thomas Janka, and Nikolaos Stergioulas.
\newblock {Neutron-star radius constraints from GW170817 and future
  detections}.
\newblock {\em Astrophys. J.}, 850(2):L34, 2017.

\bibitem{Radice:2017lry}
David Radice, Albino Perego, Francesco Zappa, and Sebastiano Bernuzzi.
\newblock {GW170817: Joint Constraint on the Neutron Star Equation of State
  from Multimessenger Observations}.
\newblock {\em Astrophys. J.}, 852(2):L29, 2018.

\bibitem{Bauswein:2013jpa}
A.~Bauswein, T.~W. Baumgarte, and H.~T. Janka.
\newblock {Prompt merger collapse and the maximum mass of neutron stars}.
\newblock {\em Phys. Rev. Lett.}, 111(13):131101, 2013.

\bibitem{Abbott17a}
B.~P. {Abbott}, R.~{Abbott}, T.~D. {Abbott}, F.~{Acernese}, K.~{Ackley},
  C.~{Adams}, T.~{Adams}, P.~{Addesso}, R.~X. {Adhikari}, V.~B. {Adya}, and
  et~al.
\newblock {GW170817: Observation of Gravitational Waves from a Binary Neutron
  Star Inspiral}.
\newblock {\em Physical Review Letters}, 119(16):161101, October 2017.

\bibitem{De:2018uhw}
Soumi De, Daniel Finstad, James~M. Lattimer, Duncan~A. Brown, Edo Berger, and
  Christopher~M. Biwer.
\newblock {Constraining the nuclear equation of state with GW170817}.
\newblock 2018.

\bibitem{Most:2018hfd}
Elias~R. Most, Lukas~R. Weih, Luciano Rezzolla, and Jürgen Schaffner-Bielich.
\newblock {New constraints on radii and tidal deformabilities of neutron stars
  from GW170817}.
\newblock {\em Phys. Rev. Lett.}, 120(26):261103, 2018.

\bibitem{Hotokezaka2013a}
K.~{Hotokezaka}, K.~{Kyutoku}, M.~{Tanaka}, K.~{Kiuchi}, Y.~{Sekiguchi},
  M.~{Shibata}, and S.~{Wanajo}.
\newblock {Progenitor Models of the Electromagnetic Transient Associated with
  the Short Gamma Ray Burst 130603B}.
\newblock {\em ApJL}, 778:L16, November 2013.

\bibitem{Dessart2009}
L.~{Dessart}, C.~D. {Ott}, A.~{Burrows}, S.~{Rosswog}, and E.~{Livne}.
\newblock {Neutrino Signatures and the Neutrino-Driven Wind in Binary Neutron
  Star Mergers}.
\newblock {\em ApJ}, 690:1681--1705, January 2009.

\bibitem{Siegel2014}
D.~M. {Siegel}, R.~{Ciolfi}, and L.~{Rezzolla}.
\newblock {Magnetically Driven Winds from Differentially Rotating Neutron Stars
  and X-Ray Afterglows of Short Gamma-Ray Bursts}.
\newblock {\em ApJL}, 785:L6, April 2014.

\bibitem{Ciolfi2017}
R.~{Ciolfi}, W.~{Kastaun}, B.~{Giacomazzo}, A.~{Endrizzi}, D.~M. {Siegel}, and
  R.~{Perna}.
\newblock {General relativistic magnetohydrodynamic simulations of binary
  neutron star mergers forming a long-lived neutron star}.
\newblock {\em \prd}, 95(6):063016, March 2017.

\bibitem{Fernandez2013}
R.~{Fern{\'a}ndez} and B.~D. {Metzger}.
\newblock {Delayed outflows from black hole accretion tori following neutron
  star binary coalescence}.
\newblock {\em MNRAS}, 435:502--517, October 2013.

\bibitem{just:2015}
O.~{Just}, A.~{Bauswein}, R.~{Ardevol Pulpillo}, S.~{Goriely}, and H.-T.
  {Janka}.
\newblock {Comprehensive nucleosynthesis analysis for ejecta of compact binary
  mergers}.
\newblock {\em MNRAS}, 448:541--567, March 2015.

\bibitem{Siegel2017a}
D.~M. {Siegel} and B.~D. {Metzger}.
\newblock {Three-Dimensional General-Relativistic Magnetohydrodynamic
  Simulations of Remnant Accretion Disks from Neutron Star Mergers: Outflows
  and r -Process Nucleosynthesis}.
\newblock {\em \prl}, 119(23):231102, December 2017.

\bibitem{Villar2017}
V.~A. {Villar}, J.~{Guillochon}, E.~{Berger}, B.~D. {Metzger}, P.~S.
  {Cowperthwaite}, M.~{Nicholl}, K.~D. {Alexander}, P.~K. {Blanchard},
  R.~{Chornock}, T.~{Eftekhari}, W.~{Fong}, R.~{Margutti}, and P.~K.~G.
  {Williams}.
\newblock {The Combined Ultraviolet, Optical, and Near-infrared Light Curves of
  the Kilonova Associated with the Binary Neutron Star Merger GW170817: Unified
  Data Set, Analytic Models, and Physical Implications}.
\newblock {\em ApJL}, 851:L21, December 2017.

\bibitem{Fernandez2018}
R.~{Fern{\'a}ndez}, A.~{Tchekhovskoy}, E.~{Quataert}, F.~{Foucart}, and
  D.~{Kasen}.
\newblock {Long-term GRMHD Simulations of Neutron Star Merger Accretion Disks:
  Implications for Electromagnetic Counterparts}.
\newblock {\em ArXiv e-prints}, August 2018.

\bibitem{Wu2016}
M.-R. {Wu}, R.~{Fern{\'a}ndez}, G.~{Mart{\'{\i}}nez-Pinedo}, and B.~D.
  {Metzger}.
\newblock {Production of the entire range of r-process nuclides by black hole
  accretion disc outflows from neutron star mergers}.
\newblock {\em \mnras}, 463:2323--2334, December 2016.

\bibitem{neilsen:2014}
D.~{Neilsen}, S.~L. {Liebling}, M.~{Anderson}, L.~{Lehner}, E.~{O'Connor}, and
  C.~{Palenzuela}.
\newblock {Magnetized neutron stars with realistic equations of state and
  neutrino cooling}.
\newblock {\em \prd}, 89(10):104029, May 2014.

\bibitem{deaton:2013}
M.~B. {Deaton}, M.~D. {Duez}, F.~{Foucart}, E.~{O'Connor}, C.~D. {Ott}, L.~E.
  {Kidder}, C.~D. {Muhlberger}, M.~A. {Scheel}, and B.~{Szilagyi}.
\newblock {Black Hole-Neutron Star Mergers with a Hot Nuclear Equation of
  State: Outflow and Neutrino-cooled Disk for a Low-mass, High-spin Case}.
\newblock {\em The Astrophysical Journal}, 776:47, October 2013.

\bibitem{perego:2014}
A.~{Perego}, S.~{Rosswog}, R.~M. {Cabez{\'o}n}, O.~{Korobkin},
  R.~{K{\"a}ppeli}, A.~{Arcones}, and M.~{Liebend{\"o}rfer}.
\newblock {Neutrino-driven winds from neutron star merger remnants}.
\newblock {\em MNRAS}, 443:3134--3156, October 2014.

\bibitem{shibata:2011}
M.~{Shibata}, K.~{Kiuchi}, Y.~{Sekiguchi}, and Y.~{Suwa}.
\newblock {Truncated Moment Formalism for Radiation Hydrodynamics in Numerical
  Relativity}.
\newblock {\em Progress of Theoretical Physics}, 125:1255--1287, June 2011.

\bibitem{foucart:2015}
F.~{Foucart}, E.~{O'Connor}, L.~{Roberts}, M.~D. {Duez}, R.~{Haas}, L.~E.
  {Kidder}, C.~D. {Ott}, H.~P. {Pfeiffer}, M.~A. {Scheel}, and B.~{Szilagyi}.
\newblock {Post-merger evolution of a neutron star-black hole binary with
  neutrino transport}.
\newblock {\em \prd}, 91(12):124021, June 2015.

\bibitem{shibata:2012}
M.~{Shibata} and Y.~{Sekiguchi}.
\newblock {Radiation Magnetohydrodynamics for Black Hole-Torus System in Full
  General Relativity: A Step toward Physical Simulation}.
\newblock {\em Progress of Theoretical Physics}, 127:535--559, March 2012.

\bibitem{correlations_CJH}
C.~J. Horowitz, O.~L. Caballero, Zidu Lin, Evan O'Connor, and A.~Schwenk.
\newblock Neutrino-nucleon scattering in supernova matter from the virial
  expansion.
\newblock {\em Phys. Rev. C}, 95:025801, Feb 2017.

\bibitem{correlations2_CJH}
Zidu Lin and C.~J. Horowitz.
\newblock Neutrino scattering in supernovae and the universal spin correlations
  of a unitary gas.
\newblock {\em Phys. Rev. C}, 96:055804, Nov 2017.

\bibitem{muons_CJH}
R.~Bollig, H.-T. Janka, A.~Lohs, G.~Mart\'{\i}nez-Pinedo, C.~J. Horowitz, and
  T.~Melson.
\newblock Muon creation in supernova matter facilitates neutrino-driven
  explosions.
\newblock {\em Phys. Rev. Lett.}, 119:242702, Dec 2017.

\bibitem{weakmag_CJH}
C.~J. Horowitz.
\newblock Weak magnetism for antineutrinos in supernovae.
\newblock {\em Phys. Rev. D}, 65:043001, Jan 2002.

\bibitem{bindingE_CJH}
C.~J. Horowitz, G.~Shen, Evan O'Connor, and Christian~D. Ott.
\newblock Charged-current neutrino interactions in core-collapse supernovae in
  a virial expansion.
\newblock {\em Phys. Rev. C}, 86:065806, Dec 2012.

\bibitem{coherent2_CJH}
D.~Akimov, J.~B. Albert, P.~An, C.~Awe, P.~S. Barbeau, B.~Becker, V.~Belov,
  A.~Brown, A.~Bolozdynya, B.~Cabrera-Palmer, M.~Cervantes, J.~I. Collar, R.~J.
  Cooper, R.~L. Cooper, C.~Cuesta, D.~J. Dean, J.~A. Detwiler, A.~Eberhardt,
  Y.~Efremenko, S.~R. Elliott, E.~M. Erkela, L.~Fabris, M.~Febbraro, N.~E.
  Fields, W.~Fox, Z.~Fu, A.~Galindo-Uribarri, M.~P. Green, M.~Hai, M.~R. Heath,
  S.~Hedges, D.~Hornback, T.~W. Hossbach, E.~B. Iverson, L.~J. Kaufman, S.~Ki,
  S.~R. Klein, A.~Khromov, A.~Konovalov, M.~Kremer, A.~Kumpan, C.~Leadbetter,
  L.~Li, W.~Lu, K.~Mann, D.~M. Markoff, K.~Miller, H.~Moreno, P.~E. Mueller,
  J.~Newby, J.~L. Orrell, C.~T. Overman, D.~S. Parno, S.~Penttila,
  G.~Perumpilly, H.~Ray, J.~Raybern, D.~Reyna, G.~C. Rich, D.~Rimal, D.~Rudik,
  K.~Scholberg, B.~J. Scholz, G.~Sinev, W.~M. Snow, V.~Sosnovtsev, A.~Shakirov,
  S.~Suchyta, B.~Suh, R.~Tayloe, R.~T. Thornton, I.~Tolstukhin, J.~Vanderwerp,
  R.~L. Varner, C.~J. Virtue, Z.~Wan, J.~Yoo, C.-H. Yu, A.~Zawada,
  J.~Zettlemoyer, and A.~M. Zderic.
\newblock Observation of coherent elastic neutrino-nucleus scattering.
\newblock {\em Science}, 2017.

\bibitem{coherent_CJH}
C.~J. Horowitz, K.~J. Coakley, and D.~N. McKinsey.
\newblock Supernova observation via neutrino-nucleus elastic scattering in the
  clean detector.
\newblock {\em Phys. Rev. D}, 68:023005, Jul 2003.

\bibitem{Metzger2010}
B.~D. {Metzger}, G.~{Mart{\'{\i}}nez-Pinedo}, S.~{Darbha}, E.~{Quataert},
  A.~{Arcones}, D.~{Kasen}, R.~{Thomas}, P.~{Nugent}, I.~V. {Panov}, and N.~T.
  {Zinner}.
\newblock {Electromagnetic counterparts of compact object mergers powered by
  the radioactive decay of r-process nuclei}.
\newblock {\em \mnras}, 406:2650--2662, August 2010.

\bibitem{Roberts2011}
L.~F. {Roberts}, D.~{Kasen}, W.~H. {Lee}, and E.~{Ramirez-Ruiz}.
\newblock {Electromagnetic Transients Powered by Nuclear Decay in the Tidal
  Tails of Coalescing Compact Binaries}.
\newblock {\em ApJL}, 736:L21, July 2011.

\bibitem{Kasen2013}
D.~{Kasen}, N.~R. {Badnell}, and J.~{Barnes}.
\newblock {Opacities and Spectra of the r-process Ejecta from Neutron Star
  Mergers}.
\newblock {\em ApJ}, 774:25, September 2013.

\bibitem{Tanaka2013}
M.~{Tanaka} and K.~{Hotokezaka}.
\newblock {Radiative Transfer Simulations of Neutron Star Merger Ejecta}.
\newblock {\em ApJ}, 775:113, October 2013.

\bibitem{Wollaeger2018}
R.~T. {Wollaeger}, O.~{Korobkin}, C.~J. {Fontes}, S.~K. {Rosswog}, W.~P.
  {Even}, C.~L. {Fryer}, J.~{Sollerman}, A.~L. {Hungerford}, D.~R. {van
  Rossum}, and A.~B. {Wollaber}.
\newblock {Impact of ejecta morphology and composition on the electromagnetic
  signatures of neutron star mergers}.
\newblock {\em MNRAS}, 478:3298--3334, August 2018.

\bibitem{Abbott17b}
B.~P. {Abbott}, R.~{Abbott}, T.~D. {Abbott}, F.~{Acernese}, K.~{Ackley},
  C.~{Adams}, T.~{Adams}, P.~{Addesso}, R.~X. {Adhikari}, V.~B. {Adya}, and
  et~al.
\newblock {Multi-messenger Observations of a Binary Neutron Star Merger}.
\newblock {\em The Astrophysical Journal Letters}, 848:L12, October 2017.

\bibitem{Pian2017}
E.~{Pian}, P.~{D'Avanzo}, S.~{Benetti}, M.~{Branchesi}, E.~{Brocato},
  S.~{Campana}, E.~{Cappellaro}, S.~{Covino}, V.~{D'Elia}, J.~P.~U. {Fynbo},
  F.~{Getman}, G.~{Ghirlanda}, G.~{Ghisellini}, A.~{Grado}, G.~{Greco},
  J.~{Hjorth}, C.~{Kouveliotou}, A.~{Levan}, L.~{Limatola}, D.~{Malesani},
  P.~A. {Mazzali}, A.~{Melandri}, P.~{Moeller}, L.~{Nicastro}, E.~{Palazzi},
  S.~{Piranomonte}, A.~{Rossi}, O.~S. {Salafia}, J.~{Selsing}, G.~{Stratta},
  M.~{Tanaka}, N.~R. {Tanvir}, L.~{Tomasella}, D.~{Watson}, S.~{Yang},
  L.~{Amati}, L.~A. {Antonelli}, S.~{Ascenzi}, M.~G. {Bernardini}, M.~{Boer},
  F.~{Bufano}, A.~{Bulgarelli}, M.~{Capaccioli}, P.~{Casella}, A.~J.
  {Castro-Tirado}, E.~{Chassande-Mottin}, R.~{Ciolfi}, C.~M. {Copperwheat},
  M.~{Dadina}, G.~{De Cesare}, A.~{di Paola}, Y.~Z. {Fan}, B.~{Gendre},
  G.~{Giuffrida}, A.~{Giunta}, L.~K. {Hunt}, G.~L. {Israel}, Z.-P. {Jin}, M.~M.
  {Kasliwal}, S.~{Klose}, M.~{Lisi}, F.~{Longo}, E.~{Maiorano}, M.~{Mapelli},
  N.~{Masetti}, L.~{Nava}, B.~{Patricelli}, D.~{Perley}, A.~{Pescalli},
  T.~{Piran}, A.~{Possenti}, L.~{Pulone}, M.~{Razzano}, R.~{Salvaterra},
  P.~{Schipani}, M.~{Spera}, A.~{Stamerra}, L.~{Stella}, G.~{Tagliaferri},
  V.~{Testa}, E.~{Troja}, M.~{Turatto}, S.~D. {Vergani}, and D.~{Vergani}.
\newblock {Spectroscopic identification of r-process nucleosynthesis in a
  double neutron-star merger}.
\newblock {\em Nature}, 551:67--70, November 2017.

\bibitem{tanvir17}
N.~R. {Tanvir}, A.~J. {Levan}, C.~{Gonz{\'a}lez-Fern{\'a}ndez}, O.~{Korobkin},
  I.~{Mandel}, S.~{Rosswog}, J.~{Hjorth}, P.~{D'Avanzo}, A.~S. {Fruchter},
  C.~L. {Fryer}, T.~{Kangas}, B.~{Milvang-Jensen}, S.~{Rosetti}, D.~{Steeghs},
  R.~T. {Wollaeger}, Z.~{Cano}, C.~M. {Copperwheat}, S.~{Covino}, V.~{D'Elia},
  A.~{de Ugarte Postigo}, P.~A. {Evans}, W.~P. {Even}, S.~{Fairhurst},
  R.~{Figuera Jaimes}, C.~J. {Fontes}, Y.~I. {Fujii}, J.~P.~U. {Fynbo}, B.~P.
  {Gompertz}, J.~{Greiner}, G.~{Hodosan}, M.~J. {Irwin}, P.~{Jakobsson}, U.~G.
  {J{\o}rgensen}, D.~A. {Kann}, J.~D. {Lyman}, D.~{Malesani}, R.~G. {McMahon},
  A.~{Melandri}, P.~T. {O'Brien}, J.~P. {Osborne}, E.~{Palazzi}, D.~A.
  {Perley}, E.~{Pian}, S.~{Piranomonte}, M.~{Rabus}, E.~{Rol}, A.~{Rowlinson},
  S.~{Schulze}, P.~{Sutton}, C.~C. {Th{\"o}ne}, K.~{Ulaczyk}, D.~{Watson},
  K.~{Wiersema}, and R.~A.~M.~J. {Wijers}.
\newblock {The Emergence of a Lanthanide-rich Kilonova Following the Merger of
  Two Neutron Stars}.
\newblock {\em Astrophysical Journal Letters}, 848:L27, October 2017.

\bibitem{cowperthwaite17}
P.~S. {Cowperthwaite}, E.~{Berger}, V.~A. {Villar}, B.~D. {Metzger},
  M.~{Nicholl}, R.~{Chornock}, P.~K. {Blanchard}, W.~{Fong}, R.~{Margutti},
  M.~{Soares-Santos}, K.~D. {Alexander}, S.~{Allam}, J.~{Annis}, D.~{Brout},
  D.~A. {Brown}, R.~E. {Butler}, H.-Y. {Chen}, H.~T. {Diehl}, Z.~{Doctor},
  M.~R. {Drout}, T.~{Eftekhari}, B.~{Farr}, D.~A. {Finley}, R.~J. {Foley},
  J.~A. {Frieman}, C.~L. {Fryer}, J.~{Garc{\'{\i}}a-Bellido}, M.~S.~S. {Gill},
  J.~{Guillochon}, K.~{Herner}, D.~E. {Holz}, D.~{Kasen}, R.~{Kessler},
  J.~{Marriner}, T.~{Matheson}, E.~H. {Neilsen}, Jr., E.~{Quataert},
  A.~{Palmese}, A.~{Rest}, M.~{Sako}, D.~M. {Scolnic}, N.~{Smith}, D.~L.
  {Tucker}, P.~K.~G. {Williams}, E.~{Balbinot}, J.~L. {Carlin}, E.~R. {Cook},
  F.~{Durret}, T.~S. {Li}, P.~A.~A. {Lopes}, A.~C.~C. {Louren{\c c}o}, J.~L.
  {Marshall}, G.~E. {Medina}, J.~{Muir}, R.~R. {Mu{\~n}oz}, M.~{Sauseda}, D.~J.
  {Schlegel}, L.~F. {Secco}, A.~K. {Vivas}, W.~{Wester}, A.~{Zenteno},
  Y.~{Zhang}, T.~M.~C. {Abbott}, M.~{Banerji}, K.~{Bechtol},
  A.~{Benoit-L{\'e}vy}, E.~{Bertin}, E.~{Buckley-Geer}, D.~L. {Burke},
  D.~{Capozzi}, A.~{Carnero Rosell}, M.~{Carrasco Kind}, F.~J. {Castander},
  M.~{Crocce}, C.~E. {Cunha}, C.~B. {D'Andrea}, L.~N. {da Costa}, C.~{Davis},
  D.~L. {DePoy}, S.~{Desai}, J.~P. {Dietrich}, A.~{Drlica-Wagner}, T.~F.
  {Eifler}, A.~E. {Evrard}, E.~{Fernandez}, B.~{Flaugher}, P.~{Fosalba},
  E.~{Gaztanaga}, D.~W. {Gerdes}, T.~{Giannantonio}, D.~A. {Goldstein},
  D.~{Gruen}, R.~A. {Gruendl}, G.~{Gutierrez}, K.~{Honscheid}, B.~{Jain}, D.~J.
  {James}, T.~{Jeltema}, M.~W.~G. {Johnson}, M.~D. {Johnson}, S.~{Kent},
  E.~{Krause}, R.~{Kron}, K.~{Kuehn}, N.~{Nuropatkin}, O.~{Lahav}, M.~{Lima},
  H.~{Lin}, M.~A.~G. {Maia}, M.~{March}, P.~{Martini}, R.~G. {McMahon},
  F.~{Menanteau}, C.~J. {Miller}, R.~{Miquel}, J.~J. {Mohr}, E.~{Neilsen},
  R.~C. {Nichol}, R.~L.~C. {Ogando}, A.~A. {Plazas}, N.~{Roe}, A.~K. {Romer},
  A.~{Roodman}, E.~S. {Rykoff}, E.~{Sanchez}, V.~{Scarpine}, R.~{Schindler},
  M.~{Schubnell}, I.~{Sevilla-Noarbe}, M.~{Smith}, R.~C. {Smith},
  F.~{Sobreira}, E.~{Suchyta}, M.~E.~C. {Swanson}, G.~{Tarle}, D.~{Thomas},
  R.~C. {Thomas}, M.~A. {Troxel}, V.~{Vikram}, A.~R. {Walker}, R.~H.
  {Wechsler}, J.~{Weller}, B.~{Yanny}, and J.~{Zuntz}.
\newblock {The Electromagnetic Counterpart of the Binary Neutron Star Merger
  LIGO/Virgo GW170817. II. UV, Optical, and Near-infrared Light Curves and
  Comparison to Kilonova Models}.
\newblock {\em Astrophysical Journal Letters}, 848:L17, October 2017.

\bibitem{Chornock2017}
R.~{Chornock}, E.~{Berger}, D.~{Kasen}, P.~S. {Cowperthwaite}, M.~{Nicholl},
  V.~A. {Villar}, K.~D. {Alexander}, P.~K. {Blanchard}, T.~{Eftekhari},
  W.~{Fong}, R.~{Margutti}, P.~K.~G. {Williams}, J.~{Annis}, D.~{Brout}, D.~A.
  {Brown}, H.-Y. {Chen}, M.~R. {Drout}, B.~{Farr}, R.~J. {Foley}, J.~A.
  {Frieman}, C.~L. {Fryer}, K.~{Herner}, D.~E. {Holz}, R.~{Kessler},
  T.~{Matheson}, B.~D. {Metzger}, E.~{Quataert}, A.~{Rest}, M.~{Sako}, D.~M.
  {Scolnic}, N.~{Smith}, and M.~{Soares-Santos}.
\newblock {The Electromagnetic Counterpart of the Binary Neutron Star Merger
  LIGO/Virgo GW170817. IV. Detection of Near-infrared Signatures of r-process
  Nucleosynthesis with Gemini-South}.
\newblock {\em ApJL}, 848:L19, October 2017.

\bibitem{drout17}
M.~R. {Drout}, A.~L. {Piro}, B.~J. {Shappee}, C.~D. {Kilpatrick}, J.~D.
  {Simon}, C.~{Contreras}, D.~A. {Coulter}, R.~J. {Foley}, M.~R. {Siebert},
  N.~{Morrell}, K.~{Boutsia}, F.~{Di Mille}, T.~W.-S. {Holoien}, D.~{Kasen},
  J.~A. {Kollmeier}, B.~F. {Madore}, A.~J. {Monson}, A.~{Murguia-Berthier},
  Y.-C. {Pan}, J.~X. {Prochaska}, E.~{Ramirez-Ruiz}, A.~{Rest}, C.~{Adams},
  K.~{Alatalo}, E.~{Ba{\~n}ados}, J.~{Baughman}, T.~C. {Beers}, R.~A.
  {Bernstein}, T.~{Bitsakis}, A.~{Campillay}, T.~T. {Hansen}, C.~R. {Higgs},
  A.~P. {Ji}, G.~{Maravelias}, J.~L. {Marshall}, C.~M. {Bidin}, J.~L. {Prieto},
  K.~C. {Rasmussen}, C.~{Rojas-Bravo}, A.~L. {Strom}, N.~{Ulloa},
  J.~{Vargas-Gonz{\'a}lez}, Z.~{Wan}, and D.~D. {Whitten}.
\newblock {Light curves of the neutron star merger GW170817/SSS17a:
  Implications for r-process nucleosynthesis}.
\newblock {\em Science}, 358:1570--1574, December 2017.

\bibitem{Tanaka2017}
M.~{Tanaka}, Y.~{Utsumi}, P.~A. {Mazzali}, N.~{Tominaga}, M.~{Yoshida},
  Y.~{Sekiguchi}, T.~{Morokuma}, K.~{Motohara}, K.~{Ohta}, K.~S. {Kawabata},
  F.~{Abe}, K.~{Aoki}, Y.~{Asakura}, S.~{Baar}, S.~{Barway}, I.~A. {Bond},
  M.~{Doi}, T.~{Fujiyoshi}, H.~{Furusawa}, S.~{Honda}, Y.~{Itoh},
  M.~{Kawabata}, N.~{Kawai}, J.~H. {Kim}, C.-H. {Lee}, S.~{Miyazaki},
  K.~{Morihana}, H.~{Nagashima}, T.~{Nagayama}, T.~{Nakaoka}, F.~{Nakata},
  R.~{Ohsawa}, T.~{Ohshima}, H.~{Okita}, T.~{Saito}, T.~{Sumi}, A.~{Tajitsu},
  J.~{Takahashi}, M.~{Takayama}, Y.~{Tamura}, I.~{Tanaka}, T.~{Terai}, P.~J.
  {Tristram}, N.~{Yasuda}, and T.~{Zenko}.
\newblock {Kilonova from post-merger ejecta as an optical and near-Infrared
  counterpart of GW170817}.
\newblock {\em PASJ}, 69:102, December 2017.

\bibitem{Shibata2017}
M.~{Shibata}, S.~{Fujibayashi}, K.~{Hotokezaka}, K.~{Kiuchi}, K.~{Kyutoku},
  Y.~{Sekiguchi}, and M.~{Tanaka}.
\newblock {Modeling GW170817 based on numerical relativity and its
  implications}.
\newblock {\em PRD}, 96(12):123012, December 2017.

\bibitem{Perego2017}
A.~{Perego}, D.~{Radice}, and S.~{Bernuzzi}.
\newblock {AT 2017gfo: An Anisotropic and Three-component Kilonova Counterpart
  of GW170817}.
\newblock {\em ApJL}, 850:L37, December 2017.

\bibitem{Fujibayashi2018}
S.~{Fujibayashi}, K.~{Kiuchi}, N.~{Nishimura}, Y.~{Sekiguchi}, and
  M.~{Shibata}.
\newblock {Mass Ejection from the Remnant of a Binary Neutron Star Merger:
  Viscous-radiation Hydrodynamics Study}.
\newblock {\em The Astrophysical Journal}, 860:64, June 2018.

\bibitem{Grossman2014}
D.~{Grossman}, O.~{Korobkin}, S.~{Rosswog}, and T.~{Piran}.
\newblock {The long-term evolution of neutron star merger remnants - II.
  Radioactively powered transients}.
\newblock {\em MNRAS}, 439:757--770, March 2014.

\bibitem{Martin2015}
D.~{Martin}, A.~{Perego}, A.~{Arcones}, F.-K. {Thielemann}, O.~{Korobkin}, and
  S.~{Rosswog}.
\newblock {Neutrino-driven Winds in the Aftermath of a Neutron Star Merger:
  Nucleosynthesis and Electromagnetic Transients}.
\newblock {\em ApJ}, 813:2, November 2015.

\bibitem{Barnes2016}
J.~{Barnes}, D.~{Kasen}, M.-R. {Wu}, and G.~{Mart{\'{\i}}nez-Pinedo}.
\newblock {Radioactivity and Thermalization in the Ejecta of Compact Object
  Mergers and Their Impact on Kilonova Light Curves}.
\newblock {\em The Astrophysical Journal}, 829:110, October 2016.

\bibitem{Kawaguchi2018}
K.~{Kawaguchi}, M.~{Shibata}, and M.~{Tanaka}.
\newblock {Radiative transfer simulation for the optical and near-infrared
  electromagnetic counterparts to GW170817}.
\newblock {\em ArXiv e-prints}, June 2018.

\bibitem{rosswog18}
S.~{Rosswog}, J.~{Sollerman}, U.~{Feindt}, A.~{Goobar}, O.~{Korobkin},
  C.~{Fremling}, and M.~{Kasliwal}.
\newblock {The first direct double neutron star merger detection: implications
  for cosmic nucleosynthesis}.
\newblock {\em ArXiv e-prints}, October 2017.

\bibitem{kasliwal17}
M.~M. {Kasliwal}, E.~{Nakar}, L.~P. {Singer}, D.~L. {Kaplan}, D.~O. {Cook},
  A.~{Van Sistine}, R.~M. {Lau}, C.~{Fremling}, O.~{Gottlieb}, J.~E. {Jencson},
  S.~M. {Adams}, U.~{Feindt}, K.~{Hotokezaka}, S.~{Ghosh}, D.~A. {Perley},
  P.-C. {Yu}, T.~{Piran}, J.~R. {Allison}, G.~C. {Anupama},
  A.~{Balasubramanian}, K.~W {Bannister}, J.~{Bally}, J.~{Barnes}, S.~{Barway},
  E.~{Bellm}, V.~{Bhalerao}, D.~{Bhattacharya}, N.~{Blagorodnova}, J.~S.
  {Bloom}, P.~R. {Brady}, C.~{Cannella}, D.~{Chatterjee}, S.~B. {Cenko}, B.~E.
  {Cobb}, C.~{Copperwheat}, A.~{Corsi}, K.~{De}, D.~{Dobie}, S.~W.~K. {Emery},
  P.~A. {Evans}, O.~D. {Fox}, D.~A. {Frail}, C.~{Frohmaier}, A.~{Goobar},
  G.~{Hallinan}, F.~{Harrison}, G.~{Helou}, T.~{Hinderer}, A.~Y.~Q. {Ho},
  A.~{Horesh}, W.-H. {Ip}, R.~{Itoh}, D.~{Kasen}, H.~{Kim}, N.~P.~M. {Kuin},
  T.~{Kupfer}, C.~{Lynch}, K.~{Madsen}, P.~A. {Mazzali}, A.~A. {Miller},
  K.~{Mooley}, T.~{Murphy}, C.-C. {Ngeow}, D.~{Nichols}, S.~{Nissanke},
  P.~{Nugent}, E.~O. {Ofek}, H.~{Qi}, R.~M. {Quimby}, S.~{Rosswog}, F.~{Rusu},
  E.~M. {Sadler}, P.~{Schmidt}, J.~{Sollerman}, I.~{Steele}, A.~R.
  {Williamson}, Y.~{Xu}, L.~{Yan}, Y.~{Yatsu}, C.~{Zhang}, and W.~{Zhao}.
\newblock {Illuminating Gravitational Waves: A Concordant Picture of Photons
  from a Neutron Star Merger}.
\newblock {\em Science in press, available via doi:10.1126/science.aap9455},
  October 2017.

\bibitem{pinto00a}
P.~A. {Pinto} and R.~G. {Eastman}.
\newblock {The Physics of Type IA Supernova Light Curves. I. Analytic Results
  and Time Dependence}.
\newblock {\em \apj}, 530:744--756, February 2000.

\bibitem{rosswog17}
S.~{Rosswog}, U.~{Feindt}, O.~{Korobkin}, M.-R. {Wu}, J.~{Sollerman},
  A.~{Goobar}, and G.~{Martinez-Pinedo}.
\newblock {Detectability of compact binary merger macronovae}.
\newblock {\em Classical and Quantum Gravity}, 34(10):104001, May 2017.

\bibitem{richers:2015}
S.~{Richers}, D.~{Kasen}, E.~{O'Connor}, R.~{Fern{\'a}ndez}, and C.~D. {Ott}.
\newblock {Monte Carlo Neutrino Transport through Remnant Disks from Neutron
  Star Mergers}.
\newblock {\em The Astrophysical Journal}, 813:38, November 2015.

\bibitem{foucart:2018}
F.~{Foucart}, M.~D. {Duez}, L.~E. {Kidder}, R.~{Nguyen}, H.~P. {Pfeiffer}, and
  M.~A. {Scheel}.
\newblock {Evaluating radiation transport errors in merger simulations using a
  Monte-Carlo algorithm}.
\newblock {\em ArXiv e-prints}, June 2018.

\bibitem{zhu:2016}
Y.~L. {Zhu}, A.~{Perego}, and G.~C. {McLaughlin}.
\newblock {Matter-neutrino resonance transitions above a neutron star merger
  remnant}.
\newblock {\em \prd}, 94(10):105006, November 2016.

\bibitem{frensel:2017}
M.~{Frensel}, M.-R. {Wu}, C.~{Volpe}, and A.~{Perego}.
\newblock {Neutrino flavor evolution in binary neutron star merger remnants}.
\newblock {\em \prd}, 95(2):023011, January 2017.

\bibitem{holmbeck18}
E.~M. {Holmbeck}, T.~C. {Beers}, I.~U. {Roederer}, V.~M. {Placco}, T.~T.
  {Hansen}, C.~M. {Sakari}, C.~{Sneden}, C.~{Liu}, Y.~S. {Lee}, J.~J. {Cowan},
  and A.~{Frebel}.
\newblock {The R-Process Alliance: 2MASS J09544277+5246414, the Most
  Actinide-enhanced R-II Star Known}.
\newblock {\em ApJL}, 859:L24, June 2018.

\bibitem{Schatz2002}
H.~{Schatz}, R.~{Toenjes}, B.~{Pfeiffer}, T.~C. {Beers}, J.~J. {Cowan},
  V.~{Hill}, and K.-L. {Kratz}.
\newblock {Thorium and Uranium Chronometers Applied to CS 31082-001}.
\newblock {\em The Astrophysical Journal}, 579:626--638, November 2002.

\bibitem{farouqi2010}
K.~{Farouqi}, K.-L. {Kratz}, B.~{Pfeiffer}, T.~{Rauscher}, F.-K. {Thielemann},
  and J.~W. {Truran}.
\newblock {Charged-particle and Neutron-capture Processes in the High-entropy
  Wind of Core-collapse Supernovae}.
\newblock {\em \apj}, 712:1359--1377, April 2010.

\bibitem{wanajo2002}
S.~{Wanajo}, N.~{Itoh}, Y.~{Ishimaru}, S.~{Nozawa}, and T.~C. {Beers}.
\newblock {The r-Process in the Neutrino Winds of Core-Collapse Supernovae and
  U-Th Cosmochronology}.
\newblock {\em \apj}, 577:853--865, October 2002.

\bibitem{2015ARA&A..53..631F}
A.~{Frebel} and J.~E. {Norris}.
\newblock {Near-Field Cosmology with Extremely Metal-Poor Stars}.
\newblock {\em Annual Review of Astronomy and Astrophysics}, 53:631--688,
  August 2015.

\bibitem{Frebel2018}
A.~{Frebel}.
\newblock {From Nuclei to the Cosmos: Tracing Heavy-Element Production with the
  Oldest Stars}.
\newblock {\em ArXiv e-prints}, June 2018.

\bibitem{2016A&A...586A..49B}
C.~{Battistini} and T.~{Bensby}.
\newblock {The origin and evolution of r- and s-process elements in the Milky
  Way stellar disk}.
\newblock {\em Astronomy \& Astrophysics}, 586:A49, February 2016.

\bibitem{2018MNRAS.478.4513B}
S.~{Buder}, M.~{Asplund}, L.~{Duong}, J.~{Kos}, K.~{Lind}, M.~K. {Ness},
  S.~{Sharma}, J.~{Bland-Hawthorn}, A.~R. {Casey}, G.~M. {De Silva},
  V.~{D'Orazi}, K.~C. {Freeman}, G.~F. {Lewis}, J.~{Lin}, S.~L. {Martell},
  K.~J. {Schlesinger}, J.~D. {Simpson}, D.~B. {Zucker}, T.~{Zwitter}, A.~M.
  {Amarsi}, B.~{Anguiano}, D.~{Carollo}, L.~{Casagrande}, K.~{{\v C}otar},
  P.~L. {Cottrell}, G.~{Da Costa}, X.~D. {Gao}, M.~R. {Hayden}, J.~{Horner},
  M.~J. {Ireland}, P.~R. {Kafle}, U.~{Munari}, D.~M. {Nataf}, T.~{Nordlander},
  D.~{Stello}, Y.-S. {Ting}, G.~{Traven}, F.~{Watson}, R.~A. {Wittenmyer},
  R.~F.~G. {Wyse}, D.~{Yong}, J.~C. {Zinn}, and M.~{{\v Z}erjal}.
\newblock {The GALAH Survey: second data release}.
\newblock {\em \mnras}, 478:4513--4552, August 2018.

\bibitem{2014ARA&A..52...43B}
E.~{Berger}.
\newblock {Short-Duration Gamma-Ray Bursts}.
\newblock {\em Annual Review of Astronomy and Astrophysics}, 52:43--105, August
  2014.

\bibitem{2017ApJ...848L..23F}
W.~{Fong}, E.~{Berger}, P.~K. {Blanchard}, R.~{Margutti}, P.~S.
  {Cowperthwaite}, R.~{Chornock}, K.~D. {Alexander}, B.~D. {Metzger}, V.~A.
  {Villar}, M.~{Nicholl}, T.~{Eftekhari}, P.~K.~G. {Williams}, J.~{Annis},
  D.~{Brout}, D.~A. {Brown}, H.-Y. {Chen}, Z.~{Doctor}, H.~T. {Diehl}, D.~E.
  {Holz}, A.~{Rest}, M.~{Sako}, and M.~{Soares-Santos}.
\newblock {The Electromagnetic Counterpart of the Binary Neutron Star Merger
  LIGO/Virgo GW170817. VIII. A Comparison to Cosmological Short-duration
  Gamma-Ray Bursts}.
\newblock {\em The Astrophysical Journal Letter}, 848:L23, October 2017.

\bibitem{2017arXiv170607053B}
K.~{Belczynski}, J.~{Klencki}, G.~{Meynet}, C.~L. {Fryer}, D.~A. {Brown},
  M.~{Chruslinska}, W.~{Gladysz}, R.~{O'Shaughnessy}, T.~{Bulik}, E.~{Berti},
  D.~E. {Holz}, D.~{Gerosa}, M.~{Giersz}, S.~{Ekstrom}, C.~{Georgy},
  A.~{Askar}, D.~{Wysocki}, and J.-P. {Lasota}.
\newblock {The origin of low spin of black holes in LIGO/Virgo mergers}.
\newblock {\em ArXiv e-prints}, June 2017.

\bibitem{2003PASA...20..401G}
B.~K. {Gibson}, Y.~{Fenner}, A.~{Renda}, D.~{Kawata}, and H.-c. {Lee}.
\newblock {Galactic Chemical Evolution}.
\newblock {\em Publications of the Astronomical Society of Australia},
  20:401--415, 2003.

\bibitem{2008EAS....32..311P}
N.~{Prantzos}.
\newblock {An Introduction to Galactic Chemical Evolution}.
\newblock In C.~{Charbonnel} and J.-P. {Zahn}, editors, {\em EAS Publications
  Series}, volume~32 of {\em EAS Publications Series}, pages 311--356, November
  2008.

\bibitem{2013ARA&A..51..457N}
K.~{Nomoto}, C.~{Kobayashi}, and N.~{Tominaga}.
\newblock {Nucleosynthesis in Stars and the Chemical Enrichment of Galaxies}.
\newblock {\em Annual Review of Astronomy and Astrophysics}, 51:457--509,
  August 2013.

\bibitem{2014SAAS...37..145M}
F.~{Matteucci}.
\newblock {Chemical Evolution of the Milky Way and Its Satellites}.
\newblock {\em The Origin of the Galaxy and Local Group, Saas-Fee Advanced
  Course, Volume 37.~ISBN 978-3-642-41719-1.~Springer-Verlag Berlin Heidelberg,
  2014, p.~145}, 37:145, 2014.

\bibitem{2017ApJ...836..230C}
B.~{C{\^o}t{\'e}}, K.~{Belczynski}, C.~L. {Fryer}, C.~{Ritter}, A.~{Paul},
  B.~{Wehmeyer}, and B.~W. {O'Shea}.
\newblock {Advanced LIGO Constraints on Neutron Star Mergers and r-process
  Sites}.
\newblock {\em The Astrophysical Journal}, 836:230, February 2017.

\bibitem{2018arXiv180101141H}
K.~{Hotokezaka}, P.~{Beniamini}, and T.~{Piran}.
\newblock {Neutron Star Mergers as sites of r-process Nucleosynthesis and Short
  Gamma-Ray Bursts}.
\newblock {\em ArXiv e-prints}, January 2018.

\bibitem{2018ApJ...859...67C}
B.~{C{\^o}t{\'e}}, D.~W. {Silvia}, B.~W. {O'Shea}, B.~{Smith}, and J.~H.
  {Wise}.
\newblock {Validating Semi-analytic Models of High-redshift Galaxy Formation
  Using Radiation Hydrodynamical Simulations}.
\newblock {\em \apj}, 859:67, May 2018.

\bibitem{2016ApJ...824...82C}
B.~{C{\^o}t{\'e}}, C.~{Ritter}, B.~W. {O'Shea}, F.~{Herwig}, M.~{Pignatari},
  S.~{Jones}, and C.~L. {Fryer}.
\newblock {Uncertainties in Galactic Chemical Evolution Models}.
\newblock {\em \apj}, 824:82, June 2016.

\bibitem{2017ApJ...835..128C}
B.~{C{\^o}t{\'e}}, B.~W. {O'Shea}, C.~{Ritter}, F.~{Herwig}, and K.~A. {Venn}.
\newblock {The Impact of Modeling Assumptions in Galactic Chemical Evolution
  Models}.
\newblock {\em \apj}, 835:128, February 2017.

\bibitem{2012ApJ...760..112G}
F.~A. {G{\'o}mez}, C.~E. {Coleman-Smith}, B.~W. {O'Shea}, J.~{Tumlinson}, and
  R.~L. {Wolpert}.
\newblock {Characterizing the Formation History of Milky Way like Stellar Halos
  with Model Emulators}.
\newblock {\em \apj}, 760:112, December 2012.

\bibitem{2014ApJ...787...20G}
F.~A. {G{\'o}mez}, C.~E. {Coleman-Smith}, B.~W. {O'Shea}, J.~{Tumlinson}, and
  R.~L. {Wolpert}.
\newblock {Dissecting Galaxy Formation Models with Sensitivity Analysis--a New
  Approach to Constrain the Milky Way Formation History}.
\newblock {\em \apj}, 787:20, May 2014.

\bibitem{2017nuco.confb0203C}
B.~{C{\^o}t{\'e}}, C.~{Ritter}, F.~{Herwig}, B.~W. {O'Shea}, M.~{Pignatari},
  D.~{Silvia}, S.~{Jones}, and C.~L. {Fryer}.
\newblock {JINA-NuGrid Galactic Chemical Evolution Pipeline}.
\newblock In S.~{Kubono}, T.~{Kajino}, S.~{Nishimura}, T.~{Isobe},
  S.~{Nagataki}, T.~{Shima}, and Y.~{Takeda}, editors, {\em 14th International
  Symposium on Nuclei in the Cosmos (NIC2016)}, page 020203, 2017.

\bibitem{Roederer2009}
I.~U. {Roederer}, K.-L. {Kratz}, A.~{Frebel}, N.~{Christlieb}, B.~{Pfeiffer},
  J.~J. {Cowan}, and C.~{Sneden}.
\newblock {The End of Nucleosynthesis: Production of Lead and Thorium in the
  Early Galaxy}.
\newblock {\em The Astrophysical Journal}, 698:1963--1980, June 2009.

\bibitem{Mashonkina2014}
L.~{Mashonkina}, N.~{Christlieb}, and K.~{Eriksson}.
\newblock {The Hamburg/ESO R-process Enhanced Star survey (HERES). X. HE
  2252-4225, one more r-process enhanced and actinide-boost halo star}.
\newblock {\em Astronomy \& Astrophysics}, 569:A43, September 2014.

\bibitem{JiFrebel2018}
A.~P. {Ji} and A.~{Frebel}.
\newblock {From Actinides to Zinc: Using the Full Abundance Pattern of the
  Brightest Star in Reticulum II to Distinguish between Different r-process
  Sites}.
\newblock {\em The Astrophysical Journal}, 856:138, April 2018.

\bibitem{Eichler2015}
M.~{Eichler}, A.~{Arcones}, A.~{Kelic}, O.~{Korobkin}, K.~{Langanke},
  T.~{Marketin}, G.~{Martinez-Pinedo}, I.~{Panov}, T.~{Rauscher}, S.~{Rosswog},
  C.~{Winteler}, N.~T. {Zinner}, and F.-K. {Thielemann}.
\newblock {The Role of Fission in Neutron Star Mergers and Its Impact on the
  r-Process Peaks}.
\newblock {\em The Astrophysical Journal}, 808:30, July 2015.

\bibitem{Goriely2015}
S.~{Goriely}.
\newblock {The fundamental role of fission during r-process nucleosynthesis in
  neutron star mergers}.
\newblock {\em European Physical Journal A}, 51:22, February 2015.

\bibitem{Rosswog2013}
S.~{Rosswog}, T.~{Piran}, and E.~{Nakar}.
\newblock {The multimessenger picture of compact object encounters: binary
  mergers versus dynamical collisions}.
\newblock {\em MNRAS}, 430:2585--2604, April 2013.

\bibitem{Winteler2012}
C.~{Winteler}, R.~{K{\"a}ppeli}, A.~{Perego}, A.~{Arcones}, N.~{Vasset},
  N.~{Nishimura}, M.~{Liebend{\"o}rfer}, and F.-K. {Thielemann}.
\newblock {Magnetorotationally Driven Supernovae as the Origin of Early Galaxy
  r-process Elements?}
\newblock {\em The Astrophysical Journal Letters}, 750:L22, May 2012.

\bibitem{Kelic2009}
A.~{Kelic}, M.~{Valentina Ricciardi}, and K.-H. {Schmidt}.
\newblock {ABLA07 - towards a complete description of the decay channels of a
  nuclear system from spontaneous fission to multifragmentation}.
\newblock {\em ArXiv e-prints}, June 2009.

\bibitem{Martin.etal:2015}
D.~{Martin}, A.~{Perego}, A.~{Arcones}, F.-K. {Thielemann}, O.~{Korobkin}, and
  S.~{Rosswog}.
\newblock {Neutrino-driven Winds in the Aftermath of a Neutron Star Merger:
  Nucleosynthesis and Electromagnetic Transients}.
\newblock {\em ApJ}, 813:2, November 2015.

\bibitem{Hansen.etal:2014}
C.~J. {Hansen}, F.~{Montes}, and A.~{Arcones}.
\newblock {How Many Nucleosynthesis Processes Exist at Low Metallicity?}
\newblock {\em ApJ}, 797:123, December 2014.

\bibitem{Nishimura2017}
N.~{Nishimura}, H.~{Sawai}, T.~{Takiwaki}, S.~{Yamada}, and F.-K. {Thielemann}.
\newblock {The Intermediate r-process in Core-collapse Supernovae Driven by the
  Magneto-rotational Instability}.
\newblock {\em ApJL}, 836:L21, February 2017.

\bibitem{Moesta2017}
P.~{M{\"o}sta}, L.~F. {Roberts}, G.~{Halevi}, C.~D. {Ott}, J.~{Lippuner},
  R.~{Haas}, and E.~{Schnetter}.
\newblock {R-process Nucleosynthesis from Three-Dimensional Magnetorotational
  Core-Collapse Supernovae}.
\newblock {\em ArXiv e-prints}, December 2017.

\bibitem{ObergaulingerAloy2017}
M.~{Obergaulinger} and M.~{\'A}. {Aloy}.
\newblock {Protomagnetar and black hole formation in high-mass stars}.
\newblock {\em MNRAS}, 469:L43--L47, July 2017.

\bibitem{hanke:2013}
F.~{Hanke}, B.~{M{\"u}ller}, A.~{Wongwathanarat}, A.~{Marek}, and H.-T.
  {Janka}.
\newblock {SASI Activity in Three-dimensional Neutrino-hydrodynamics
  Simulations of Supernova Cores}.
\newblock {\em \apj}, 770:66, June 2013.

\bibitem{lentz:2015}
E.~J. {Lentz}, S.~W. {Bruenn}, W.~R. {Hix}, A.~{Mezzacappa}, O.~E.~B. {Messer},
  E.~{Endeve}, J.~M. {Blondin}, J.~A. {Harris}, P.~{Marronetti}, and K.~N.
  {Yakunin}.
\newblock {Three-dimensional Core-collapse Supernova Simulated Using a 15 M
  $_{\odot}$ Progenitor}.
\newblock {\em \apj Letters}, 807:L31, July 2015.

\bibitem{roberts:2016}
L.~F. {Roberts}, C.~D. {Ott}, R.~{Haas}, E.~P. {O'Connor}, P.~{Diener}, and
  E.~{Schnetter}.
\newblock {General-Relativistic Three-Dimensional Multi-group Neutrino
  Radiation-Hydrodynamics Simulations of Core-Collapse Supernovae}.
\newblock {\em \apj}, 831:98, November 2016.

\bibitem{melson:2015a}
T.~{Melson}, H.-T. {Janka}, R.~{Bollig}, F.~{Hanke}, A.~{Marek}, and
  B.~{M{\"u}ller}.
\newblock {Neutrino-driven Explosion of a 20 Solar-mass Star in Three
  Dimensions Enabled by Strange-quark Contributions to Neutrino-Nucleon
  Scattering}.
\newblock {\em \apj Letters}, 808:L42, August 2015.

\bibitem{summa:2018}
A.~{Summa}, H.-T. {Janka}, T.~{Melson}, and A.~{Marek}.
\newblock {Rotation-supported Neutrino-driven Supernova Explosions in Three
  Dimensions and the Critical Luminosity Condition}.
\newblock {\em \apj}, 852:28, January 2018.

\bibitem{oconnor:2018b}
E.~{O'Connor} and S.~{Couch}.
\newblock {Exploring Fundamentally Three-dimensional Phenomena in High-fidelity
  Simulations of Core-collapse Supernovae}.
\newblock {\em submitted to \apj, arXiv:1807.07579}, July 2018.

\bibitem{tamborra:2013}
I.~{Tamborra}, F.~{Hanke}, B.~{M{\"u}ller}, H.-T. {Janka}, and G.~{Raffelt}.
\newblock {Neutrino Signature of Supernova Hydrodynamical Instabilities in
  Three Dimensions}.
\newblock {\em Physical Review Letters}, 111(12):121104, September 2013.

\bibitem{couch:2014}
S.~M. {Couch} and E.~P. {O'Connor}.
\newblock {High-resolution Three-dimensional Simulations of Core-collapse
  Supernovae in Multiple Progenitors}.
\newblock {\em \apj}, 785:123, April 2014.

\bibitem{blondin:2003}
J.~M. {Blondin}, A.~{Mezzacappa}, and C.~{DeMarino}.
\newblock {Stability of Standing Accretion Shocks, with an Eye toward
  Core-Collapse Supernovae}.
\newblock {\em \apj}, 584:971--980, February 2003.

\bibitem{murphy:2013}
J.~W. {Murphy}, J.~C. {Dolence}, and A.~{Burrows}.
\newblock {The Dominance of Neutrino-driven Convection in Core-collapse
  Supernovae}.
\newblock {\em \apj}, 771:52, July 2013.

\bibitem{couch:2015a}
S.~M. {Couch} and C.~D. {Ott}.
\newblock {The Role of Turbulence in Neutrino-driven Core-collapse Supernova
  Explosions}.
\newblock {\em \apj}, 799:5, January 2015.

\bibitem{couch:2013b}
S.~M. {Couch} and C.~D. {Ott}.
\newblock {Revival of the Stalled Core-collapse Supernova Shock Triggered by
  Precollapse Asphericity in the Progenitor Star}.
\newblock {\em \apj Letters}, 778:L7, November 2013.

\bibitem{couch:2015}
S.~M. {Couch}, E.~{Chatzopoulos}, W.~D. {Arnett}, and F.~X. {Timmes}.
\newblock {The Three-dimensional Evolution to Core Collapse of a Massive Star}.
\newblock {\em \apj Letters}, 808:L21, July 2015.

\bibitem{muller:2015}
B.~{M{\"u}ller} and H.-T. {Janka}.
\newblock {Non-radial instabilities and progenitor asphericities in
  core-collapse supernovae}.
\newblock {\em \mnras}, 448:2141--2174, April 2015.

\bibitem{muller:2016a}
B.~{M{\"u}ller}, M.~{Viallet}, A.~{Heger}, and H.-T. {Janka}.
\newblock {The Last Minutes of Oxygen Shell Burning in a Massive Star}.
\newblock {\em \apj}, 833:124, December 2016.

\bibitem{muller:2017}
B.~{M{\"u}ller}, T.~{Melson}, A.~{Heger}, and H.-T. {Janka}.
\newblock {Supernova simulations from a 3D progenitor model - Impact of
  perturbations and evolution of explosion properties}.
\newblock {\em \mnras}, 472:491--513, November 2017.

\bibitem{tamborra:2014}
I.~{Tamborra}, F.~{Hanke}, H.-T. {Janka}, B.~{M{\"u}ller}, G.~G. {Raffelt}, and
  A.~{Marek}.
\newblock {Self-sustained Asymmetry of Lepton-number Emission: A New Phenomenon
  during the Supernova Shock-accretion Phase in Three Dimensions}.
\newblock {\em \apj}, 792:96, September 2014.

\bibitem{bruenn:2016}
S.~W. {Bruenn}, E.~J. {Lentz}, W.~R. {Hix}, A.~{Mezzacappa}, J.~A. {Harris},
  O.~E.~B. {Messer}, E.~{Endeve}, J.~M. {Blondin}, M.~A. {Chertkow}, E.~J.
  {Lingerfelt}, P.~{Marronetti}, and K.~N. {Yakunin}.
\newblock {The Development of Explosions in Axisymmetric Ab Initio
  Core-collapse Supernova Simulations of 12-25 M Stars}.
\newblock {\em \apj}, 818:123, February 2016.

\bibitem{summa:2016}
A.~{Summa}, F.~{Hanke}, H.-T. {Janka}, T.~{Melson}, A.~{Marek}, and
  B.~{M{\"u}ller}.
\newblock {Progenitor-dependent Explosion Dynamics in Self-consistent,
  Axisymmetric Simulations of Neutrino-driven Core-collapse Supernovae}.
\newblock {\em \apj}, 825:6, July 2016.

\bibitem{burrows:2016}
A.~{Burrows}, D.~{Vartanyan}, J.~C. {Dolence}, M.~A. {Skinner}, and
  D.~{Radice}.
\newblock {Crucial Physical Dependencies of the Core-Collapse Supernova
  Mechanism}.
\newblock {\em ArXiv e-prints}, November 2016.

\bibitem{oconnor:2018}
E.~P. {O'Connor} and S.~M. {Couch}.
\newblock {Two-dimensional Core-collapse Supernova Explosions Aided by General
  Relativity with Multidimensional Neutrino Transport}.
\newblock {\em \apj}, 854:63, February 2018.

\bibitem{oconnor:2018a}
E.~{O'Connor}, R.~{Bollig}, A.~{Burrows}, S.~{Couch}, T.~{Fischer}, H.-T.
  {Janka}, K.~{Kotake}, E.~J. {Lentz}, M.~{Liebend{\"o}rfer}, O.~E.~B.
  {Messer}, A.~{Mezzacappa}, T.~{Takiwaki}, and D.~{Vartanyan}.
\newblock {Global Comparison of Core-Collapse Supernova Simulations in
  Spherical Symmetry}.
\newblock {\em ArXiv e-prints}, June 2018.

\bibitem{pan:2018a}
K.-C. {Pan}, C.~{Mattes}, E.~P. {O'Connor}, S.~M. {Couch}, A.~{Perego}, and
  A.~{Arcones}.
\newblock {The Impact of Different Neutrino Transport Methods on
  Multidimensional Core-collapse Supernova Simulations}.
\newblock {\em ArXiv e-prints}, June 2018.

\bibitem{just:2018}
O.~{Just}, R.~{Bollig}, H.-T. {Janka}, M.~{Obergaulinger}, R.~{Glas}, and
  S.~{Nagataki}.
\newblock {Core-collapse supernova simulations in one and two dimensions:
  comparison of codes and approximations}.
\newblock {\em ArXiv e-prints}, May 2018.

\bibitem{cabezon:2018}
R.~M. {Cabez{\'o}n}, K.-C. {Pan}, M.~{Liebend{\"o}rfer}, T.~{Kuroda},
  K.~{Ebinger}, O.~{Heinimann}, F.-K. {Thielemann}, and A.~{Perego}.
\newblock {Core-collapse supernovae in the hall of mirrors. A three-dimensional
  code-comparison project}.
\newblock {\em ArXiv e-prints}, June 2018.

\bibitem{Reb97}
R.~Surman, J.~Engel, {\relax J.R}~Bennett, and {\relax B.S}~Meyer.
\newblock {\em Phys. Rev. Lett.}, 79:1809, 1997.

\bibitem{Matt12}
{\relax M.R}~Mumpower, {\relax G.C}~McLaughlin, and R.~Surman.
\newblock {\em Phys. Rev. C}, 85:045801, 2012.

\bibitem{OrfordVassh2018}
R.~{Orford}, N.~{Vassh}, J.~A. {Clark}, G.~C. {McLaughlin}, M.~R. {Mumpower},
  G.~{Savard}, R.~{Surman}, A.~{Aprahamian}, F.~{Buchinger}, M.~T. {Burkey},
  D.~A. {Gorelov}, T.~Y. {Hirsh}, J.~W. {Klimes}, G.~E. {Morgan}, A.~{Nystrom},
  and K.~S. {Sharma}.
\newblock {Precision Mass Measurements of Neutron-Rich Neodymium and Samarium
  Isotopes and Their Role in Understanding Rare-Earth Peak Formation}.
\newblock {\em Physical Review Letters}, 120(26):262702, June 2018.

\bibitem{REMM1}
{\relax M.R}~Mumpower, {\relax G.C}~McLaughlin, R.~Surman, and {\relax
  A.W}~Steiner.
\newblock {\em Astrophys. J}, 833:282, 2016.

\bibitem{REMM2}
{\relax M.R}~Mumpower, {\relax G.C}~McLaughlin, R.~Surman, and {\relax
  A.W}~Steiner.
\newblock {\em J. Phys. G. Nucl. Partic.}, 44:034003, 2017.

\bibitem{solardata}
H.~Palme and H.~Beer.
\newblock {\em in Landolt B{\"o}rnstein, New Series, Group VI}, volume~3.
\newblock Springer, Berlin, 1993.

\bibitem{goriely99}
S.~Goriely.
\newblock Uncertainties in the solar system r-abundance distribution.
\newblock {\em Astron. Astrophys.}, 342:881--891, 1999.

\bibitem{sekiguchi:2012}
Yuichiro {Sekiguchi}, Kenta {Kiuchi}, Koutarou {Kyutoku}, and Masaru {Shibata}.
\newblock {Current status of numerical-relativity simulations in Kyoto}.
\newblock {\em Progress of Theoretical and Experimental Physics}, 2012:01A304,
  October 2012.

\bibitem{rprocPBH}
G.~M. {Fuller}, A.~{Kusenko}, and V.~{Takhistov}.
\newblock {Primordial Black Holes and r -Process Nucleosynthesis}.
\newblock {\em Physical Review Letters}, 119(6):061101, August 2017.

\bibitem{Hansen2018}
T.~T. {Hansen}, E.~M. {Holmbeck}, T.~C. {Beers}, V.~M. {Placco}, I.~U.
  {Roederer}, A.~{Frebel}, C.~M. {Sakari}, J.~D. {Simon}, and I.~B. {Thompson}.
\newblock {The R-process Alliance: First Release from the Southern Search for
  R-process-enhanced Stars in the Galactic Halo}.
\newblock {\em The Astrophysical Journal}, 858:92, May 2018.

\bibitem{Mashonkina2011}
L.~{Mashonkina}, T.~{Gehren}, J.-R. {Shi}, A.~J. {Korn}, and F.~{Grupp}.
\newblock {A non-LTE study of neutral and singly-ionized iron line spectra in
  1D models of the Sun and selected late-type stars}.
\newblock {\em Astronomy and Astrophysics}, 528:A87, April 2011.

\bibitem{Ezzeddine2017}
R.~{Ezzeddine}, A.~{Frebel}, and B.~{Plez}.
\newblock {Ultra-metal-poor Stars: Spectroscopic Determination of Stellar
  Atmospheric Parameters Using Iron Non-LTE Line Abundances}.
\newblock {\em The Astrophysical Journal}, 847:142, October 2017.

\bibitem{Gustafsson1975}
B.~{Gustafsson}, R.~A. {Bell}, K.~{Eriksson}, and A.~{Nordlund}.
\newblock {A grid of model atmospheres for metal-deficient giant stars. I}.
\newblock {\em Astronomy and Astrophysics}, 42:407--432, September 1975.

\bibitem{Gustafsson2008}
B.~{Gustafsson}, B.~{Edvardsson}, K.~{Eriksson}, U.~G. {J{\o}rgensen},
  {\AA}.~{Nordlund}, and B.~{Plez}.
\newblock {A grid of MARCS model atmospheres for late-type stars. I. Methods
  and general properties}.
\newblock {\em Astronomy and Astrophysics}, 486:951--970, August 2008.

\bibitem{hansen18}
T.~T. {Hansen}, E.~M. {Holmbeck}, T.~C. {Beers}, V.~M. {Placco}, I.~U.
  {Roederer}, A.~{Frebel}, C.~M. {Sakari}, J.~D. {Simon}, and I.~B. {Thompson}.
\newblock {The $R$-Process Alliance: First Release from the Southern Search for
  $r$-Process-Enhanced Stars in the Galactic Halo}.
\newblock {\em ArXiv e-prints}, April 2018.

\bibitem{sakari18}
C.~M. {Sakari}, V.~M. {Placco}, T.~{Hansen}, E.~M. {Holmbeck}, T.~C. {Beers},
  A.~{Frebel}, I.~U. {Roederer}, K.~A. {Venn}, G.~{Wallerstein}, C.~E. {Davis},
  E.~M. {Farrell}, and D.~{Yong}.
\newblock {The r-process Pattern of a Bright, Highly r-process-enhanced
  Metal-poor Halo Star at [Fe/H]=-2}.
\newblock {\em ApJL}, 854:L20, February 2018.

\bibitem{placco17}
V.~M. {Placco}, E.~M. {Holmbeck}, A.~{Frebel}, T.~C. {Beers}, R.~A. {Surman},
  A.~P. {Ji}, R.~{Ezzeddine}, S.~D. {Points}, C.~C. {Kaleida}, T.~T. {Hansen},
  C.~M. {Sakari}, and A.~R. {Casey}.
\newblock {RAVE J203843.2-002333: The First Highly R-process-enhanced Star
  Identified in the RAVE Survey}.
\newblock {\em ApJ}, 844:18, July 2017.

\bibitem{cain18}
M.~{Cain}, A.~{Frebel}, M.~{Gull}, A.~P. {Ji}, V.~M. {Placco}, T.~C. {Beers},
  J.~{Melendez}, R.~{Ezzeddine}, A.~R. {Casey}, T.~T. {Hansen}, I.~U.
  {Roederer}, and C.~{Sakari}.
\newblock {The R-Process Alliance: Chemical Abundances for a Trio of
  R-Process-Enhanced Stars -- One Strong, One Moderate, One Mild}.
\newblock {\em ArXiv e-prints}, July 2018.

\bibitem{gull18}
M.~{Gull}, A.~{Frebel}, M.~G. {Cain}, V.~M. {Placco}, A.~P. {Ji}, C.~{Abate},
  R.~{Ezzeddine}, A.~I. {Karakas}, T.~T. {Hansen}, C.~{Sakari}, E.~M.
  {Holmbeck}, R.~M. {Santucci}, A.~R. {Casey}, and T.~C. {Beers}.
\newblock {The R-Process Alliance: Discovery of the first metal-poor star with
  a combined r- and s-process element signature}.
\newblock {\em ArXiv e-prints}, June 2018.

\bibitem{Ji2016}
A.~P. {Ji}, A.~{Frebel}, A.~{Chiti}, and J.~D. {Simon}.
\newblock {R-process enrichment from a single event in an ancient dwarf
  galaxy}.
\newblock {\em \nat}, 531:610--613, March 2016.

\bibitem{roederer16ret}
I.~U. {Roederer}, M.~{Mateo}, J.~I. {Bailey}, III, Y.~{Song}, E.~F. {Bell},
  J.~D. {Crane}, S.~{Loebman}, D.~L. {Nidever}, E.~W. {Olszewski}, S.~A.
  {Shectman}, I.~B. {Thompson}, M.~{Valluri}, and M.~G. {Walker}.
\newblock {Detailed Chemical Abundances in the r-process-rich Ultra-faint Dwarf
  Galaxy Reticulum 2}.
\newblock {\em Astronomical Journal}, 151:82, March 2016.

\bibitem{kasen17}
D.~{Kasen}, B.~{Metzger}, J.~{Barnes}, E.~{Quataert}, and E.~{Ramirez-Ruiz}.
\newblock {Origin of the heavy elements in binary neutron-star mergers from a
  gravitational-wave event}.
\newblock {\em Nature}, 551:80--84, November 2017.

\bibitem{roederer18kinematics}
I.~U. {Roederer}, K.~{Hattori}, and M.~{Valluri}.
\newblock {\em AAS Journals, submitted}, 2018.

\bibitem{beniamini16b}
P.~{Beniamini}, K.~{Hotokezaka}, and T.~{Piran}.
\newblock {r-process Production Sites as Inferred from Eu Abundances in Dwarf
  Galaxies}.
\newblock {\em Astrophysical Journal}, 832:149, December 2016.

\bibitem{Sneden09}
C.~{Sneden}, J.~E. {Lawler}, J.~J. {Cowan}, I.~I. {Ivans}, and E.~A. {Den
  Hartog}.
\newblock {New Rare Earth Element Abundance Distributions for the Sun and Five
  r-Process-Rich Very Metal-Poor Stars}.
\newblock {\em The Astrophysical Journal Supplement}, 182:80--96, May 2009.

\bibitem{Lawler09}
J.~E. {Lawler}, C.~{Sneden}, J.~J. {Cowan}, I.~I. {Ivans}, and E.~A. {den
  Hartog}.
\newblock {VizieR Online Data Catalog: Transition probabilities of rare earth
  elements (Lawler+, 2009)}.
\newblock {\em VizieR Online Data Catalog}, 218, November 2009.

\bibitem{Waxman.ea.2017.knEmission}
E.~{Waxman}, E.~{Ofek}, D.~{Kushnir}, and A.~{Gal-Yam}.
\newblock {Constraints on the ejecta of the GW170817 neutron-star merger from
  its electromagnetic emission}.
\newblock {\em ArXiv e-prints}, November 2017.

\bibitem{KasenBarnes.2018.btherm}
D.~{Kasen} and J.~{Barnes}.
\newblock {Radioactive Heating and Late Time Kilonova Light Curves}.
\newblock {\em ArXiv e-prints}, July 2018.

\bibitem{Hotokezaka.ea.2016.gtherm}
K.~{Hotokezaka}, S.~{Wanajo}, M.~{Tanaka}, A.~{Bamba}, Y.~{Terada}, and
  T.~{Piran}.
\newblock {Radioactive decay products in neutron star merger ejecta: heating
  efficiency and {$\gamma$}-ray emission}.
\newblock {\em \mnras}, 459:35--43, June 2016.

\bibitem{baade1956supernovae}
Walter Baade, GR~Burbidge, F~Hoyle, EM~Burbidge, RF~Christy, and WA~Fowler.
\newblock Supernovae and californium 254.
\newblock {\em Publications of the Astronomical Society of the Pacific},
  68(403):296--300, 1956.

\bibitem{fields1956transplutonium}
PR~Fields, MH~Studier, H~Diamond, JF~Mech, MG~Inghram, GL~Pyle, CM~Stevens,
  S~Fried, WM~Manning, A~Ghiorso, et~al.
\newblock Transplutonium elements in thermonuclear test debris.
\newblock {\em Physical Review}, 102:180, 1956.

\bibitem{phillips1963spontaneous}
L~Phillips, RC~Gatti, R~Brandt, and SG~Thompson.
\newblock Spontaneous fission half lives of 254cf, 255fm, and 250cf.
\newblock {\em Journal of Inorganic and Nuclear Chemistry}, 25(9):1085--1087,
  1963.

\bibitem{beers2005}
T.~C. {Beers} and N.~{Christlieb}.
\newblock {The Discovery and Analysis of Very Metal-Poor Stars in the Galaxy}.
\newblock {\em \araa}, 43:531--580, September 2005.

\bibitem{yoon2018}
J.~{Yoon}, T.~C. {Beers}, S.~{Dietz}, Y.~S. {Lee}, V.~M. {Placco}, G.~{Da
  Costa}, S.~{Keller}, C.~I. {Owen}, and M.~{Sharma}.
\newblock {Galactic Archeology with the AEGIS Survey: The Evolution of Carbon
  and Iron in the Galactic Halo}.
\newblock {\em ArXiv e-prints}, June 2018.

\bibitem{herwig2005}
F.~{Herwig}.
\newblock {Evolution of Asymptotic Giant Branch Stars}.
\newblock {\em \araa}, 43:435--479, September 2005.

\bibitem{bisterzo2011}
S.~{Bisterzo}, R.~{Gallino}, O.~{Straniero}, S.~{Cristallo}, and
  F.~{K{\"a}ppeler}.
\newblock {The s-process in low-metallicity stars - II. Interpretation of
  high-resolution spectroscopic observations with asymptotic giant branch
  models}.
\newblock {\em \mnras}, 418:284--319, November 2011.

\bibitem{abate2013}
C.~{Abate}, O.~R. {Pols}, R.~G. {Izzard}, S.~S. {Mohamed}, and S.~E. {de Mink}.
\newblock {Wind Roche-lobe overflow: Application to carbon-enhanced metal-poor
  stars}.
\newblock {\em \aap}, 552:A26, April 2013.

\bibitem{hampel2016}
M.~{Hampel}, R.~J. {Stancliffe}, M.~{Lugaro}, and B.~S. {Meyer}.
\newblock {The Intermediate Neutron-capture Process and Carbon-enhanced
  Metal-poor Stars}.
\newblock {\em \apj}, 831:171, November 2016.

\bibitem{meynet2010}
G.~{Meynet}, R.~{Hirschi}, S.~{Ekstrom}, A.~{Maeder}, C.~{Georgy},
  P.~{Eggenberger}, and C.~{Chiappini}.
\newblock {Are C-rich ultra iron-poor stars also He-rich?}
\newblock {\em \aap}, 521:A30, October 2010.

\bibitem{nomoto2013}
K.~{Nomoto}, C.~{Kobayashi}, and N.~{Tominaga}.
\newblock {Nucleosynthesis in Stars and the Chemical Enrichment of Galaxies}.
\newblock {\em \araa}, 51:457--509, August 2013.

\bibitem{yoon2016}
J.~{Yoon}, T.~C. {Beers}, V.~M. {Placco}, K.~C. {Rasmussen}, D.~{Carollo},
  S.~{He}, T.~T. {Hansen}, I.~U. {Roederer}, and J.~{Zeanah}.
\newblock {Observational Constraints on First-star Nucleosynthesis. I. Evidence
  for Multiple Progenitors of CEMP-No Stars}.
\newblock {\em \apj}, 833:20, December 2016.

\bibitem{wilson1927}
Edwin~B. Wilson.
\newblock Probable inference, the law of succession, and statistical inference.
\newblock {\em Journal of the American Statistical Association},
  22(158):209--212, 1927.

\bibitem{placco2014c}
V.~M. {Placco}, A.~{Frebel}, T.~C. {Beers}, and R.~J. {Stancliffe}.
\newblock {Carbon-enhanced Metal-poor Star Frequencies in the Galaxy:
  Corrections for the Effect of Evolutionary Status on Carbon Abundances}.
\newblock {\em \apj}, 797:21, December 2014.

\bibitem{Fattoyev18}
F.~J. {Fattoyev}, J.~{Piekarewicz}, and C.~J. {Horowitz}.
\newblock {Neutron Skins and Neutron Stars in the Multimessenger Era}.
\newblock {\em Physical Review Letters}, 120(17):172702, April 2018.

\bibitem{Suh17}
I.-S. {Suh}, G.~J. {Mathews}, J.~R. {Haywood}, and N.~Q. {Lan}.
\newblock {Analysis of the Conformally Flat Approximation for Binary Neutron
  Star Initial Conditions}.
\newblock {\em Advances in Astronomy}, 2017:612703, 2017.

\bibitem{Lattimer12}
J.~M. {Lattimer}.
\newblock {The Nuclear Equation of State and Neutron Star Masses}.
\newblock {\em Annual Review of Nuclear and Particle Science}, 62:485--515,
  November 2012.

\bibitem{Smartt17}
S.~J. {Smartt}, T.-W. {Chen}, A.~{Jerkstrand}, M.~{Coughlin}, E.~{Kankare},
  S.~A. {Sim}, M.~{Fraser}, C.~{Inserra}, K.~{Maguire}, K.~C. {Chambers}, M.~E.
  {Huber}, T.~{Kr{\"u}hler}, G.~{Leloudas}, M.~{Magee}, L.~J. {Shingles}, K.~W.
  {Smith}, D.~R. {Young}, J.~{Tonry}, R.~{Kotak}, A.~{Gal-Yam}, J.~D. {Lyman},
  D.~S. {Homan}, C.~{Agliozzo}, J.~P. {Anderson}, C.~R. {Angus}, C.~{Ashall},
  C.~{Barbarino}, F.~E. {Bauer}, M.~{Berton}, M.~T. {Botticella}, M.~{Bulla},
  J.~{Bulger}, G.~{Cannizzaro}, Z.~{Cano}, R.~{Cartier}, A.~{Cikota},
  P.~{Clark}, A.~{De Cia}, M.~{Della Valle}, L.~{Denneau}, M.~{Dennefeld},
  L.~{Dessart}, G.~{Dimitriadis}, N.~{Elias-Rosa}, R.~E. {Firth},
  H.~{Flewelling}, A.~{Fl{\"o}rs}, A.~{Franckowiak}, C.~{Frohmaier},
  L.~{Galbany}, S.~{Gonz{\'a}lez-Gait{\'a}n}, J.~{Greiner}, M.~{Gromadzki},
  A.~N. {Guelbenzu}, C.~P. {Guti{\'e}rrez}, A.~{Hamanowicz}, L.~{Hanlon},
  J.~{Harmanen}, K.~E. {Heintz}, A.~{Heinze}, M.-S. {Hernandez}, S.~T.
  {Hodgkin}, I.~M. {Hook}, L.~{Izzo}, P.~A. {James}, P.~G. {Jonker}, W.~E.
  {Kerzendorf}, S.~{Klose}, Z.~{Kostrzewa-Rutkowska}, M.~{Kowalski},
  M.~{Kromer}, H.~{Kuncarayakti}, A.~{Lawrence}, T.~B. {Lowe}, E.~A. {Magnier},
  I.~{Manulis}, A.~{Martin-Carrillo}, S.~{Mattila}, O.~{McBrien},
  A.~{M{\"u}ller}, J.~{Nordin}, D.~{O'Neill}, F.~{Onori}, J.~T. {Palmerio},
  A.~{Pastorello}, F.~{Patat}, G.~{Pignata}, P.~{Podsiadlowski}, M.~L. {Pumo},
  S.~J. {Prentice}, A.~{Rau}, A.~{Razza}, A.~{Rest}, T.~{Reynolds}, R.~{Roy},
  A.~J. {Ruiter}, K.~A. {Rybicki}, L.~{Salmon}, P.~{Schady}, A.~S.~B.
  {Schultz}, T.~{Schweyer}, I.~R. {Seitenzahl}, M.~{Smith}, J.~{Sollerman},
  B.~{Stalder}, C.~W. {Stubbs}, M.~{Sullivan}, H.~{Szegedi}, F.~{Taddia},
  S.~{Taubenberger}, G.~{Terreran}, B.~{van Soelen}, J.~{Vos}, R.~J.
  {Wainscoat}, N.~A. {Walton}, C.~{Waters}, H.~{Weiland}, M.~{Willman},
  P.~{Wiseman}, D.~E. {Wright}, {\L}.~{Wyrzykowski}, and O.~{Yaron}.
\newblock {A kilonova as the electromagnetic counterpart to a
  gravitational-wave source}.
\newblock {\em \nat}, 551:75--79, November 2017.

\bibitem{Shibagaki16}
S.~{Shibagaki}, T.~{Kajino}, G.~J. {Mathews}, S.~{Chiba}, S.~{Nishimura}, and
  G.~{Lorusso}.
\newblock {Relative Contributions of the Weak, Main, and Fission-recycling
  r-process}.
\newblock {\em The Astrophysical Journal}, 816:79, January 2016.

\bibitem{Flanagan:2007ix}
Eanna~E. Flanagan and Tanja Hinderer.
\newblock {Constraining neutron star tidal Love numbers with gravitational wave
  detectors}.
\newblock {\em Phys. Rev. D}, 77:021502, 2008.

\bibitem{Hinderer:2007mb}
Tanja Hinderer.
\newblock {Tidal Love numbers of neutron stars}.
\newblock {\em Astrophys. J.}, 677:1216--1220, 2008.

\bibitem{LIGO:2017qsa}
Benjamin~P. Abbott et~al.
\newblock {GW170817: Observation of Gravitational Waves from a Binary Neutron
  Star Inspiral}.
\newblock {\em Phys. Rev. Lett.}, 119(16):161101, 2017.

\bibitem{Yagi:2016bkt}
Kent Yagi and Nicolás Yunes.
\newblock {Approximate Universal Relations for Neutron Stars and Quark Stars}.
\newblock {\em Phys. Rept.}, 681:1--72, 2017.

\bibitem{Chatziioannou:2018vzf}
Katerina Chatziioannou, Carl-Johan Haster, and Aaron Zimmerman.
\newblock {Measuring the neutron star tidal deformability with
  equation-of-state-independent relations and gravitational waves}.
\newblock {\em Phys. Rev. D}, 97(10):104036, 2018.

\bibitem{LIGO:2018exr}
B.~P. Abbott et~al.
\newblock {GW170817: Measurements of neutron star radii and equation of state}.
\newblock 2018.

\bibitem{LIGO:2018wiz}
B.~P. Abbott et~al.
\newblock {Properties of the binary neutron star merger GW170817}.
\newblock 2018.

\bibitem{Hinderer:2009ca}
Tanja Hinderer, Benjamin~D. Lackey, Ryan~N. Lang, and Jocelyn~S. Read.
\newblock {Tidal deformability of neutron stars with realistic equations of
  state and their gravitational wave signatures in binary inspiral}.
\newblock {\em Phys. Rev. D}, 81:123016, 2010.

\bibitem{Most:2018eaw}
Elias~R. Most, L.~Jens Papenfort, Veronica Dexheimer, Matthias Hanauske, Stefan
  Schramm, Horst Stöcker, and Luciano Rezzolla.
\newblock {Signatures of quark-hadron phase transitions in general-relativistic
  neutron-star mergers}.
\newblock 2018.

\bibitem{Pehlivan:2011hp}
Y.~Pehlivan, A.~B. Balantekin, Toshitaka Kajino, and Takashi Yoshida.
\newblock {Invariants of Collective Neutrino Oscillations}.
\newblock {\em Phys. Rev.}, D84:065008, 2011.

\bibitem{Birol:2018qhx}
Savas Birol, Y.~Pehlivan, A.~B. Balantekin, and T.~Kajino.
\newblock {Neutrino Spectral Split in the Exact Many Body Formalism}.
\newblock 2018.

\bibitem{Athanassopoulos2004}
S.~Athanassopoulos, E.~Mavrommatis, K.A. Gernoth, and J.W. Clark.
\newblock Nuclear mass systematics using neural networks.
\newblock {\em Nucl. Phys. A}, 743:222 -- 235, 2004.

\bibitem{Bayram2017}
Tuncay Bayram and Serkan Akkoyun.
\newblock An approach to adjustment of relativistic mean field model
  parameters.
\newblock {\em EPJ Web Conf.}, 146:12033, 2017.

\bibitem{Yuan2016}
Cenxi Yuan.
\newblock Uncertainty decomposition method and its application to the liquid
  drop model.
\newblock {\em Phys. Rev. C}, 93:034310, Mar 2016.

\bibitem{Utama16}
R.~Utama, J.~Piekarewicz, and H.~B. Prosper.
\newblock Nuclear mass predictions for the crustal composition of neutron
  stars: A bayesian neural network approach.
\newblock {\em Phys. Rev. C}, 93:014311, Jan 2016.

\bibitem{Utama17}
R.~Utama and J.~Piekarewicz.
\newblock Refining mass formulas for astrophysical applications: A bayesian
  neural network approach.
\newblock {\em Phys. Rev. C}, 96:044308, Oct 2017.

\bibitem{Utama18}
R.~Utama and J.~Piekarewicz.
\newblock Validating neural-network refinements of nuclear mass models.
\newblock {\em Phys. Rev. C}, 97:014306, Jan 2018.

\bibitem{Niu2018}
Z.M. Niu and H.Z. Liang.
\newblock Nuclear mass predictions based on bayesian neural network approach
  with pairing and shell effects.
\newblock {\em Phys. Lett. B}, 778:48 -- 53, 2018.

\bibitem{Neufcourt2018}
L.~Neufcourt, Y.~Cao, W.~Nazarewicz, and F.~Viens.
\newblock Bayesian approach to model-based extrapolation of nuclear
  observables.
\newblock {\em arXiv:1806.00552}, 2018.

\bibitem{bolsterli1972}
M.~Bolsterli, E.~O. Fiset, J.~R. Nix, and J.~L. Norton.
\newblock {New} {Calculation} of {Fission} {Barriers} for {Heavy} and
  {Superheavy} {Nuclei}.
\newblock {\em Phys. Rev. C}, 5(3):1050, 1972.

\bibitem{brack1972}
Matthias Brack, Jens Damgaard, A.~S. Jensen, H.~C. Pauli, V.~M. Strutinsky, and
  C.~Y. Wong.
\newblock {Funny} hills: {The} shell-correction approach to nuclear shell
  effects and its applications to the fission process.
\newblock {\em Rev. Mod. Phys.}, 44(2):320, 1972.

\bibitem{bjornholm1980}
S.~Bj{\o}rnholm and J.~E. Lynn.
\newblock {The} double-humped fission barrier.
\newblock {\em Rev. Mod. Phys.}, 52(4):725, 1980.

\bibitem{randrup2011}
J.~Randrup, P.~Möller, and A.~J. Sierk.
\newblock {Fission-fragment} mass distributions from strongly damped shape
  evolution.
\newblock {\em Phys. Rev. C}, 84(3):034613, 2011.

\bibitem{randrup2011a}
J{\o}rgen Randrup and Peter M{\o}ller.
\newblock {Brownian} {Shape} {Motion} on {Five-Dimensional} {Potential-Energy}
  {Surfaces:} {Nuclear} {Fission-Fragment} {Mass} {Distributions}.
\newblock {\em Phys. Rev. Lett.}, 106(13):132503, 2011.

\bibitem{randrup2013}
J.~Randrup and P.~Möller.
\newblock {Energy} dependence of fission-fragment mass distributions from
  strongly damped shape evolution.
\newblock {\em Phys. Rev. C}, 88(6):064606, 2013.

\bibitem{ishizuka2017}
Chikako Ishizuka, Mark~D. Usang, Fedir~A. Ivanyuk, Joachim~A. Maruhn, Katsuhisa
  Nishio, and Satoshi Chiba.
\newblock {Four-dimensional} {Langevin} approach to low-energy nuclear fission
  of {$^{236}\mathbf{U}$}.
\newblock {\em Phys. Rev. C}, 96(6):064616, 2017.

\bibitem{usang2017}
M.~D. Usang, F.~A. Ivanyuk, C.~Ishizuka, and S.~Chiba.
\newblock {Analysis} of the total kinetic energy of fission fragments with the
  {Langevin} equation.
\newblock {\em Phys. Rev. C}, 96(6):064617, 2017.

\bibitem{ward2017}
D.~E. Ward, B.~G. Carlsson, T.~Døssing, P.~M{\o}ller, J.~Randrup, and
  S.~Aberg.
\newblock {Nuclear} shape evolution based on microscopic level densities.
\newblock {\em Phys. Rev. C}, 95(2):024618, 2017.

\bibitem{schunck2016}
N~Schunck and L~M Robledo.
\newblock {Microscopic} theory of nuclear fission: a review.
\newblock {\em Rep. Prog. Phys.}, 79(11):116301, 2016.

\bibitem{regnier2016}
D.~Regnier, N.~Dubray, N.~Schunck, and M.~Verrière.
\newblock {Fission} fragment charge and mass distributions in {$^{239}$Pu(n},f)
  in the adiabatic nuclear energy density functional theory.
\newblock {\em Phys. Rev. C}, 93(5):054611, 2016.

\bibitem{zdeb2017}
A.~Zdeb, A.~Dobrowolski, and M.~Warda.
\newblock {Fission} dynamics of {Cf}252.
\newblock {\em Phys. Rev. C}, 95(5):054608, 2017.

\bibitem{sadhukhan2016}
Jhilam Sadhukhan, Witold Nazarewicz, and Nicolas Schunck.
\newblock {Microscopic} modeling of mass and charge distributions in the
  spontaneous fission of {$^{240}\mathrm{Pu}$}.
\newblock {\em Phys. Rev. C}, 93(1):011304, 2016.

\bibitem{sadhukhan2017}
Jhilam Sadhukhan, Chunli Zhang, Witold Nazarewicz, and Nicolas Schunck.
\newblock {Formation} and distribution of fragments in the spontaneous fission
  of {${}^{\mathbf{240}}\mathbf{Pu}$}.
\newblock {\em Phys. Rev. C}, 96(6):061301, 2017.

\bibitem{staszczak2013}
A.~Staszczak, A.~Baran, and W.~Nazarewicz.
\newblock {Spontaneous} fission modes and lifetimes of superheavy elements in
  the nuclear density functional theory.
\newblock {\em Phys. Rev. C}, 87(2):024320, 2013.

\bibitem{baran2015}
A.~Baran, M.~Kowal, P.~G. Reinhard, L.~M. Robledo, A.~Staszczak, and M.~Warda.
\newblock {Fission} barriers and probabilities of spontaneous fission for
  elements with {Z}$\geq$100.
\newblock {\em Nucl. Phys. A}, 944:442, 2015.

\bibitem{goddard2015}
Philip Goddard, Paul Stevenson, and Arnau Rios.
\newblock {Fission} dynamics within time-dependent {Hartree-Fock:}
  {Deformation-induced} fission.
\newblock {\em Phys. Rev. C}, 92(5):054610, 2015.

\bibitem{scamps2015}
Guillaume Scamps, Cédric Simenel, and Denis Lacroix.
\newblock {Superfluid} dynamics of {$^{258}$Fm} fission.
\newblock {\em Phys. Rev. C}, 92(1):011602, 2015.

\bibitem{goddard2016}
Philip Goddard, Paul Stevenson, and Arnau Rios.
\newblock {Fission} dynamics within time-dependent {Hartree-Fock.} {II.}
  {Boost-induced} fission.
\newblock {\em Phys. Rev. C}, 93(1):014620, 2016.

\bibitem{bulgac2016}
Aurel Bulgac, Piotr Magierski, Kenneth~J. Roche, and Ionel Stetcu.
\newblock {Induced} {Fission} of {$^{240}\mathrm{Pu}$} within a {Real-Time}
  {Microscopic} {Framework}.
\newblock {\em Phys. Rev. Lett.}, 116(12):122504, 2016.

\bibitem{tanimura2017}
Yusuke Tanimura, Denis Lacroix, and Sakir Ayik.
\newblock {Microscopic} {Phase-Space} {Exploration} {Modeling} of {Fm}258
  {Spontaneous} {Fission}.
\newblock {\em Phys. Rev. Lett.}, 118(15):152501, 2017.

\bibitem{rodriguez-guzman2014}
R.~Rodr{\'i}guez-Guzmán and L.~M. Robledo.
\newblock {Microscopic} description of fission in {Uranium} isotopes with the
  {Gogny} energy density functional.
\newblock {\em Phys. Rev. C}, 89(5):054310, 2014.

\bibitem{rodriguez-guzman2014a}
R.~Rodr{\i}guez-Guzmán and L.~M. Robledo.
\newblock {Microscopic} description of fission in neutron-rich {Plutonium}
  isotopes with the {Gogny-D1M} energy density functional.
\newblock {\em Eur. Phys. J. A}, 50(9):142, 2014.

\bibitem{giuliani2018}
Samuel~A. Giuliani, Gabriel Mart\'{\i}nez-Pinedo, and Luis~M. Robledo.
\newblock Fission properties of superheavy nuclei for $r$-process calculations.
\newblock {\em Phys. Rev. C}, 97:034323, 2018.

\bibitem{Mumpower2018a}
M.~R. {Mumpower}, T.~{Kawano}, T.~M. {Sprouse}, N.~{Vassh}, E.~M. {Holmbeck},
  R.~{Surman}, and P.~{Moller}.
\newblock {$\beta$-delayed fission in $r$-process nucleosynthesis}.
\newblock {\em ArXiv e-prints}, February 2018.

\bibitem{Panov2010}
I.~V. Panov, I.~Yu. Korneev, T.~Rauscher, G.~Mart{\'{i}}nez-Pinedo,
  A.~Keli{\'{c}}-Heil, N.~T. Zinner, and F.-K. Thielemann.
\newblock {Neutron-induced astrophysical reaction rates for translead nuclei}.
\newblock {\em Astron. Astrophys.}, 513:14, apr 2010.

\bibitem{Giuliani2017}
S.~A. Giuliani, G.~Mart{\'{i}}nez-Pinedo, L.~M. Robledo, and M.-R. Wu.
\newblock {R-Process calculations with a microscopic description of the fission
  process}.
\newblock {\em Acta Phys. Pol. B}, 48(3), 2017.

\bibitem{RingSchuck}
P.~Ring and Schuck P.
\newblock {\em The nuclear many-body problem}.
\newblock Springer-Verlag, New-York, 1980.

\bibitem{Litvinova2007}
E.~Litvinova, P.~Ring, and V.~Tselyaev.
\newblock {\em Phys. Rev. C}, 75:064308, 2007.

\bibitem{Litvinova2008}
E.~Litvinova, P.~Ring, and V.~Tselyaev.
\newblock {\em Phys. Rev. C}, 78:014312, 2008.

\bibitem{Marketin2012}
T.~Marketin, E.~Litvinova, D.~Vretenar, and P.~Ring.
\newblock {\em Phys. Lett. B}, 706:477, 2012.

\bibitem{Robin2016}
C.~Robin and E.~Litvinova.
\newblock {\em The Eur. Phys. J. A}, 52:205, 2016.

\bibitem{Robin2018}
C.~Robin and E.~Litvinova.
\newblock 2018.
\newblock arXiv:1806.01409 [nucl-th].

\bibitem{Litvinova2009}
E.~Litvinova, H.P. Loens, K.~Langanke, G.~Mart\'inez-Pinedo, T.~Rauscher,
  P.~Ring, F.-K. Thielemann, and V.~Tselyaev.
\newblock {\em Nucl. Phys. A}, 823:26, 2009.

\bibitem{RobinINPC16}
Caroline Robin and Elena Litvinova.
\newblock Quasiparticle-vibration coupling effects on nuclear transitions of
  astrophysical interest.
\newblock {\em AIP Conference Proceedings}, 1912(1):020014, 2017.

\bibitem{RobinNMP17}
C.~Robin and E.~Litvinova.
\newblock {\em AIP Conf. Proc.}, 1912(1):020014, 2017.

\bibitem{LitvinovaWibowo2018}
Elena Litvinova and Herlik Wibowo.
\newblock Finite-temperature relativistic nuclear field theory: An application
  to the dipole response.
\newblock {\em Phys. Rev. Lett.}, 121:082501, Aug 2018.

\bibitem{Mus14}
M.~T. Mustonen, T.~Shafer, Z.~Zenginerler, and J.~Engel.
\newblock {Finite Amplitude Method for Charge-Changing Transitions in
  Axially-Deformed Nuclei}.
\newblock {\em Phys. Rev.}, C90:024308, 2014.

\bibitem{Nak07}
Takashi Nakatsukasa, Tsunenori Inakura, and Kazuhiro Yabana.
\newblock {Finite amplitude method for the RPA solution}.
\newblock {\em Phys. Rev.}, C76:024318, 2007.

\bibitem{Avo11}
Paolo Avogadro and Takashi Nakatsukasa.
\newblock {Finite amplitude method for the quasi-particle-random-phase
  approximation}.
\newblock {\em Phys. Rev.}, C84:014314, 2011.

\bibitem{Mus16}
M.~T. Mustonen and J.~Engel.
\newblock {Global description of $\beta$ decay in even-even nuclei with the
  axially-deformed Skyrme finite-amplitude method}.
\newblock {\em Phys. Rev.}, C93(1):014304, 2016.

\bibitem{Sha16}
T.~Shafer, J.~Engel, C.~Fröhlich, G.~C. McLaughlin, M.~Mumpower, and
  R.~Surman.
\newblock {$\beta$ decay of deformed r -process nuclei near A=80 and A=160 ,
  including odd- A and odd-odd nuclei, with the Skyrme finite-amplitude
  method}.
\newblock {\em Phys. Rev.}, C94(5):055802, 2016.

\bibitem{Moore2013}
I.D. Moore, T.~Eronen, D.~Gorelov, J.~Hakala, A.~Jokinen, A.~Kankainen, V.S.
  Kolhinen, J.~Koponen, H.~Penttil/"a, I.~Pohjalainen, M.~Reponen, J.~Rissanen,
  A.~Saastamoinen, S.~Rinta-Antila, V.~Sonnenschein, and J.~/"Ayst/"o.
\newblock Towards commissioning the new igisol-4 facility.
\newblock {\em Nucl. Instrum. Meth. Phys. Res. B}, 317:208 -- 213, 2013.

\bibitem{Eronen2012}
T.~Eronen, V.~S. Kolhinen, V.~V. Elomaa, D.~Gorelov, U.~Hager, J.~Hakala,
  A.~Jokinen, A.~Kankainen, P.~Karvonen, S.~Kopecky, I.~D. Moore,
  H.~Penttil{\"a}, S.~Rahaman, S.~Rinta-Antila, J.~Rissanen, A.~Saastamoinen,
  J.~Szerypo, C.~Weber, and J.~{\"A}yst{\"o}.
\newblock Jyfltrap: a penning trap for precision mass spectroscopy and isobaric
  purification.
\newblock {\em Eur. Phys. J. A}, 48(4):46, Apr 2012.

\bibitem{Vilen2018}
M.~Vilen, J.~M. Kelly, A.~Kankainen, M.~Brodeur, A.~Aprahamian, L.~Canete,
  T.~Eronen, A.~Jokinen, T.~Kuta, I.~D. Moore, M.~R. Mumpower, D.~A.
  Nesterenko, H.~Penttil\"a, I.~Pohjalainen, W.~S. Porter, S.~Rinta-Antila,
  R.~Surman, A.~Voss, and J.~\"Ayst\"o.
\newblock Precision mass measurements on neutron-rich rare-earth isotopes at
  jyfltrap: Reduced neutron pairing and implications for $r$-process
  calculations.
\newblock {\em Phys. Rev. Lett.}, 120:262701, Jun 2018.

\bibitem{Surman1997}
Rebecca Surman, Jonathan Engel, Jonathan~R. Bennett, and Bradley~S. Meyer.
\newblock Source of the rare-earth element peak in $\mathit{r}$-process
  nucleosynthesis.
\newblock {\em Phys. Rev. Lett.}, 79:1809--1812, Sep 1997.

\bibitem{Mumpower2012}
Matthew~R. Mumpower, G.~C. McLaughlin, and Rebecca Surman.
\newblock Formation of the rare-earth peak: Gaining insight into late-time
  $r$-process dynamics.
\newblock {\em Phys. Rev. C}, 85:045801, Apr 2012.

\bibitem{Eronen2016}
Tommi Eronen, Anu Kankainen, and Juha \"{A}yst\"{o}.
\newblock Ion traps in nuclear physics—recent results and achievements.
\newblock {\em Progr. Part. Nucl. Phys.}, 91:259 -- 293, 2016.

\bibitem{Caballero2018}
R.~{Caballero-Folch}, I.~{Dillmann}, J.~{Agramunt}, J.~L. {Ta{\'{\i}}n},
  A.~{Algora}, J.~{Aysto}, F.~{Calvi{\~n}o}, L.~{Canete}, G.~{Cort{\`e}s},
  C.~{Domingo-Pardo}, T.~{Eronen}, E.~{Ganioglu}, W.~{Gelletly}, D.~{Gorelov},
  V.~{Guadilla}, J.~{Hakala}, A.~{Jokinen}, A.~{Kankainen}, V.~{Kolhinen},
  J.~{Koponen}, M.~{Marta}, E.~{Mendoza}, A.~{Montaner-Piz{\'a}}, I.~{Moore},
  C.~R. {Nobs}, S.~E.~A. {Orrigo}, H.~{Penttila}, I.~{Pohjalainen},
  J.~{Reinikainen}, A.~{Riego}, S.~{Rinta-Antila}, B.~{Rubio},
  P.~{Salvador-Casti{\~n}eira}, V.~{Simutkin}, A.~{Tarife{\~n}o-Saldivia},
  A.~{Tolosa}, and A.~{Voss}.
\newblock {First determination of $\beta$-delayed multiple neutron emission
  beyond A = 100 through direct neutron measurement: The P$\_{2n}$ value of
  $^{136}$Sb}.
\newblock {\em ArXiv e-prints}, March 2018.

\bibitem{Tain2015}
J.~L. Tain, E.~Valencia, A.~Algora, J.~Agramunt, B.~Rubio, S.~Rice,
  W.~Gelletly, P.~Regan, A.-A. Zakari-Issoufou, M.~Fallot, A.~Porta,
  J.~Rissanen, T.~Eronen, J.~\"Ayst\"o, L.~Batist, M.~Bowry, V.~M. Bui,
  R.~Caballero-Folch, D.~Cano-Ott, V.-V. Elomaa, E.~Estevez, G.~F. Farrelly,
  A.~R. Garcia, B.~Gomez-Hornillos, V.~Gorlychev, J.~Hakala, M.~D. Jordan,
  A.~Jokinen, V.~S. Kolhinen, F.~G. Kondev, T.~Mart\'{\i}nez, E.~Mendoza,
  I.~Moore, H.~Penttil\"a, Zs. Podoly\'ak, M.~Reponen, V.~Sonnenschein, and
  A.~A. Sonzogni.
\newblock Enhanced $\ensuremath{\gamma}$-ray emission from neutron unbound
  states populated in $\ensuremath{\beta}$ decay.
\newblock {\em Phys. Rev. Lett.}, 115:062502, Aug 2015.

\bibitem{Martinez2014}
T.~Martinez, D.~Cano-Ott, J.~Castilla, A.R. Garcia, J.~Marin, G.~Martinez,
  E.~Mendoza, C.~Santos, F.J. Tera, D.~Villamarin, J.~Agramunt, A.~Algora,
  C.~Domingo, M.D. Jordan, B.~Rubio, J.L. Taín, C.~Bhattacharya, K.~Banerjee,
  S.~Bhattacharya, P.~Roy, J.K. Meena, S.~Kundu, G.~Mukherjee, T.K. Ghosh, T.K.
  Rana, R.~Pandey, A.~Saxena, B.~Behera, H.~Penttil\"{a}, A.~Jokinen,
  S.~Rinta-Antila, C.~Guerrero, and M.C. Ovejero.
\newblock Monster: a tof spectrometer for beta-delayed neutron spectroscopy.
\newblock {\em Nucl. Data Sheets}, 120:78 -- 80, 2014.

\bibitem{sa08nimb}
G.~{Savard}, S.~{Baker}, C.~{Davids}, A.~F. {Levand}, E.~F. {Moore}, R.~C.
  {Pardo}, R.~{Vondrasek}, B.~J. {Zabransky}, and G.~{Zinkann}.
\newblock {Radioactive beams from gas catchers: The CARIBU facility}.
\newblock {\em Nuclear Instruments and Methods in Physics Research B},
  266:4086--4091, October 2008.

\bibitem{hi16nimb}
T.~Y. {Hirsh}, N.~{Paul}, M.~{Burkey}, A.~{Aprahamian}, F.~{Buchinger},
  S.~{Caldwell}, J.~A. {Clark}, A.~F. {Levand}, L.~L. {Ying}, S.~T. {Marley},
  G.~E. {Morgan}, A.~{Nystrom}, R.~{Orford}, A.~P. {Galv{\'a}n}, J.~{Rohrer},
  G.~{Savard}, K.~S. {Sharma}, and K.~{Siegl}.
\newblock {First operation and mass separation with the CARIBU MR-TOF}.
\newblock {\em Nuclear Instruments and Methods in Physics Research B},
  376:229--232, June 2016.

\bibitem{sc13prl}
J.~{Van Schelt}, D.~{Lascar}, G.~{Savard}, J.~A. {Clark}, P.~F. {Bertone},
  S.~{Caldwell}, A.~{Chaudhuri}, A.~F. {Levand}, G.~{Li}, G.~E. {Morgan},
  R.~{Orford}, R.~E. {Segel}, K.~S. {Sharma}, and M.~G. {Sternberg}.
\newblock {First Results from the CARIBU Facility: Mass Measurements on the
  r-Process Path}.
\newblock {\em Physical Review Letters}, 111(6):061102, August 2013.

\bibitem{el13prl}
S.~{Eliseev}, K.~{Blaum}, M.~{Block}, C.~{Droese}, M.~{Goncharov}, E.~{Minaya
  Ramirez}, D.~A. {Nesterenko}, Y.~N. {Novikov}, and L.~{Schweikhard}.
\newblock {Phase-Imaging Ion-Cyclotron-Resonance Measurements for Short-Lived
  Nuclides}.
\newblock {\em Physical Review Letters}, 110(8):082501, February 2013.

\bibitem{Kurtukian2014}
T.~Kurtukian-Nieto, J.~Benlliure, K.-H. Schmidt, L.~Audouin, F.~Becker,
  B.~Blank, E.~Casarejos, F.~Farget, M.~Fern\'andez-Ord\'o\~nez, J.~Giovinazzo,
  D.~Henzlova, B.~Jurado, J.~Pereira, and O.~Yordanov.
\newblock Production cross sections of heavy neutron-rich nuclei approaching
  the nucleosynthesis r-process path around $a=195$.
\newblock {\em Phys. Rev. C}, 89:024616, Feb 2014.

\bibitem{Watanabe2015}
Y.~X. Watanabe, Y.~H. Kim, S.~C. Jeong, Y.~Hirayama, N.~Imai, H.~Ishiyama,
  H.~S. Jung, H.~Miyatake, S.~Choi, J.~S. Song, E.~Clement, G.~de~France,
  A.~Navin, M.~Rejmund, C.~Schmitt, G.~Pollarolo, L.~Corradi, E.~Fioretto,
  D.~Montanari, M.~Niikura, D.~Suzuki, H.~Nishibata, and J.~Takatsu.
\newblock Pathway for the production of neutron-rich isotopes around the
  $n=126$ shell closure.
\newblock {\em Phys. Rev. Lett.}, 115:172503, Oct 2015.

\bibitem{Schultz2016}
B.E. Schultz, J.M. Kelly, C.~Nicoloff, J.~Long, S.~Ryan, and M.~Brodeur.
\newblock Construction and simulation of a multi-reflection time-of-flight mass
  spectrometer at the university of notre dame.
\newblock {\em Nuclear Instruments and Methods in Physics Research Section B:
  Beam Interactions with Materials and Atoms}, 376:251 -- 255, 2016.
\newblock Proceedings of the XVIIth International Conference on Electromagnetic
  Isotope Separators and Related Topics (EMIS2015), Grand Rapids, MI, U.S.A.,
  11-15 May 2015.

\bibitem{Kozub2012}
R.~L. Kozub, G.~Arbanas, A.~S. Adekola, D.~W. Bardayan, J.~C. Blackmon, K.~Y.
  Chae, K.~A. Chipps, J.~A. Cizewski, L.~Erikson, R.~Hatarik, W.~R. Hix, K.~L.
  Jones, W.~Krolas, J.~F. Liang, Z.~Ma, C.~Matei, B.~H. Moazen, C.~D. Nesaraja,
  S.~D. Pain, D.~Shapira, J.~F. Shriner, M.~S. Smith, and T.~P. Swan.
\newblock Neutron single particle structure in $^{131}\mathrm{Sn}$ and direct
  neutron capture cross sections.
\newblock {\em Phys. Rev. Lett.}, 109:172501, Oct 2012.

\bibitem{Manning2018}
B.~Manning~et al.
\newblock Direct neutron capture on tin isotopes near the ${N}=82$ shell
  closure.
\newblock {\em submitted}, 2018.

\bibitem{Chiba2008}
S.~Chiba, H.~Koura, T.~Hayakawa, T.~Maruyama, T.~Kawano, and T.~Kajino.
\newblock Direct and semi-direct capture in low-energy
  ($n,\ensuremath{\gamma}$) reactions of neutron-rich tin isotopes and its
  implications for $r$-process nucleosynthesis.
\newblock {\em Phys. Rev. C}, 77:015809, Jan 2008.

\bibitem{Ratkiewicz2018}
A.~Ratkiewicz~et al.
\newblock Towards neutron capture on exotic nuclei: Demonstrating $(d,p\gamma)$
  as a surrogate reaction for $(n,\gamma)$.
\newblock {\em submitted to Phys.\ Rev.\ Lett.}, 2018.

\bibitem{Potel2015}
G.~Potel, F.~M. Nunes, and I.~J. Thompson.
\newblock Establishing a theory for deuteron-induced surrogate reactions.
\newblock {\em Phys. Rev. C}, 92:034611, Sep 2015.

\bibitem{Escher2018}
J.~E. {Escher}, J.~T. {Burke}, R.~J. {Casperson}, R.~O. {Hughes}, and N.~D.
  {Scielzo}.
\newblock {One-nucleon pickup reactions and compound-nuclear decays}.
\newblock In {\em European Physical Journal Web of Conferences}, volume 178,
  page 03002, May 2018.

\bibitem{Pain2017}
S.D. Pain, A.~Ratkiewicz, T.~Baugher, M.~Febbraro, A.~Lepailleur, A.D.
  Ayangeakaa, J.~Allen, J.T. Anderson, D.W. Bardayan, J.C. Blackmon,
  R.~Blanchard, S.~Burcher, M.P. Carpenter, S.M. Cha, K.Y. Chae, K.A. Chipps,
  J.A. Cizewski, A.~Engelhardt, H.~Garland, K.L. Jones, R.L. Kozub, E.J. Lee,
  M.R. Hall, O.~Hall, J.~Hu, P.D. O’Malley, I.~Marsh, B.C. Rasco,
  D.~Santiago-Gonzales, D.~Seweryniak, S.~Shadrick, H.~Sims, K.~Smith, M.S.
  Smith, P.-L. Tai, P.~Thompson, C.~Thornsberry, R.L. Varner, D.~Walter, G.L.
  Wilson, and S.~Zhu.
\newblock Direct reaction measurements using {GODDESS}.
\newblock {\em Physics Procedia}, 90:455 -- 462, 2017.
\newblock Conference on the Application of Accelerators in Research and
  Industry, CAARI 2016, 30 October to 4 November 2016, Ft. Worth, TX, USA.

\bibitem{Lepailleur2018}
A.~Lepailleur~et al.
\newblock $^{134}${X}e$(d,p\gamma)$ with {GODDESS}.
\newblock {\em in preparation}, 2018.

\bibitem{Surman2014}
R.~Surman, M.~Mumpower, R.~Sinclair, K.~L. Jones, W.~R. Hix, and G.~C.
  McLaughlin.
\newblock Sensitivity studies for the weak r process: neutron capture rates.
\newblock {\em AIP Advances}, 4(4):041008, 2014.

\bibitem{Spy14}
A.~Spyrou, S.N. Liddick, A.C. Larsen, M.~Guttormsen, K.~Cooper, A.C. Dombos,
  D.J. Morrissey, F.~Naqvi, G.~Perdikakis, S.J. Quinn, T.~Renstr{\o}m, J.A.
  Rodriguez, A.~Simon, C.S. Sumithrarachchi, and R.G.T. Zegers.
\newblock {\em Physical Review Letters}, 113:232502, 2014.

\bibitem{Gut87}
M.~Guttormsen, T.~Rams{\o}y, and J.~Rekstad.
\newblock {\em Nucl. Instrum. Methods}, 255:518, 1987.

\bibitem{Lid16}
S.~Liddick, A.~Spyrou, B.P. Crider, F.~Naqvi, A.~C. Larsen, M.~Guttormsen,
  M.~Mumpower, R.~Surman, G.~Perdikakis, D.L. Bleuel, A.~Couture, L.~Crespo
  Campo, A.C. Dombos, R.~Lewis, S.~Mosby, S.~Nikas, C.J. Prokop,
  T.~Renstr{\o}m, B.~Rubio, S.~Siem, and S.J. Quinn.
\newblock {\em Physical Review Letters}, 116:242502, 2016.

\bibitem{Spy17}
A.~Spyrou, A.~C. Larsen, S.~Liddick, F.~Naqvi, B.P. Crider, A.C. Dombos,
  M.~Guttormsen, D.L. Bleuel, A.~Couture, L.~Crespo Campo, R.~Lewis, S.~Mosby,
  M.~Mumpower, G.~Perdikakis, C.J. Prokop, S.J. Quinn, T.~Renstr{\o}m, S.~Siem,
  and R.~Surman.
\newblock {\em Journal of Physics G}, 44:044002, 2017.

\bibitem{Lew18}
R.~Lewis, S.~Liddick, A.~Spyrou, B.P. Crider, F.~Naqvi, A.~C. Larsen,
  M.~Guttormsen, M.~Mumpower, R.~Surman, G.~Perdikakis, D.L. Bleuel,
  A.~Couture, L.~Crespo Campo, A.C. Dombos, S.~Mosby, S.~Nikas, C.J. Prokop,
  T.~Renstr{\o}m, B.~Rubio, S.~Siem, and S.J. Quinn.
\newblock {\em Physical Review C}, In preparation, 2018.

\bibitem{2012Ozawa}
A.~Ozawa, T.~Uesaka, M.~Wakasugi, and the Rare-RI Ring~Collaboration.
\newblock The rare-ri ring.
\newblock {\em Progress of Theoretical and Experimental Physics},
  2012(1):03C009, 2012.

\bibitem{2015Atanasov}
D.~Atanasov, P.~Ascher, K.~Blaum, R.~B. Cakirli, T.~E. Cocolios, S.~George,
  S.~Goriely, F.~Herfurth, H.-T. Janka, O.~Just, M.~Kowalska, S.~Kreim,
  D.~Kisler, Yu.~A. Litvinov, D.~Lunney, V.~Manea, D.~Neidherr, M.~Rosenbusch,
  L.~Schweikhard, A.~Welker, F.~Wienholtz, R.~N. Wolf, and K.~Zuber.
\newblock Precision mass measurements of $^{129--131}\mathrm{Cd}$ and their
  impact on stellar nucleosynthesis via the rapid neutron capture process.
\newblock {\em Phys. Rev. Lett.}, 115:232501, Dec 2015.

\bibitem{2015Lorusso}
G.~Lorusso, S.~Nishimura, Z.~Y. Xu, A.~Jungclaus, Y.~Shimizu, G.~S. Simpson,
  P.-A. S\"oderstr\"om, H.~Watanabe, F.~Browne, P.~Doornenbal, G.~Gey, H.~S.
  Jung, B.~Meyer, T.~Sumikama, J.~Taprogge, Zs. Vajta, J.~Wu, H.~Baba,
  G.~Benzoni, K.~Y. Chae, F.~C.~L. Crespi, N.~Fukuda, R.~Gernh\"auser,
  N.~Inabe, T.~Isobe, T.~Kajino, D.~Kameda, G.~D. Kim, Y.-K. Kim,
  I.~Kojouharov, F.~G. Kondev, T.~Kubo, N.~Kurz, Y.~K. Kwon, G.~J. Lane, Z.~Li,
  A.~Montaner-Piz\'a, K.~Moschner, F.~Naqvi, M.~Niikura, H.~Nishibata,
  A.~Odahara, R.~Orlandi, Z.~Patel, Zs. Podoly\'ak, H.~Sakurai, H.~Schaffner,
  P.~Schury, S.~Shibagaki, K.~Steiger, H.~Suzuki, H.~Takeda, A.~Wendt, A.~Yagi,
  and K.~Yoshinaga.
\newblock $\ensuremath{\beta}$-decay half-lives of 110 neutron-rich nuclei
  across the $n=82$ shell gap: Implications for the mechanism and universality
  of the astrophysical $r$ process.
\newblock {\em Phys. Rev. Lett.}, 114:192501, May 2015.

\bibitem{2018Shimizu}
Yohei Shimizu, Toshiyuki Kubo, Naoki Fukuda, Naohito Inabe, Daisuke Kameda,
  Hiromi Sato, Hiroshi Suzuki, Hiroyuki Takeda, Koichi Yoshida, Giuseppe
  Lorusso, Hiroshi Watanabe, Gary~S. Simpson, Andrea Jungclaus, Hidetada Baba,
  Frank Browne, Pieter Doornenbal, Guillaunme Gey, Tadaaki Isobe, Zhihuan Li,
  Shunji Nishimura, P.-A S\"oderstr\"om, Toshiyuki Sumikama, Jan Taprogge,
  Zsolt Vajta, Jin Wu, Zhengyu Xu, Atsuko Odahara, Ayumi Yagi, Hiroki
  Nishibata, Radomira Lozeva, Changbum Moon, and HyoSoon Jung.
\newblock Observation of new neutron-rich isotopes among fission fragments from
  in-flight fission of 345 mev/nucleon 238u: Search for new isotopes conducted
  concurrently with decay measurement campaigns.
\newblock {\em Journal of the Physical Society of Japan}, 87(1):014203, 2018.

\bibitem{MOLLER20161}
P.~Moller, A.J. Sierk, T.~Ichikawa, and H.~Sagawa.
\newblock Nuclear ground-state masses and deformations: Frdm(2012).
\newblock {\em Atomic Data and Nuclear Data Tables}, 109-110:1 -- 204, 2016.

\end{thebibliography}

\end{document}